\documentclass{cernyrep} 
\usepackage[bookmarks, colorlinks=true, linktoc=page, pdftex, linkcolor=red, citecolor=blue, urlcolor=red]{hyperref}
\pdfoutput=1
\usepackage{texnames}
\usepackage[T1]{fontenc}
\usepackage{wrapfig}

\newcommand\vect[1]{\ensuremath{\mathbf{#1}}}

\def\Lag{\ensuremath{\mathcal{L}}}
\def\Ham{\ensuremath{\mathcal{H}}}
\def\hc{\ensuremath{\mathrm{h.c.}}}
\def\Act{\ensuremath{\mathcal{A}}}

\def\echarge{\ensuremath{\mathit{e}}}
\newcommand\op[1]{\ensuremath{\hat{#1}}}
\newcommand\bra[1]{\ensuremath{\langle #1|}}
\newcommand\ket[1]{\ensuremath{|#1 \rangle}}
\newcommand\braket[2]{\ensuremath{\langle #1|#2 \rangle}}
\newcommand\commutator[1]{\ensuremath{\left[ #1 \right]}}

\newcommand{\normord}[1]{:\mathrel{#1}:}

\begin{document}

\title{Quantum Field Theory and the Electroweak Standard Model\footnote{Lectures 
given at the European School of High-Energy Physics (ESHEP), 
June 2018, Maratea, Italy}
}
 
\author{A.V. Bednyakov}

\institute{BLTP JINR, Dubna, Russia}

\begin{abstract}
These lecture notes cover the basics of Quantum Field Theory (QFT) and peculiarities in the construction of the Electroweak (EW) sector of the Standard Model (SM). 
In addition, the present status, issues, and prospects of the SM are discussed.
\end{abstract}

\maketitle % this produces the title block
 
\section{Introduction}
\label{sec:what_is_SM}

The Standard Model (SM)~\cite{Glashow:1961tr,Weinberg:1967tq,Salam:1968rm} was established in the mid-1970s. Its success is incredible: even after almost half a century, no significant deviations from the SM predictions have been found. 
\begin{center}
\emph{But what is the SM?}
\end{center}
After the discovery\cite{Aad:2012tfa,Chatrchyan:2012xdj} of the Higgs boson at the LHC, it is fair to give the following \emph{short} answer \cite{Oerter:2006iy}:
\begin{center}
	\emph{ The Absolutely Amazing Theory of Almost Everything.}
\end{center}
There are many excellent lectures (\eg \cite{Kleiss:2008zz,Iliopoulos:2013rna,Boos:2016nwm,Arbuzov:2018fza}) and textbooks
(\eg \cite{Peskin_Schroeder, Okun:2014}) that can provide a lot of convincing arguments for such a fancy name. In this course we are not able to cover all the aspects of the SM, but just review some basic facts and underlying principles of the model emphasizing salient features of the latter.

\begin{figure}
\centering\includegraphics[width=.7\linewidth]{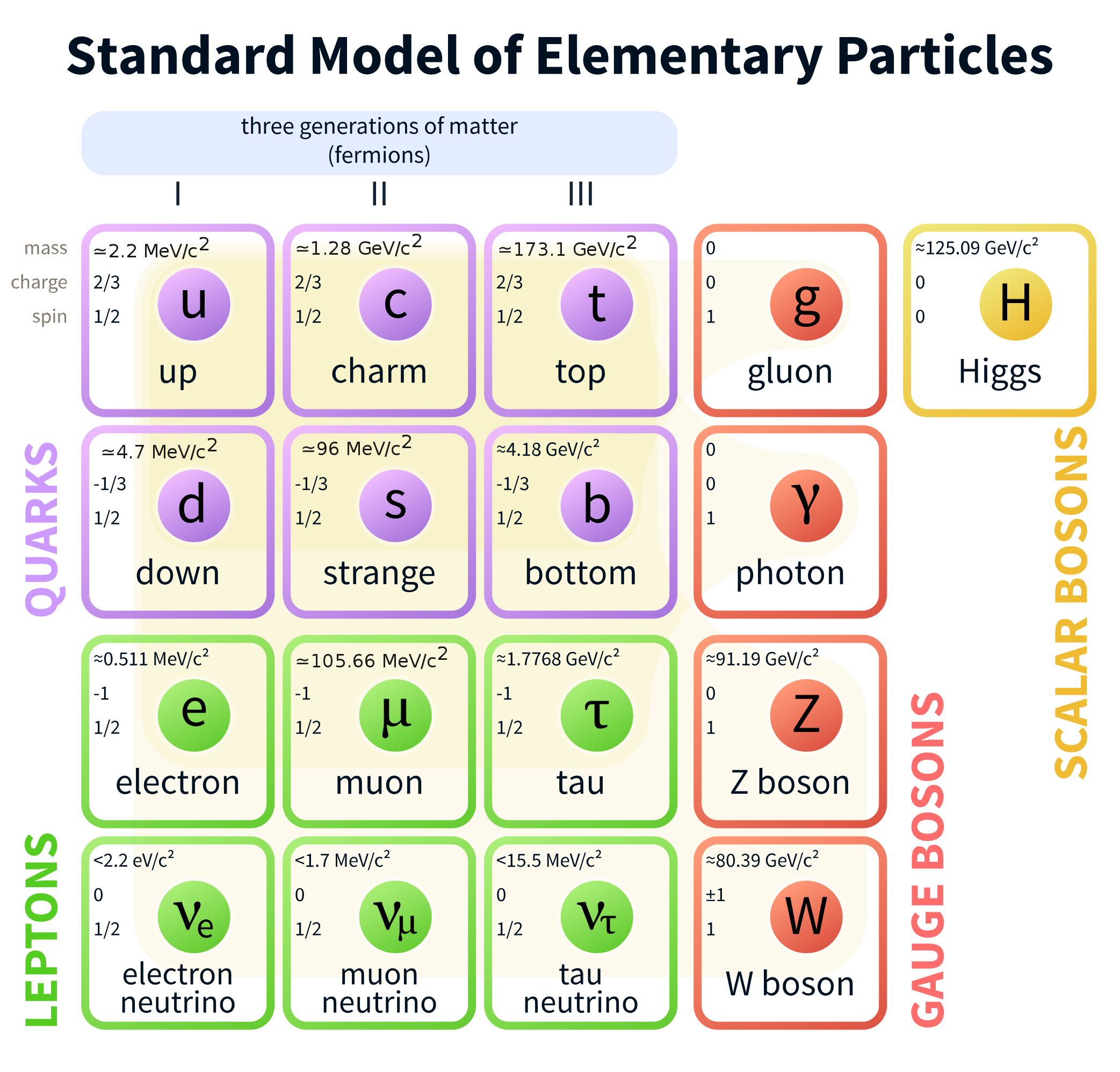}
\caption{Particle content of the Standard Model. 
	Courtesy to Wikipedia.}
	\label{fig:SM_field_content}
\end{figure}

Let us start with a brief overview of the SM particle content (see Fig.~\ref{fig:SM_field_content}).  One usually distinguishes fermions (half-integer spin) from bosons (integer spin). 
Traditionally, fermions are associated with ``matter'', while bosons take the role of ``force carriers'' that mediate interactions between spin-1/2 particles. 
In the SM, there are three \emph{generations} involving two types of fermions - \emph{quarks} and \emph{leptons}. In total, we have
\begin{itemize}
				\item  6 quarks of different flavour ($q=u,d,c,s,t,b$),
				\item  3 charged ($l=e,\mu,\tau$) and 3 neutral ($\nu_l = \nu_e, \nu_\mu, \nu_\tau$)
								leptons.
\end{itemize}
All of them participate in the weak interactions. 
Both quarks $q$ and charged leptons $l$ take part in the electromagnetic interactions.
In addition,  
quarks carry a \emph{colour} charge 
and are influenced by the strong force. 
In the SM the above-mentioned interactions are mediated 
by the exchange of spin-1 (or vector) bosons:
\begin{itemize}
				\item 8 gluons are responsible for the strong force between quarks;
				\item 4 electroweak bosons mediate the electromagnetic (photon - $\gamma$) 
								and weak ($Z,W^\pm$) interactions.
\end{itemize}
There is also a famous spin-0 Higgs boson $h$, which plays an important role in the construction of the SM. It turns out that only gluons and photons ($\gamma$) are assumed to be massless.\footnote{Initially neutrinos $\nu_l$ were assumed to be massless in the SM but experiments show that it is not the case.}
All other \emph{elementary} particles are massive due to the Higgs  mechanism.

In the SM the properties of the particle interactions can be read off the SM \emph{Lagrangian} $\Lag_{SM}$. One can find its compact version on the famous CERN T-shirt. However, there is a lot of structure behind the short expression and it is \emph{Quantum Field Theory} or QFT (see, \eg textbooks \cite{Peskin_Schroeder, Bogoliubov_Shirkov, Ryder:1985wq, Weinberg_12, Zee:2003mt}) that allows us to derive the full Lagrangian and  understand why the T-shirt Lagrangian is unique in a sense.

The form of $\Lag_{SM}$ is \emph{restricted} by various kinds of (postulated) \emph{symmetries}.
Moreover, the SM is a \emph{renormalizable} model. 
The latter fact allows us to
use \emph{perturbation theory} (PT)
to
provide high-precision predictions for thousands and thousands observables and verify the model experimentally.
All these peculiarities will be discussed during the lectures, which have the following structure.  

We begin by introducing quantum fields in Sec.~\ref{sec:quantum_fields} as the key objects of the relativistic quantum theory of particles. 
Then we discuss (global) symmetries in Sec.~\ref{sec:global_symmtries} 
and emphasize the relation between symmetries and particle properties.
We switch from free to interacting fields in Sec.~\ref{sec:interactions} and
give a brief overview of techniques used to perform calculations in QFT models.
We introduce gauge (or local) symmetries in Sec.~\ref{sec:gauge_symmetries}
and discuss how they are realized in the SM (Sec.~\ref{sec:ew_sm}).
The experimental status of the SM can be found in Sec.~\ref{sec:exp_test}.
Final remarks and conclusions are provided in Sec.~\ref{sec:conclusion}.

\section{From particles to quantum fields}
\label{sec:quantum_fields}
Before we begin our discussion of quantum fields let us set up our notation. We work in natural units with the speed of light $c=1$ and the (reduced) Planck constant $\hbar=1$. In this way, all the quantities in particle physics are expressed in powers of electron-Volts (eV). To recover ordinary units, the following conversion constants can be used:
\begin{align}
				\left[
				\raisebox{-0.5\height}{\includegraphics[width=0.1\linewidth]{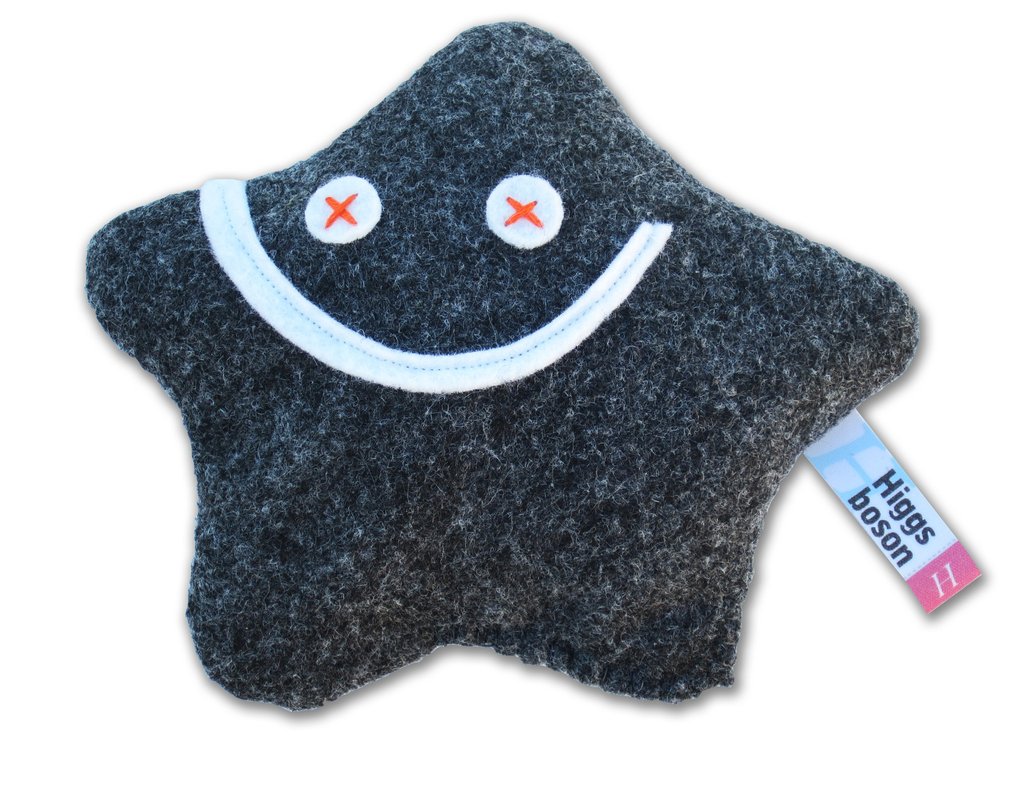}} \right] \quad 
				\hbar  \simeq  6.58 %2 119 514(40) 
				\cdot 10^{-22}~\text{MeV} \cdot s, \qquad \hbar c \simeq 1.97 %326 9788(12) 
				\cdot 10^{-14}~\text{GeV}\cdot\text{cm} \quad
				\left[
				\raisebox{-0.5\height}{\includegraphics[width=0.1\linewidth]{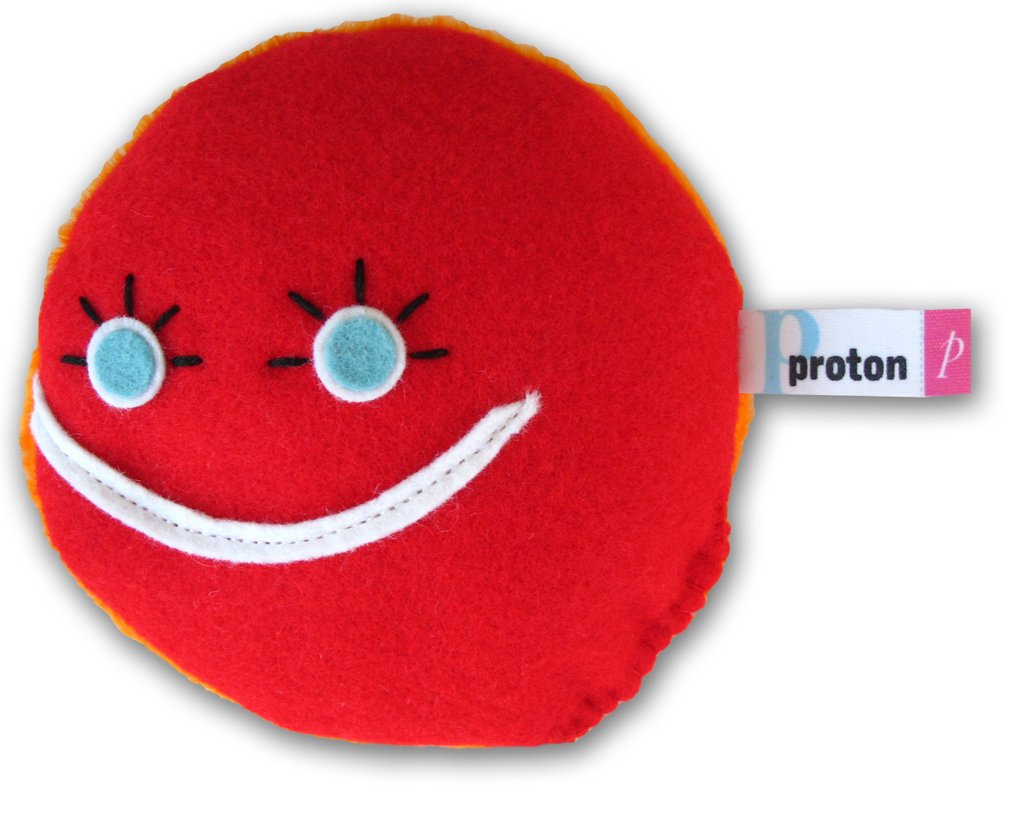}}
\right].
\end{align}
In High-Energy Physics (HEP) we routinely deal with particles traveling at speed $v\lesssim c$.
As a consequence, we require that our theory should respect Lorentz \emph{symmetry} that leaves a scalar product\footnote{Summation over repeated indices is implied.} 
\begin{align}
		px \equiv  p_\mu x_\mu = g_{\mu\nu} p_\mu  x_\nu 
		= p_0 x_0 - \vect{p} \cdot \vect{x}, \quad g_{\mu\nu} = \mbox{diag}(1,-1,-1,-1) 
	\end{align}
of any four-vectors, \eg space-time coordinates $x_\mu$ and energy-momenta $p_\mu$
	\begin{align*}
					x_\mu &= \{x_0,\vect{x}\},~\mbox{with time } t\equiv x_0, \\
					p_\mu & = \{p_0,\vect{p}\},~\mbox{with energy } E\equiv p_0,
	\end{align*}
	invariant under rotations and boosts parametrized by $\Lambda_{\mu\nu}$:
\begin{align}
			x_\mu \to x'_\mu = \Lambda_{\mu\nu} x_\nu,
			\qquad x_\mu x_\mu = x'_\mu x'_\mu \Rightarrow 
			\Lambda_{\mu\alpha} \Lambda_{\mu\beta} = g_{\alpha\beta}
			\end{align}

			It is this requirement that forces us to use QFT as a theory of \emph{relativistic} particles. Relativistic quantum mechanics (RQM) describing a fixed number of particles turns out to be inconsistent.  Indeed, from the energy-momentum relation for a free relativistic particle
\begin{align*}
				E^2 = \vect{p}^2 + m^2 \quad (\text{instead of } E=\frac{\vect{p}^2}{2m}\text{ in the non-relativistic case}),	
\end{align*}
			and the \emph{correspondence} principle 
				\begin{align*}
								E  \to i \frac{\partial}{\partial t},
					\qquad 
					\vect{p}  \to - i \vect{\nabla}
				\end{align*}
				one obtains a relativistic analog of the Shr\"odinger equation  - the Klein-Gordon (KG) equation 
				\begin{align}
								\left(\partial_t^2 - \vect{\nabla}^2 + m^2\right)\phi(t,\vect{x}) = 0
								\quad (\text{instead of } i\partial_t \psi  = -\frac{\vect{\nabla}^2}{2 m}\psi)	
				\label{eq:KG}
				\end{align}
				for a wave-function $\phi(t, \vect{x})\equiv \langle \vect{x} \ket{\phi(t)} $.
			It has two plane-wave solutions for
	any three-dimensional $\vect{p}$:	
	\begin{align}
					\phi_{\vect{p}}(t,\vect{x}) = e^{-i E t + \vect{p} \vect{x}}, \quad \mbox{with }E= \pm \omega_p, \quad \omega_p = + \sqrt{\vect{p}^2 + m^2}.
\label{eq:KG_spectrum}
				\end{align}
				One can see that the spectrum \eqref{eq:KG_spectrum} is not bounded from below. Another manifestation of this problem is the fact that for a \emph{general} wave-packet solution
		\begin{align}
			\phi(t,\vect{x}) = \frac{1}{(2\pi)^{3/2}} \int \frac{d \vect p}{\sqrt{2 \omega_p}} 
			\left[ a(\vect{p}) 
            e^{-i \omega_p t + i \vect{p} \vect{x}} 
			+ b(\vect{p}) 
    e^{+i \omega_p t - i \vect{p} \vect{x}} 
    \right]
\end{align}
we are not able to introduce a \emph{positive-definite} probability density  $\rho$ 
			\begin{align}
				%\partial_\mu j_\mu & = 0, \quad j_\mu =  i (\phi^* \partial_\mu \phi - \phi \partial_\mu\phi^*) \nonumber \\
				\rho & \equiv  j_0 = i \left( \phi^* \partial_t \phi - \phi \partial_t \phi^*\right) \Rightarrow 2 E \mbox{ for } \phi \propto e^{- i E t}, 
			\end{align}
			required to interpret $\phi$ as a wave-function of a single particle.  
			Of course, one can try to 
			impose the positive-energy condition $[b(\vect{p})\equiv 0]$ but it is not stable under \emph{interactions}. A single-particle interpretation fails to account for the appearance of negative-energy modes and we need a new formalism to deal with such situations. Moreover, in RQM space coordinates play a role of dynamical variables and are represented by operators, while time is an evolution parameter. Obviously, a \emph{consistent} relativistic theory should treat space and time on equal footing.

			In order to circumvent these difficulties,  one can re-interpret $\phi(\vect{x}, t)$ satisfying \eqref{eq:KG} as a \emph{quantum} field, \ie an \emph{operator}\footnote{We use the Heisenberg picture, in which operators $\mathcal{O}_H(t)$ depend on time, while
			in the Schr\"odinger picture it is the states that evolve:
			$\bra{\psi(t)} \mathcal{O}_S \ket{\psi(t)} = \bra{\psi} \mathcal{O}_H(t) \ket{\psi}$ with $\mathcal{O}_S = \mathcal{O}_H(t=0), \ket{\psi} = \ket{\psi(t=0)}$.}
			$\hat \phi(\vect{x},t)$. % that represent infinitely many dynamical variables. 
			The space coordinates $\vect{x}$ can be treated as a \emph{label} for infinitely many dynamical variables and we are free to choose a system of reference, in which we evolve these variables. As a consequence, a single field can account for an infinite number of particles, which are treated as field \emph{excitations}. 
			In the QFT notation the solution of the KG equation%\footnote{For brevity $\op{\phi} \to \phi$.} 
			($p_0 = \omega_p$) can be rewritten 
		\begin{align}
						\phi(x) & = \frac{1}{(2\pi)^{3/2}} \int \frac{d\vect p}{\sqrt{2 \omega_p}} 
			\left[ a^-_\vect{p}
						e^{-i p x} 
			+ b^+_\vect{p} 
    		e^{+i p x } 
    \right]
		\label{eq:fsf_operator}
		\end{align}
		as a linear combination of \emph{operators}
		$a^\pm_\vect{p}$ and $b^\pm_\vect{p}$ obeying
\begin{align}
						& a^-_\vect{p} a^+_\vect{p'} - a^+_\vect{p'} a^-_\vect{p} \equiv \left[ a^-_\vect{p}, a^+_\vect{p'} \right] = \delta^3(\vect{p} - \vect{p'}), \quad
						\left[ b^-_\vect{p}, b^+_\vect{p'} \right]  =  \delta^3(\vect{p} - \vect{p'}). \quad  
		\end{align}
		All other commutators are zero, \eg $\left[ a^\pm_\vect{p}, a^\pm_\vect{p'} \right]  = 0$. 
	The operators satisfy $a^\pm_\vect{p}=(a^{\mp}_\vect{p})^\dagger$ and $b^\pm_\vect{p} = (b^\mp_\vect{p})^\dagger$, and for $a^\pm_\vect{p}\equiv b^\pm_\vect{p}$ the field is hermitian $\phi^\dagger(x) = \phi(x)$.

	The operator \eqref{eq:fsf_operator} needs some space to act on and in QFT we consider the \emph{Fock} space. It consists of a \emph{vacuum} $|0\rangle$, which is \emph{annihilated} by $a^-_\vect{p}$ (and $b^-_\vect{p}$)	
		for every $\vect{p}$ 
\begin{align*}
						\langle 0 | 0 \rangle & = 1, 
						\quad a^-_\vect{p} |0\rangle =0, %= b^-_\vect{p} |0\rangle =0, 
						\quad ~ \langle 0|a^+_\vect{p} = (a^-_\vect{p} |0\rangle)^\dagger = 0,
						 %~ \langle 0|b^+_\vect{p} = (b^-_\vect{p} |0\rangle)^\dagger = 0,
						\label{eq:scalar_com_rel}
		\end{align*}
		and field excitations. The latter are \emph{created} from the vacuum by acting with $a^+_{\vect{k}}$ (and/or $b^+_{\vect{k}}$) , \eg
\begin{align}
								\ket{f_1 }
								& = \int d\vect k \cdot f_1(\vect{k}) a_\vect{k}^+\ket{0}  
								,	& {\text{1-particle state}}; \\
						\ket{f_2} & = 
		{\int d\vect k_1 d\vect k_2 \cdot  f_2(\vect{k}_1,\vect{k}_2) a^+_\vect{k_1} a^+_\vect{k_2}\ket{0}} & { \text{2-particle state}}, \\
		\ldots & \nonumber
\end{align}
where $f_i(\vect{k}, \ldots)$ are supposed to be square-integrable, so that, \eg $\braket{f_1}{f_1} = \int |f_1(\vect{k})|^2 d\vect{k} < \infty$. In spite of the fact that it is more appropriate to  deal with such normalizable states, in QFT we usually consider (basis) states that have definite momentum $\vect{p}$, \ie  we assume that $f_1(\vect{k}) = \delta(\vect{k} - \vect{p})$. 

The two set of operators $a^\pm$ and $b^\pm$ correspond to particles and antiparticles. From the commutation relations we deduce that $a^+_\vect{p} a^+_\vect{k} = a^+_\vect{k} a^+_\vect{p}$, so particles are \emph{not distinguishable} by construction. 
			
		The commutation relations \eqref{eq:scalar_com_rel}
		should remind us about a bunch of independent \emph{quantum harmonic oscillators}. Indeed, the corresponding Hamiltonian 
\begin{align}
		\op{\Ham}_{osc} & = 
		\sum_j\frac{1}{2} (\op{p}_j^2 + \omega_j^2 \op{x}_j^2)
			= \sum_j \frac{\omega_j}{2} \left( a^+_j a^-_j + 
										a^-_j a^+_j\right) %\quad \text{after \alert{re-ordering}} \\
						 = \sum_j\omega_j\left( \op{n}_j  + \frac{1}{2}\right)
	\label{eq:harm_osc_H}
		 %\qquad \left[\op{x}_j, \op{p}_k\right] = i \delta_{ik},~[\op{x}_j,\op{x}_k]=[\op{p}_j,\op{p}_k] = 0
\end{align}
can be expressed in terms of \emph{ladder} operators $\sqrt{2 \omega_j} a_j^\pm = (\omega_j \op{x}_j \mp i \op{p}_j)$ (no summation), which  satisfy 
		$[ a_j^-, a_k^+]  = \delta_{jk}$ similar to \Eq\eqref{eq:scalar_com_rel}. %and $\alt<2->{a_j^+}{a^+}$:
	For convenience we re-order operators entering into $\op{\Ham}_{osc}$ and introduce 
	$\op{n}_j = a_j^+ a_j^-$
	that counts energy quanta  $\op{n}_j\ket{n_j} = n_j \ket{n_j}$. 
	A direct consequence of the re-ordering is the fact that the lowest possible state (vacuum $\ket{0}$) has non-zero energy, which is equal to the sum of  zero-point energies $\sum_j\omega_j/2$ of all oscillators. 

	We can make the analogy between a (free) field and harmonic oscillators more pronounced if we put our field in a box of size $L$. In this case,  the energy $\omega_p$  and momentum $\vect{p}$ are quantized
\begin{align*}
				\vect{p} & \to \vect{p}_{\vect{j}} = (2\pi/L) \vect{j},\quad
			\omega_p  \to \omega_\vect{j} = \sqrt{ (2\pi/L)^2 \vect{j}^2 + m^2},	
				\quad \vect{j} = (j_1,j_2,j_3),\, j_i \in \mathbb{Z}.  %\left[ a^-_\vect{p}, a^+_\vect{p'} \right] = \delta^3(\vect{p} - \vect{p'})	
		\end{align*}
		The corresponding $\op{\Ham}_{osc}$ \eqref{eq:harm_osc_H} can be used to deduce the (QFT) Hamiltonian (by taking the limit $L\to\infty$):
		\begin{align*}
						\op{\mathcal{H}}_{part} = \lim_{L\to\infty} \underbrace{\left[\left(\frac{2\pi}{L}\right)^3 \sum_\vect{j}\right]}_{\int d \vect{p}} 
						\omega_{\vect{j}}
						\left[ \vphantom{ \left(\frac{2\pi}{L}\right)^3} \right.
										\left.
										\underbrace{\left(\frac{L}{2\pi}\right)^\frac{3}{2} a^+_\vect{j}}_{a^+_\vect{p}} 
										\underbrace{\left(\frac{L}{2\pi}\right)^\frac{3}{2} a^-_\vect{j}}_{a^-_\vect{p}} + 
						\frac{1}{2} \underbrace{\left(\frac{L}{2\pi}\right)^3}_{\delta(\vect{0})}  \right.
\left. \vphantom{ \left(\frac{2\pi}{L}\right)^3} \right].
		\end{align*}
		Since our field \eqref{eq:fsf_operator} involves two kinds of ladder operators, we have 
		\begin{align}
						\op{\Ham} & = \op{\Ham}_{part} + \op{\Ham}_{antipart} %= 
						= \int d\vect{p} \, \omega_p 
										\left[n_\vect{p} + \bar n_\vect{p} \right] + 
						\int d\vect{p} \, \omega_p \delta(\vect{0})
\label{eq:free_qft_H}
		\end{align}
		with $\bar n_\vect{p}\equiv b^+_\vect{p} b^-_\vect{p}$ and
		$n_\vect{p} \equiv a^+_\vect{p} a^-_\vect{p}$.
		The interpretation of the first term is straightforward: ($\bar n_\vect{p}$) $n_\vect{p}$  \emph{counts} (anti-)particles with definite momentum $\vect{p}$ %in a multi-particle state 
		and there is a sum over the corresponding energies. The second term in \Eq\eqref{eq:free_qft_H} looks disturbing. It is associated with \emph{infinite} vacuum (no particles) energy:
\begin{align*}
				E_0 = \bra{0}\op{\Ham}\ket{0} =  {\int d\vect{p} \, \omega_p \delta(\vect{0})}.
\end{align*}
Actually, there are two kinds of infinities in $E_0$:
		\begin{itemize}
						\item \makebox[1.8cm][l]{InfraRed}  (large distances, $L\to \infty$) due to $L^3 \to (2\pi)^3 \delta(\vect{0})$;
						\item \makebox[1.8cm][l]{UltraViolet} (small distances, $\vect{p},\omega_p \to \infty$).
		\end{itemize}
		One usually ``solves'' this problem by 
	introducing \emph{normal-ordered} Hamiltonians, \eg 
						\[
						\normord{\op{\Ham}_{osc}} 
		=  \frac{\omega_j}{2}\left(\normord{a_j^+ a_j^- + a_j^- a_j^+}\right) 
		= \omega_j \normord{a_j^+ a_j^-} = \omega_j a_j^+ a_j^-.  
		\]
		With $\normord{\op{\Ham}}$ we measure all energies with respect to the vacuum 
		$\mathcal{\op{H}} \to \normord{\op{\Ham}} = \op{\Ham} - \bra{0} \mathcal{\op{H}} \ket{0}$ and ignore (non-trivial) dynamics of the latter.  
In what follows we assume that operators are normal-ordered by default.

		It is easy to check that 
		$[\op{\Ham}, a^\pm_{\vect{p}}] = \pm \omega_\vect{p} a^\pm_{\vect{p}}$ and 
	  $[\op{\Ham}, b^\pm_{\vect{p}}] = \pm \omega_\vect{p} b^\pm_{\vect{p}}$. As a consequence,   	single-particle states with definite momentum $\vect{p}$
		\begin{align}
						\ket{\vect{p}} & = a^+_\vect{p} \ket{0}, \quad
						\op{\mathcal{H}} \ket{\vect{p}} = \omega_p \ket{\vect{p}},  \qquad
						\ket{\bar{\vect{p}}} = b^+_{\vect{p}} \ket{0}, \quad 
						\op{\mathcal{H}} \ket{\bar{ \vect{p}}}  = \omega_p \ket{\bar{\vect{p}}}
		\end{align}
		are eigenvectors of the Hamiltonian with \emph{positive} energies and we avoid introduction of negative energies in our formalism from the very beginning. 
		One can generalize Eq.~\eqref{eq:harm_osc_H} and  ``construct'' the momentum $\op{\vect{P}}$ 		
		and charge $\op{Q}$ operators\footnote{It is worth pointing here that by construction both $\op{Q}$ and $\op{\vect{P}}$ do not depend on time and commute. In the next section, we look at this fact from a different perspective and connect it to various symmetries.}: %, which have the following properties: 
		\begin{align}
						\op{\vect{P}} & = \int d\vect{p}
						\,\vect{p} \left[ n_\vect{p} + \bar n_\vect{p} \right], \quad 
						\op{\vect{P}} \ket{0} = 0 \ket{0},\qquad
						\op{\vect{P}} \ket{\vect{p}} = \vect{p} \ket{\vect{p}}
						\qquad
		\op{\vect{P}} \ket{\vect{p}} = 
		\vect{p} \ket{\bar{\vect{p}}},
		\label{eq:3_mom_op}
		\\
						\op{Q} & = 
						\phantom{\vect{p}}
						\int d\vect{p}
						\left[ n_\vect{p} - \bar n_\vect{p} \right], \quad 
		\op{Q} \ket{0}  = 0 \ket{0},
						\qquad
						\op{Q} \ket{\vect{p}} = + \ket{\vect{p}}
						\qquad
						\op{Q} \ket{\bar{\vect{p}}} 
						= - \ket{\bar{\vect{p}}}.
		\label{eq:charge_op}
		\end{align}
		The charge operator $\op{Q}$ distinguishes particles from anti-particles. One can
		show that the field $\phi^\dagger$ ($\phi$) increases (decreases) the charge of a  state 
				\begin{align*}
								\commutator{\op{Q},\phi^\dagger (x)} = + \phi^\dagger (x),
								\quad
								\commutator{\op{Q},\phi (x)} = - \phi (x)
				\end{align*}
		and consider the following amplitudes: 		
	\begin{center}
		\begin{tabular}{ccc}
						$t_2 > t_1: \quad \bra{0} \underbrace{\phi(x_2)}_{a^-} \underbrace{\phi^\dagger (x_1)}_{a^+} \ket{0}$ 	
						& & 
						$t_1 > t_2: \quad \bra{0} \underbrace{\phi^\dagger (x_1)}_{b^-} \underbrace{\phi(x_2)}_{b^+} \ket{0}$ 		
						\\
		Particle (charge $+1$) 
		& & 
		Antiparticle (charge $-1$) \\ 
	  propagates from $x_1$ to $x_2$	
		& & 
	  propagates from $x_2$ to $x_1$
	\end{tabular}
	\end{center}
	Both possibilities can be taken into account in one function: 
	\begin{align}
					\underbrace{\bra{0} T [\phi(x_2) \phi^\dagger(x_1) ] \ket{0}}_{-i D_c(x-y)} & \equiv 
					\theta(t_2-t_1) \bra{0} \phi(x_2) \phi^\dagger(x_1)  \ket{0} \nonumber\\[-14pt]
				& + \theta(t_1-t_2) \bra{0} \phi^\dagger(x_1) \phi(x_2)  \ket{0},
				\label{eq:feyn_prop}
	\end{align}
	with $T$ being the \emph{time-ordering} operation ($\theta(t)=1$ for $t\geq0$ and zero otherwise). 

	Equation~\eqref{eq:feyn_prop} is nothing else but the famous Feynman propagator, which has the following momentum representation:
	\begin{align}
					%& \bra{0} T [\phi(x_2) \phi^\dagger(x_1) ] \ket{0}	
					%= \\
					%& 
					D_c(x-y) = \frac{-1}{(2\pi)^4} \int {d^4 p} \frac{e^{-i p(x - y)}}{p^2 - m^2 + i \epsilon}.
	\label{eq:feyn_prop_FT}
	\end{align}
	\begin{wrapfigure}{r}{0.5\textwidth}
				\centering\includegraphics[width=0.8\linewidth]{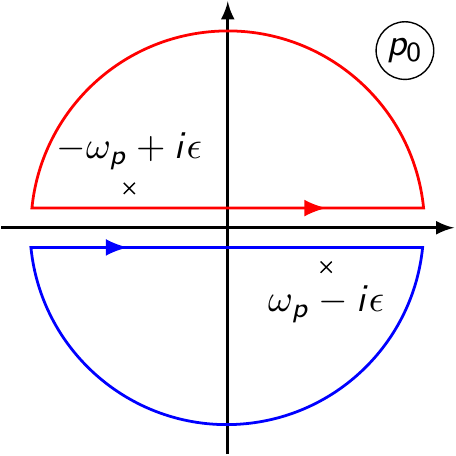}
\caption{Integration contours in $p_0$ plane.}
\label{fig:prop_complex_plane}
\end{wrapfigure}
	The {$i\epsilon$}-prescription ($\epsilon\to0$)  picks up certain poles in the complex $p_0$ plane (see Fig.~\ref{fig:prop_complex_plane}) and gives rise to the time-ordered expression \eqref{eq:feyn_prop}. 
	The propagator plays a key role in the construction of perturbation theory for interacting fields (see Sec.~\ref{sec:PT}). 

	For the moment, let us mention a couple of facts about $D_c(x)$.
	 It is a Green-function for the KG equation, \ie
	\begin{align}
					\left(\partial^2_{{x}} + m^2\right) D_c({x}-y)  
					= \delta(x-y). 
	\label{eq:KG_green_func}
	\end{align}
	This gives us an alternative way to find the expression \eqref{eq:feyn_prop_FT}. One can also see that $D_c(x-y)$ is a Lorentz and translational invariant function.

The propagator of particles can be connected to the force between two classical  static sources $J_i(\vect{x})=\delta(\vect{x}-\vect{x}_i)$ 
				located at $\vect{x}_i=(\vect{x}_1,\vect{x}_2)$. The sources disturb the vacuum $\ket{0} \to \ket{\Omega}$, since  the Hamiltonian of the system is modified $\Ham \to \Ham_0 + J \cdot \phi$.  Assuming for simplicity that
				$\phi = \phi^\dagger$, 	we can find the energy of the disturbed vacuum from 
				\begin{align*}
							\bra{\Omega} e^{-i \Ham T} \ket{\Omega} & \equiv e^{-i E_0(J) T}  \Rightarrow  \text{in the limit }T\to \infty
								\\
								&  = e^{ \frac{i^2}{2!} \int dx dy J(x) \bra{0} T(\phi(x) \phi(y)) \ket{0} J(y)}  = 
								e^{ + \frac{i}{2} \int dx dy J(x) D_c(x-y)J(y)} 
				\end{align*}
				Evaluating the integral for $J(x) = J_1(x) + J_2(x)$ and neglecting ``self-interactions``,  
				we get the contribution $\delta E_0$  to $E_0(J)$ due to interactions between two sources
				\begin{align*}
								\lim_{T\to \infty}  \delta E_0 T 
								& = - \int dx dy J_1(x) D_c(x-y) J_2(y) 
								\\
								\delta E_0 & = 
								- \int \frac{d\vect{p}}{(2\pi)^3}
								\frac{e^{+i \vect{p} (\vect{x}_1 - \vect{x}_2)}}{ \vect{p}^2 + m^2}  = {-} \frac{1}{4\pi r} e^{- m r}, \qquad r = |\vect{x_1} - \vect{x_2}|
				\end{align*}
				This is nothing else but the \emph{Yukawa} potential. %due to scalar massive field. 
				It is \emph{attractive} and \emph{falls off} exponentially over the distance scale $1/m$.  Obviously, for $m=0$ we get a Coulomb-like potential. 

\section{Symmetries and fields}
\label{sec:global_symmtries}
Let us switch to the discussion of symmetries and their role in QFT. 
A convenient way to deal with quantum fields and the symmetries of the corresponding physical systems is to consider the following \emph{Action} functional\footnote{Contrary to ordinary functions that produce numbers from numbers, a \emph{functional} takes a function and produces a number.}%, which for the considered scalar field looks like	
	\begin{align}
					\Act[\phi(x)] = \int d^4 x \, \underbrace{\Lag(\phi(x),\partial_\mu \phi)}_{\text{Lagrangian (density)}} 
					& = \int d^4 x\, \underbrace{\left( \partial_\mu \phi^\dagger \partial_\mu \phi - m^2 \phi^\dagger \phi\right)}_{\phi^\dagger \cdot  K \cdot \phi}.
					\label{eq:fsf_lag}
	\end{align}	
	To have an analogy with a mechanical system, one can rewrite $\Act[\phi]$ as 
\begin{align*}
					\Act[\phi(x)]  & = \int dt\, L(t), \qquad L = T - U, \quad H=T + U \\
					T & = \int d\vect{x} |\partial_t \phi|^2, \quad % - \text{Kinetic energy} 
					U  = \int d\vect{x} ({|\partial_\vect{x} \phi|^2} + m^2 |\phi|^2) %- \text{Potential energy} 
	\end{align*}	
	with $T$ and $U$ being kinetic and potential energy of a system of \emph{coupled} oscillators (a ``mattress'').

	Given a Lagrangian $\Lag$, one can derive the \emph{equations of motions} (EOM) via the \emph{Action Principle}. For this we consider variation of the action 
	\begin{align}
					\underbrace{
									\vphantom{\frac{\partial \mathcal{L}}{\partial \partial_\mu \phi}}
\Act[\phi'(x)] - \Act[\phi(x)]}_{\delta \Act[\phi(x)]= 0} =  
				\int d^4 x 
								\Bigg[
								\underbrace{
				\left(
												\partial_{\mu} \frac{\partial \mathcal{L}}{\partial \partial_\mu \phi}
								- \frac{\partial \mathcal{L}}{\partial \phi} 
\right)
}_{(\partial_\mu^2 + m^2) \phi = 0}
				\delta \phi
				+ \underbrace{\partial_\mu \left( 
						\frac{\partial \mathcal{L}}{\partial\partial_\mu \phi}
						\delta\phi
		\right)}_{\text{surface term=0}}
\Bigg].
	\end{align}	
	due to tiny (infinitesimal) shifts in the field $\phi'(x) = \phi(x) + \delta \phi(x)$. 
	If we require that $\delta \Act[\phi(x)]=0$ for \emph{any} variation $\delta\phi(x)$ of \emph{some} $\phi(x)$, we will immediately deduce that this can be achieved only for \emph{specific} $\phi(x)$ that satisfy EOM. 
	These \emph{particular} fields are usually called ``on-mass-shell''.    
	From the Lagrangian for our free scalar field \eqref{eq:fsf_lag} we derive the KG equation. It is related in a straightforward way to the quadratic form $K$ in \Eq\eqref{eq:fsf_lag}. Having in mind \Eq\eqref{eq:KG_green_func}, one can see that the (Feynman) propagator %of the field 
	can also be obtained by inverting $K$. This statement is easily generalized to the case of other fields. 

\begin{figure}
				\begin{center}
								\begin{tabular}{cc}
												\includegraphics[width=.3\linewidth]{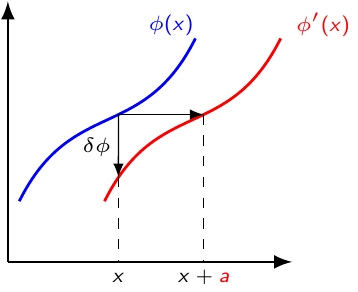} & 
												\includegraphics[width=.36\linewidth]{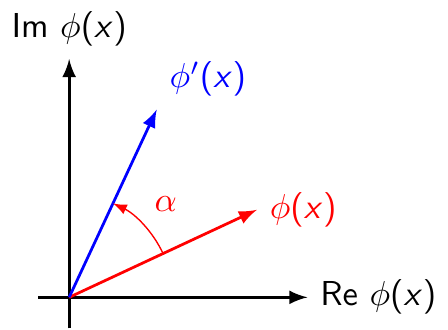} \\
$\phi'(x) = \phi(x+a)$
& 
$\phi'(x) = e^{i \alpha} \phi(x)$
								\end{tabular}
				\end{center}
				\caption{Translations (left) and phase transformations (right).}
				\label{fig:continuous_transform}
\end{figure}

 The Action functional for a physical system allows one to study \emph{Symmetries}. The latter are intimately connected with
												\emph{transformations}, which leave something \emph{invariant}.
												The transformations can be \emph{discrete}, such as   		\begin{align*}
														\text{Parity}&:         \phi'(\vect{x},t) = P \phi(\vect{x},t) = \phi(-\vect{x},t), \hfill \\
														\text{Time-reversal}&:  \phi'(\vect{x},t) = T \phi(\vect{x},t) = \phi(\vect{x},-t), \\
														\text{Charge-conjugation} &: \phi'(\vect{x},t) = C \phi(\vect{x},t) = \phi^\dagger(\vect{x},t), 
										\end{align*}
										or depend on \emph{continuous} parameters. 
										One distinguishes \emph{space-time} from \emph{internal} transformations.
										Lorentz boosts, rotations, and translations are typical examples of the former, while phase transformations belong to the latter (see Fig.~\ref{fig:continuous_transform}).  
										At the moment, we only consider \emph{global} symmetries with parameters independent of space-time coordinates and postpone the discussion of $x$-dependent or  \emph{local} (\emph{gauge}) transformations to Sec.~\ref{sec:gauge_symmetries}. 

Given $\Act[\phi]$, one can \emph{find} its \emph{symmetries}, which can be defined as  
\emph{particular} infinitesimal variations $\delta \phi(x)$ that for \emph{any} $\phi(x)$ leave $\Act[\phi]$ invariant up to a surface term (\cf~ the Action Principle)
          \begin{align*}
                  \Act[\phi'(x)] - \Act[\phi(x)] = \int d^4 x \, \partial_\mu \mathcal{K}_\mu, \quad \phi'({x}) \equiv \phi({x}) + \delta \phi({x}).
          \end{align*}
          If we compare this with the general expression
  \begin{align*}
                  \vphantom{\frac{\partial \mathcal{L}}{\partial \partial_\mu \phi}}
\Act[\phi'(x)] - \Act[\phi(x)] =  
        %0 = \delta S[\phi(x)] = 
        \int d^4 x 
                \Bigg[
                        \left(
                        \partial_{\mu} \frac{\partial \mathcal{L}}{\partial \partial_\mu \phi}
                - \frac{\partial \mathcal{L}}{\partial \phi} 
\right)
        \delta \phi            
        + \partial_\mu \left( 
            \frac{\partial \mathcal{L}}{\partial\partial_\mu \phi}
            \delta\phi
    \right)
\Bigg].
  \end{align*}
	and require in addition that $\phi$ \emph{satisfy} EOM\footnote{This requirement is crucial.}, we get a \emph{local conservation law}
\begin{equation}
        \partial_\mu J_\mu = 0, \quad
J_\mu \equiv \mathcal{K}_\mu  - \frac{\partial \Lag}{\partial\partial_\mu \phi} \delta \phi.
\label{eq:noether_1}
\end{equation}
The integration of Eq.~\eqref{eq:noether_1} over \emph{space} leads to \emph{conserved} charge:
\begin{equation}
				\frac{d}{d t} Q = 0, \qquad Q = \int d \vect{x} J_0. 
\label{eq:conserved_charge}
\end{equation}
If $\delta \phi = \rho_i \delta_i \phi$ depends on parameters $\rho_i$, we have a conservation law for every $\rho_i$. This is the essence of the \emph{first Noether theorem} \cite{Noether:1918zz}.

										A careful reader might notice that we somehow forgot about the quantum nature of our fields and in our discussion of symmetries treat them as classical objects. Let us comment on this fact. In Classical Physics \emph{symmetries} allow one to find 
										\begin{itemize} 
														\item new solutions to EOM from the given one, keeping some features of the solutions
																		(\emph{invariants}) intact;

														\item how a solution in one coordinate system (as seen by one observer) looks in other coordinates (as seen by another observer).  
										\end{itemize}
										In a quantum world a symmetry $\mathcal{S}$ guarantees that transition \emph{probabilities} $\mathcal{P}$ 
										between \emph{states} do not change upon transformation: % of the latter: %are related by the corresponding  transformation are the same: 
\begin{equation}
				\ket{A_i} \stackrel{\mathcal{S}}{\to} \ket{A'_i}, \qquad \mathcal{P}(A_i \to A_j) = \mathcal{P}(A'_i \to A'_j) , \qquad 
										|\braket{A_i}{A_j}|^2 = |\braket{A'_i}{A'_j}|^2. 
\end{equation}
										One can see that symmetries %transformations 
										can be \emph{represented} by \emph{unitary}\footnote{or anti-unitary (\eg in the case of time reversal).} operators $U$: 
\begin{equation}
										\ket{A'_i} = U \ket{A_j}, \quad  \braket{A'_i}{A'_j} = \bra{A_i} \underbrace{U^\dagger U}_{1} \ket{A_j}. 
\end{equation} 
In QFT one usually reformulates a symmetry transformation of \emph{states} as a change of \emph{operators} $\mathcal{O}_k$ via
\begin{equation}
														\bra{A_i} \mathcal{O}_k(x) \ket{A_j}  \stackrel{\mathcal{S}}{\to} \bra{A'_i} \mathcal{O}_k (x) \ket{A'_j} =
														\bra{A_i} \mathcal{O}'_k(x)\ket{A_j}, 
												%		\\
										%\bra{A_i} \mathcal{O}_k(x)\ket{A_j} & \stackrel{\mathcal{S}}{\to} \bra{A_i} \mathcal{O}_k'(x) \ket{A_j}, \quad 
														\quad \mathcal{O}_k'({x}) \equiv   U^\dagger \mathcal{O}_k({x}) U. 
\end{equation}
For example,  translational invariance leads to a relation between matrix elements of \emph{quantum} fields, \eg 
\begin{equation}
{\bra{A_i}} \phi(x) {\ket{A_j}} = 
        {\bra{A_i}} \phi'(x+a) {\ket{A_j}} =  
        {\bra{A_i}} U^\dagger (a) \phi(x+a) U(a) {\ket{A_j}}	
\end{equation}
for any states. As a consequence, any \emph{quantum} field in a translational invariant theory should satisfy 
\begin{equation}
\phi(x+a)  = U(a) \phi(x) U^\dagger(a).
\label{eq:qf_translation}
\end{equation}
One can also find similar \emph{constraints} on quantum operators due to  other symmetries. 

By means of the Noether theorem we can get almost at no cost the expressions for energy-momentum $P_\mu = \left(\Ham, \vect{P}\right)$ and charge $Q$, which we used in Sec.~\ref{sec:quantum_fields}. For example, 
$P_\mu$ is nothing else  but the conserved ``charges'', which correspond to space-time translations. 
Indeed, the Noether current in this case is just the energy-momentum tensor $T_{\mu\nu}$
\begin{align}
            \phi'(x + a) & = \phi(x), \qquad 
            \text{expand in $a$}\Rightarrow \delta \phi(x)%\equiv \phi'(x) - \phi(x) 
             = - a_\nu \partial_\nu \phi(x),  \\
            \delta \Lag(\phi(x),\partial_\mu\phi(x)) & = \partial_\nu %\mathcal{K}_\nu, \quad \mathcal{K}_\nu =  
            \left(- a_\nu \Lag\right) 
						\Rightarrow 
J_\mu  =  - a_\mu \Lag + a_\nu 
            \frac{\partial \Lag}{\partial \partial_\mu \phi} \partial_\nu \phi 
            = a_\nu T_{\mu\nu}.
    \end{align} 	
		According to \Eq~\eqref{eq:conserved_charge}, for every $a_\mu$ we have $P_\nu = \int d \vect{x} T_{0\nu}$, \ie conserved total energy-momentum. In the same way, we can apply the Noether theorem to phase transformations of our \emph{complex} field and get
		\begin{align}
                          \phi'(x) & = e^{i \alpha} \phi(x),\quad 
                          \delta\phi(x)  = \,\, i \alpha \phi(x), \quad 
													J_\mu  = i (\phi^\dagger \partial_\mu \phi - \phi \partial_\mu\phi^\dagger),\quad Q = \int d \vect{x} J_0.    
     \end{align}	

		 The corresponding quantum operators, \ie $\op{\Ham}$ \eqref{eq:harm_osc_H} or $\op{Q}$ \eqref{eq:charge_op}, are obtained\footnote{We leave this as an exercise.} (modulo ordering issues) from these (classical) expressions by plugging in quantum field %\footnote{Let us once again emphasize the crucial requirement of the Noether theorem that fields should be on-mass-shell.}
		 $\hat \phi$ from \Eq\eqref{eq:fsf_operator}.   
     %After quantisation $\phi(x,t) \to \op{\phi}(x,t)$, 
				It turns out that the charges act as \emph{generators} of symmetries, \eg for space-time translations the unitary operator from \Eq\eqref{eq:qf_translation} is given by
         \begin{equation}
								 U(a) = \exp \left(i \op{P}_\mu a_\mu\right), \qquad
                \op{\phi}(x+a) = U(a) \hat{\phi}(x) U^\dagger(a).
								\label{eq:translation_generators}
				\end{equation}
		 In addition, \emph{conserved} quantities 
		 can be used to define a convenient \emph{basis} of states, \eg we characterize our particle states by eigenvalues of $P_\mu$, and $Q$:
				\begin{equation}
                \ket{\vect{p}}\equiv \ket{\vect{p},+1},\,\ket{\bar{\vect{p}}}\equiv \ket{\vect{p},-1} \Rightarrow \op{Q} \ket{\vect{p},q} = q\ket{\vect{p},q}, \op{\vect{P}} \ket{\vect{p},q} = \vect{p} \ket{\vect{p},q}.
				\end{equation}

It is worth mentioning that some symmetries can mix fields, \eg 
\begin{align}                 
        \phi'_i(x') = S_{ij}(a) \phi_j(x)  \Rightarrow  \phi_i(x') = S_{ij}(a) U(a) \phi_j(x) U^\dagger(a), \quad x' = x'(x,a).
\end{align}
Typical examples are fields with non-zero spin: they have several components, which also change under coordinate rotations (more generally, under Lorentz transformations). Moreover, it is the Lorentz symmetry that allows us to classify fields as different \emph{representations} of the corresponding \emph{group}. 

Let us discuss this in more detail.
We can describe fields involving several degrees of freedom (per space point) 
				by adding more (and more) \emph{indices} $\phi(x) \to \Phi_\alpha^i(x)$. 
				One can split the indices into two groups: 	
				\emph{space-time} ($\alpha$) and \emph{internal} ($i$). The former are associated with space-time transformations, while the latter with transformations in the ``internal'' space:
\begin{align}
				\text{{Lorentz transform $\Lambda$}}: 
				\Phi^{'i}_\alpha({\Lambda} x) & = 
				S_{\alpha\beta}(\Lambda) \Phi^i_\beta(x), 
				\label{eq:lorentz_tranfrom_field}\\
				\text{Internal transform $a$}: 
				\phantom{\Lambda}\Phi^{'i}_\alpha(x) & = \,\, 
				U^{ij}(a)\Phi^j_\alpha(x). 
				\label{eq:internal_transform_fied}
\end{align}
A quantum field in this case can be represented as
\begin{align}
		\Phi_\alpha^i(x)  = \frac{1}{(2\pi)^{3/2}} \sum\limits_{s} %\sum\limits_{i} 
		\int \frac{d\vect p}{\sqrt{2 \omega_p}} 
		\left[ 
						u_\alpha^s(\vect{p})\, (a_\vect{p}^-)_s^i \,  
						e^{-i p x} 
						+
						v_\alpha^{*s}(\vect{p})(b^+_{\vect{p}})_s^i \, e^{+i p x } 
    \right].
		\label{eq:gen_field}
\end{align}
Here the factors $e^{\pm i p x}$ with $p_0 = \omega_\vect{p}$ (plane waves) guarantee that every component of $\Phi^i_\alpha$ satisfies the KG equation. The sum in \Eq\eqref{eq:gen_field} is over all polarization states, which are characterized by polarization ``vectors'' for particles 
$u_\alpha^s(\vect p)$ annihilated by 
$(a_\vect{p}^-)_s^i$, and anti-particles $v_\alpha^{*s} (\vect p)$ created by $(b^+_{\vect{p}})_s^i$ 
. The conjugated field $(\Phi^{i}_\alpha)^\dagger$ involves (conjugated) polarization vectors for (anti) particles that are (annihilated) created.  
Let us give a couple of examples:

\begin{itemize}
				\item Quarks are \emph{coloured} \emph{fermions} $\Psi_{{\alpha}}^{{i}}$ and, \eg $(a^-_\vect{p})^{{b}}_{s}$ annihilates the ``blue'' quark in a spin state $s$. The latter is characterized by a spinor $u_{\alpha}^s(\vect{p})$;
				\item There are \emph{eight} \emph{vector} gluons $G^{{a}}_{\mu}$. So $(a^-_\vect{p})^{{a}}_{s}$ annihilates a gluon ${a}$ in spin state $s$ having polarization 
								$u_\alpha^s(\vect{p}) \to \epsilon^s_{\mu}(\vect{p})$.
\end{itemize}
Since Lorentz symmetry plays a key role in QFT, we elaborate on some of its non-trivial representations and consider vector and fermion fields in more detail.
\subsection{Massive vector fields}
A charged Vector Field (e.g., a $W$-boson) can be written as 
\begin{align}
								W_\mu(x)  = \frac{1}{(2\pi)^{3/2}} \sum\limits_{\lambda=1}^{3} 
		\int \frac{d\vect p}{\sqrt{2 \omega_p}} 
		\left[ 
						\left( \epsilon^\lambda_\mu(\vect{p}) a^-_\lambda(\vect{p})	
						e^{-i p x} 
						+
						\epsilon^{*\lambda}_\mu(\vect{p})
						b^+_\lambda(\vect{p})\, e^{+i p x } 
		\right)
    \right].
\end{align}
				A massive spin-1 particle has %we have 
				\emph{3} independent polarization vectors, which satisfy
\begin{align*}
				p_\mu \epsilon^\lambda_\mu(\vect{p}) 
				= 0, \quad \epsilon^\lambda_\mu(\vect{p}) \epsilon_\mu^{*\lambda'}(\vect{p}) = - \delta^{\lambda\lambda'}, %\qquad p_\mu \epsilon^\lambda_\mu = 0,
				\quad \sum\limits_{\lambda=1}^{3} \epsilon^\lambda_\mu \epsilon^{*\lambda}_\nu =
				-  \left( g_{\mu\nu} - \frac{p_\mu p_\nu}{m^2} \right)\quad[ p_0 = \omega_\vect{p} ].
\end{align*}
The Feynman propagator can be found by considering  time-ordered product of two fields%(charged field) 
\begin{align}
				\bra{0} T(W_\mu(x) W^\dagger_\nu(y)) \ket{0} =     
				\frac{1}{(2\pi)^4} \int d^4 p e^{-i p(x - y)} 
				\left[
								\frac{- i\left(g_{\mu\nu} - \frac{p_\mu p_\nu}{m^2}\right)}{p^2 - m^2 + i \epsilon}  
				\right] \quad [p_0 - \text{arbitrary}]
				\label{eq:vector_prop_massive}
\end{align}
or by inverting the quadratic form of the (free) Lagrangian
\begin{align*}
				\Lag = -\frac{1}{2} W^\dagger_{\mu\nu} W_{\mu\nu} + m^2 W^\dagger_\mu W_\mu,
				\quad W_{\mu\nu} \equiv \partial_\mu W_\nu - \partial_\nu W_\mu.
\end{align*}
One can show that one of the polarization vectors $\epsilon_\mu^L\simeq p_\mu/m + \mathcal{O}(m)$ and \emph{diverges} in the limit $p_\mu\to\infty$ ($m\to0$). This indicates that one should be careful when constructing models with massive vector fields. We will return to this issue later.

\subsection{Massless vector fields}
\label{sec:massless_vectors}

				Massless (say photon) vectors are usually represented by %($\omega_p = |\vect{p}|$)
				\begin{align}
								A_\mu(x)  = \frac{1}{(2\pi)^{3/2}} \sum\limits_{\lambda=0}^{{3}} 
		\int \frac{d\vect p}{\sqrt{2 \omega_p}} 
		\left[ 
						\epsilon^\lambda_\mu(\vect{p}) a^-_\lambda(\vect{p})	
						e^{-i p x} 
						+ \hc
    \right].
		\label{eq:fvf_massless}
				\end{align}
				with 
\begin{align*}
				\epsilon^\lambda_\mu(\vect{p}) \epsilon_\mu^{*\lambda'}(\vect{p}) = g^{\lambda\lambda'}, 
				\quad
				\epsilon^\lambda_\mu(\vect{p}) \epsilon_\nu^{*\lambda}(\vect{p}) = g_{\mu\nu}, 
			%	\qquad p_\mu \epsilon^\lambda_\mu = 0,
				\quad
				[a^-_\lambda(\vect{p}), a^+_{\lambda'}(\vect{p'})] = - g_{\lambda\lambda'} \delta_{\vect{p},\vect{p}'}.
\end{align*}
The corresponding Feynman propagator can be given by 
\begin{align*}
				\bra{0} T(A_\mu(x) A_\nu(y)) \ket{0} =     
				\frac{1}{(2\pi)^4} \int d^4 p e^{-i p(x - y)} 
				\left[
								\frac{- i g_{\mu\nu} }{p^2  + i \epsilon}.
\right]
\end{align*}
In spite of the fact that we sum over four polarizations in \Eq\eqref{eq:fvf_massless} only \emph{two} of them are \emph{physical}! 
This reflects the fact that the vector-field Lagrangian in the massless case $m=0$
\begin{align*}
				\Lag = -\frac{1}{4} F_{\mu\nu} F_{\mu\nu}, \quad 
				F_{\mu\nu} \equiv \partial_\mu A_\nu - \partial_\nu A_\mu
\end{align*}
is invariant under $A_\mu \to A_\mu + \partial_\mu \alpha(x)$ for arbitrary $\alpha(x)$ (\emph{gauge} symmetry).  Additional \emph{conditions} (gauge-fixing) are needed to get rid of unphysical states.

\subsection{Fermion fields}

Spin-1/2 fermion  fields (\eg leptons) are given by\footnote{There exists a charge-conjugation matrix $C=i\gamma_2$, which relates spinors for particles $u$ and antiparticles $v$, \eg $v=C u^*$.} %($\bar{\psi} = \psi^\dagger \gamma_0$)

\begin{align*}
				\psi^{{\alpha}}(x) & 
= \frac{1}{(2\pi)^{3/2}} \int \frac{d\vect p}{\sqrt{2 \omega_p}} 
\sum_{s=1,2}
\left[ 
u^\alpha_{s}(\vect{p}) a_{s}^-(\vect{p})
						e^{-i p x} 
			+
			{v^{{\alpha}}_{s}(\vect{p})} b_{s}^+(\vect{p})
    		e^{+i p x } 
    \right],
\end{align*}
where we explicitly write the \emph{spinor} (Dirac) index $\alpha$ for $u_s$, $v_s$ and the quantum operator $\psi$.  The former satisfy the $4\times 4$ matrix (Dirac) equations 
\begin{align}
(\hat p - m) u_{s}(\vect{p}) = 0, \quad (\hat p + m) v_{s}(\vect{p}) = 0, \quad 
\hat p \equiv \gamma_\mu p_\mu, \quad {p_0 \equiv \omega_\vect{p}}  
\label{eq:dirac_eq_for_spinors}
\end{align}
and correspond to particles ($u_s$) or antiparticles ($v_s$). 
In \Eq\eqref{eq:dirac_eq_for_spinors} we use gamma-matrices
\begin{align*}
				\gamma_\mu \gamma_\nu + \gamma_\nu \gamma_\mu \equiv \left[ \gamma_\mu, \gamma_\nu \right]_{+} = 2 g_{\mu\nu}\mathbf{1}
				\quad 
				\Rightarrow
				\gamma_0^2 = \mathbf{1}, \quad \gamma_1^2 =\gamma_2^2=\gamma_3^2 = -\mathbf{1}
\end{align*}
to account for \emph{two} spin states ($s=1,2$) of particles and antiparticles. 
Fermion fields transform under the Lorentz group $x'=\Lambda x$ as (\cf~\Eq\eqref{eq:lorentz_tranfrom_field})
\begin{equation}
				\psi'(x') = \mathcal{S}_\Lambda \psi(x), \qquad \psi'(x')^\dagger = \psi(x) \mathcal{S}_\Lambda^\dagger. 
\end{equation}
		It turns out that the $4 \times 4$ matrix $\mathcal{S}_\Lambda^\dagger \neq \mathcal{S}_\Lambda^{-1}$  but $\mathcal{S}^{-1} = \gamma_0 \mathcal{S}^\dagger \gamma_0$. Due to this, it is convenient to  introduce a \emph{Dirac-conjugated} spinor $\bar{\psi}(x)\equiv \psi^\dagger \gamma_0$. The latter enters into
\begin{align*}
				\bar{\psi}'(x')\psi'(x') & = \bar{\psi}(x) \psi(x), \hspace{1.4cm} \text{Lorentz \emph{scalar}}; \\
				\bar{\psi}'(x')\gamma_\mu \psi'(x') & = \Lambda_{\mu\nu}\bar{\psi}(x)\gamma_\nu \psi(x), \quad \text{Lorentz \emph{vector}}.
\end{align*}
This allows us to convince ourselves that the Dirac Lagrangian 
\begin{align*}
				\Lag & = \bar{\psi} \left( i \hat \partial  - m \right) \psi
\end{align*}
is also a Lorentz scalar, \ie respects Lorentz symmetry.
Dirac-conjugated spinors can be used to impose Lorentz-invariant normalization on $u$ and $v$:
\begin{align*}
\bar{u}_{s}(\vect{p}) u_{r}(\vect{p}) = 2 m \delta_{rs}, 
				\qquad 
\bar{v}_{s}(\vect{p}) v_{r}(\vect{p}) = - 2 m \delta_{rs}, 
\end{align*}

An important fact about quantum fermion fields it that, contrary to the case of scalar or vector (\emph{boson}) fields, the creation/annihilation operators for fermions  $a^\pm_{s,\vect{p}}$ and antifermions $b^\pm_{s,\vect{p}}$ \emph{anticommute}: %satisfy \alert{anti}commutation relations
\begin{align*}
				\left[a^-_{r,\vect{p}},a^+_{s,\vect{p}'} \right]_{+} & = 
				\left[b^-_{r,\vect{p}},b^+_{s,\vect{p}'} \right]_{+} = \delta_{sr} \delta(\vect{p} - \vect{p'}) 
				\\
				\left[a^\pm_{r,\vect{p}},a^\pm_{s,\vect{p}'} \right]_{{+}} & = 
				\left[b^\pm_{r,\vect{p}},b^\pm_{s,\vect{p}'} \right]_{{+}} = 
				\left[a^\pm_{r,\vect{p}},b^\pm_{s,\vect{p}'} \right]_{{+}} = 0.
\end{align*}
Due to this, fermions obey the \emph{Pauli principle}, \eg $a^+_{r,\vect{p}} a^+_{r,\vect{p}}=0$.
Moreover, one can explicitly show that quantization of bosons (integer spin) with anticommutators or fermions (half-integer spin) with commutators leads to inconsistencies (violates the \emph{Spin-Statistics} theorem).

Let us continue our discussion of free fermions by emphasizing the difference between the notions of \emph{Chirality} and \emph{Helicity}.  Two independent solutions for \emph{massive} fermions ($u_{1,2}$) can be chosen to correspond to two different \emph{helicities} --- projections of spin vector $\vect{s}$ onto direction of $\vect{p}$: 
\begin{equation}
				\mathcal{H} = \vect{s}\cdot \vect{n}, \quad \vect{n} = \vect{p}/|\vect{p}|.
				\qquad
						\begin{tabular}{cc}
										&\\
										Left-Handed & Right-Handed \\
										\includegraphics[width=.1\linewidth]{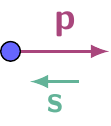}
										& 
										\includegraphics[width=.1\linewidth]{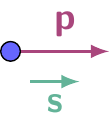}
						\end{tabular}
				\label{eq:helicity}
\end{equation}
	 In \emph{free} motion it is \emph{conserved} 
	 and serves as a good quantum number.  However, it is not a Lorentz-invariant quantity. 
	 Indeed, we can flip the sign of particle momentum by moving with speed faster than $v=|\vect{p}|/p_0$. As a consequence, $\vect{n} \to -\vect{n}$ and $\mathcal{H} \to - \mathcal{H}$.
	 However, \emph{helicity} for a \emph{massless} particle is the same for all inertial 
	 observers and coincides with \emph{chirality}, which is a \emph{Lorentz-invariant} concept. 

	\emph{By definition} Left ($\psi_L$) and Right ($\psi_R$) \emph{chiral} spinors are eigenvectors of  
	\begin{equation}
					\gamma_5  = i \gamma_0 \gamma_1 \gamma_2 \gamma_3  \Rightarrow [\gamma_\mu,\gamma_5]_+ = 0, \quad \gamma_5^2 = 1, \quad \gamma_5^\dagger = \gamma_5, 
	\end{equation}
	where 
	\begin{equation}
					\gamma_5 \psi_L  = - \psi_L, \qquad \gamma_5 \psi_R = + \psi_R.
	\end{equation}
	\emph{Any} spinor $\psi$ can be decomposed as 
\begin{equation}
				\psi = \psi_L + \psi_R, \qquad \psi_{L/R} = P_{L/R} \psi, \quad P_{L/R} = 
				\frac{1 \mp \gamma_5}{2}.
\end{equation}
			Rewriting the Dirac	Lagrangian it terms of chiral components
\begin{equation}
				\Lag = i (\underbrace{\bar{\psi}_L \hat \partial \psi_L
				+   \bar{\psi}_R \hat \partial \psi_R}_{\text{conserve chirality}}) 
				-  m (\underbrace{\bar{\psi}_L \psi_R +  \bar{\psi}_R \psi_L}_{\text{break chirality}}),
\end{equation}
	we see that, indeed, it is the mass term that mixes two chiralities. Due to this, 
	it violates \emph{chiral} symmetry corresponding to the independent rotation of left and right components
	\begin{equation}
\psi \to e^{i \gamma_5 \alpha} \psi. 
\end{equation}
	Consequently, if we drop the mass term, the symmetry of the Lagrangian is enhanced. 
	
	Up to now we were discussing the so-called Dirac mass term. For \emph{neutral} fermions (\eg neutrino) %$\psi^c = \psi$ 
there is another possibility --- a \emph{Majorana} mass. 
Since charge-conjugation applied to fermion fields,  $\psi \to \psi^c$,  \emph{flips} chirality, we can use $\psi_L^c$ in place of $\psi_R$ to write 
\begin{equation}
				\Lag = \frac{1}{2} (i \bar \psi_L \hat \partial  \psi_L - m \bar \psi_L \psi_L^c).
\end{equation}
As a consequence, to describe Majorana particles, we need only two components instead of four  since antiparticles coincide with particles in this case. At the moment, the nature of neutrinos is unclear and we refer to  \cite{Pascoli_lect} for more details.

	\section{From free to interacting fields}
\label{sec:interactions}

		 Let us summarize  what we have learned so far. If we have a Lagrangian $\Lag$ at hand, we can 
          \begin{itemize}
                  \item Derive \emph{EOM} (via the \emph{Action} Principle);
									\item Find the \emph{Symmetries} of the Action $\Act = \int d^4 x \Lag$;
                  \item Find \emph{Conserved} quantities (via the Noether Theorem).
          \end{itemize} 
 However, we usually %do not seek for symmetries of the given Action but %revert the logic and 
                \emph{start} building our models by \emph{postulating} symmetries. %symmetries (and other good properties) of the Action/Lagrangian!  
								Indeed, we assume that a general QFT Lagrangian $\Lag$ is 
          \begin{itemize}
                  \item a Lorentz (Poincare) invariant (\ie a sum of Lorentz scalars),
                  \item Local (involves a finite number of partial derivatives),
                  \item Real (hermitian) (respects unitarity=conservation of probability)
          \end{itemize}
					In addition, one can impose other symmetries and get further restrictions on the model. Having all this in mind, we can proceed further and discuss particle \emph{interactions}.

				In HEP, a typical collision/scattering experiment deals with \emph{``free''}  initial and final states and considers \emph{transitions} between these states.  To account for this in a quantum theory, one introduces the \emph{$S$-matrix} with matrix elements 
			\begin{equation}
								\mathcal{M} = \bra{\beta} S \ket{\alpha}, \qquad 
								\mathcal{M} = \delta_{\alpha\beta} + (2\pi)^4 \delta^4(p_\alpha - p_\beta) i M_{\alpha\beta}
								%S = T e^{-i\int dx \Ham_I(x)} = T e^{i \int dx \Lag_I(x)} 
			\label{eq:s_matrix_element}
				\end{equation}
			giving amplitudes
				for possible transitions between \emph{in} $\ket{\alpha}$ and \emph{out} $\ket{\beta}$ states: 
\begin{align}
				\ket{\alpha} = 
				\tilde a^+_{\vect{p}_1}... \tilde a^+_{\vect{p}_r}\ket{0}, 
				\quad
				\ket{\beta} = %(2\pi)^{3n/2} 
				\tilde a^+_{\vect{k}_1}... \tilde a^+_{\vect{k}_s}\ket{0}, 
				\quad \tilde a^+_{\vect{p}} = (2\pi)^{3/2} \sqrt{2 \omega_\vect{p}} a^+_{\vect{p}},  
				\label{eq:in_out_states}
\end{align}
where for convenience\footnote{The states created by $\tilde a^+$ are normalized in the \emph{relativistic-invariant} way.} 
 (see also \Eq\eqref{eq:Phi_ab_contraction}) we rescale our creation/annihilation operators.
Given the matrix element $M_{\alpha\beta}$, one can calculate the differential probability (per unit volume per unit time) to evolve from $\ket{\alpha}$ to $\ket{\beta}$:
\begin{equation}
				d w  =   \frac{n_1...n_r}{(2 \omega_{p_1})...
				(2 \omega_{p_r})}
				|M_{\alpha\beta}|^2 d\Phi_s, %\\ \qquad n_i - \text{particle densities} \\
\end{equation}
where $n_i$ correspond to initial-state particle densities, and an element of phase space is given by 
\begin{equation}
				d \Phi_s  = 
				(2\pi)^4 \delta^4\left(p_{in} - k_{out}\right)
				\frac{d \vect{k}_1}{(2\pi)^3 (2\omega_{k_1})}
				...	
				\frac{d \vect{k}_i}{(2\pi)^3 (2\omega_{k_i})}
				\label{eq:d_phase_space}
\end{equation}
with $p_{in} = \sum p_i$ and $k_{out} = \sum k_i$.  Since we are usually interested in processes involving one or two particles in the initial state, it is more convenient 
	to consider the differential decay width $d\Gamma$ in the rest frame of a particle with mass $m$, or cross-section $d\sigma$ of a process $2\to s$: 
\begin{align}
				d \Gamma & =  \Phi_\Gamma	|M_{1\to s}|^2 d\Phi_s, \qquad \Phi_\Gamma = \frac{1}{2 m},  \\
				d \sigma & =  \Phi_\sigma	|M|^2 d\Phi_s, \qquad \Phi_\sigma = \frac{1}{4 \sqrt{(p_1 p_2)^2 - p_1^2 p_2^2}}. 
				\label{eq:dsigma}
\end{align}
In \Eq\eqref{eq:dsigma} the factor $\Phi_\sigma$ is \emph{Lorentz-invariant} and is expressed in terms of four-momenta of initial particles $p_1$ and $p_2$.  The total width $\Gamma$  and total cross-section $\sigma$ can be obtained by integration over the
momenta of final particles restricted by energy-momentum conservation due to the four-dimensional $\delta$-function in \Eq\eqref{eq:d_phase_space}.
				
In QFT, the S-matrix is given by the time-ordered exponent
				\begin{equation}
								S = T e^{-i\int d^4x {\Ham_I}(x)} = {T} e^{i \int d^4x {\Lag_I}(x)}.
								\label{eq:s_matrix}
				\end{equation}
				involving the interaction Hamiltonian $\Ham_I$ (Lagrangian $\Lag_I$).  

				The interaction Lagrangian $\Lag_I = \Lag_{full} - \Lag_0$ is a sum of \emph{Lorentz-invariant} terms having more than \emph{two} fields and more $\partial_\mu$ than in the quadratic part $\Lag_0$, which corresponds to free particles.
				It is worth noting that in \Eq\eqref{eq:s_matrix} we treat $\Lag_I$ ($\Ham_I$)  as an operator built from \emph{free}\footnote{More precisely, operators in the \emph{interaction} picture.} quantum fields (\ie certain combinations of $a^\pm$ and $b^\pm$). 

				The \emph{time-ordering} operation, which was used to define particle propagators, is generalized in \Eq\eqref{eq:s_matrix} to account for more than two fields originating from $\Lag_I$ % is defined as 
				\begin{equation}
								{T} \Phi_1 (x_1) ... \Phi_n(x_n) = {(-1)^k} 
								\Phi_{i_1}(x_{i_1})
								...
								\Phi_{i_n}(x_{i_n}), \qquad x^0_{i_1} > ...> x^0_{i_n}.
								\label{eq:time_ordering}
				\end{equation}
				Here the factor $(-1)^k$ appears due to $k$ possible permutations of \emph{fermion} fields.

				As it was mentioned earlier, (interaction) Lagrangians should be hermitian. Any scalar combination of quantum fields can, in principle, be included in $\Lag_I$, \eg
\begin{align*}
				\Lag_I : & \quad {g} \phi^3(x), \qquad {\lambda} \phi^4(x), \qquad {y} \bar \psi(x) \psi(x) \phi(x) \\
								 &  \quad {\echarge} \bar \psi(x) \gamma_\mu \psi(x)  A_\mu(x), \qquad
				{G} \left[ (\bar \psi_1 \gamma_\mu \psi_2) \, (\bar \psi_3 \gamma_\mu \psi_4) + \hc\right] 
\end{align*}
The parameters (couplings) {$g$}, {$\lambda$}, {$\echarge$},  {$y$}, and {$G$} set the {strength} of the interactions. 
An important characteristic of any coupling in the QFT model is its \emph{dimension}, which can be deduced from the fact that Lagrangian
 has dimension $[\Lag] = 4$.
 One can notice that all the couplings (hidden) in the T-shirt Lagrangian are \emph{dimensionless}.  This fact has crucial consequences for the self-consistency of the SM model. 
				%One of the simplest, yet phenomenologically relevant, example is  $\Lag_I = - \lambda \phi^4/4!$ that describe Higgs self-interactions in the SM.

\subsection{Perturbation theory}
\label{sec:PT}

In an interacting theory it is very hard, if not impossible, to calculate the S-matrix \eqref{eq:s_matrix} exactly. Usually, we assume that the couplings in $\Lag_I$ are small allowing us to treat the terms in $\Lag_I$ as \emph{perturbations} to $\Lag_0$. 
As a consequence, we expand the T-exponent and restrict ourselves to a finite number of terms. In the simplest case of $\Lag_I = -\lambda \phi^4/4!$ we have at the $n$th order
				\begin{equation}
								\frac{i^n}{n!} 
								\left[\frac{\lambda}{4!} \right]^n
								\bra{0} \tilde a^-_{\vect{k}_1}...\tilde a^-_{\vect{k}_s}
								\int d x_1 ... d x_n
								T \left[
									\phi(x_1)^4
									...
									\phi(x_n)^4
								\right]
								\tilde a^+_{\vect{p}_1}...\tilde a^+_{\vect{p}_r} \ket{0}.
								\label{eq:PT_term}
				\end{equation}
\begin{figure}
\begin{center}
				\begin{tabular}{ccc}
				\raisebox{-0.5\height}{\includegraphics[width=0.4\linewidth]{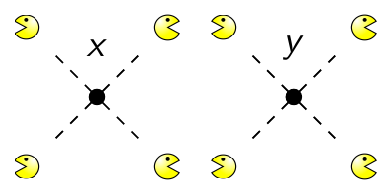}}
								& 
								$\Rightarrow$
								& 
								\raisebox{-0.49\height}{\includegraphics[width=0.4\linewidth]{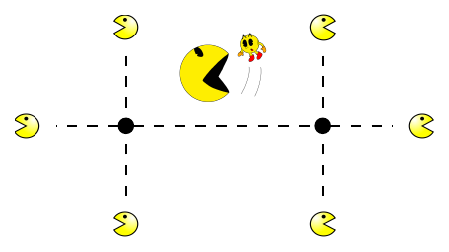}}
				\end{tabular}
\end{center}
\caption{The Wick theorem at work: one of the contributions.}
\label{fig:wick_theorem}
\end{figure}
				To proceed, one uses the \emph{Wick} theorem: 
				\begin{equation}
								T \Phi_1...\Phi_n  = \sum (-1)^\sigma 
								\bra{0} T (\Phi_{i_1} \Phi_{i_2}) \ket{0}
								...
								\bra{0} T (\Phi_{i_{m-1}} \Phi_{i_{m}}) \ket{0}
								\normord{\Phi_{i_{m+1}}...\Phi_{i_n}},
				\label{eq:wick_theorem}
				\end{equation}
				where the sum goes over all possible ways to pair the fields. The Wick theorem \eqref{eq:wick_theorem} expresses \emph{time-ordered} products of fields in terms of \emph{normal-ordered}
				ones and propagators.  				The normal-ordered operation puts \emph{all} annihilation operators originating from different $\Phi$s to the right. It also cares about fermions, \eg 
\begin{equation}
				\normord{a^-_1 a^+_2 a^-_3 a^-_4 a^+_5 a^-_6} = {(-1)}^\sigma
				a^+_2 a^+_5 a^-_1 a^-_3 a^-_4 a^-_6,
\end{equation}
where $\sigma$  correspond to the number of fermion permutations (\cf~\Eq\eqref{eq:time_ordering}).  In Fig.~\ref{fig:wick_theorem} a cartoon, which illustrates \Eq\eqref{eq:wick_theorem} for one of the contributions to $T[\Lag_I(x) \Lag_I(y)]$, is provided. 

	After application of the Wick theorem we have to calculate  
				\begin{equation}
								\bra{0} \tilde a^-_{\vect{k}_1}...\tilde a^-_{\vect{k}_s}
								\normord{\Phi_{i_{m+1}}...\Phi_{i_n}}
								\tilde a^+_{\vect{p}_1}...\tilde a^+_{\vect{p}_r} \ket{0}.
				\end{equation}
				To get a \emph{non-zero} matrix element, all $a^-(a^+)$ in the normal product of fields from the Lagrangian have to be ``killed'' by (commuted with) $a^+(a^-)$ from the initial (final) states.

				%Since all fields are \alert{linear} in $a^\pm$, one can find, e.g., 
				For our \emph{generalized} field \eqref{eq:gen_field} we have
\begin{align}
				\left[\Phi^i_\alpha(x),(a^+_{\vect{p}})^i_s\right] & = 
				\underbrace{\frac{e^{-i p x}}{(2\pi)^{3/2}\sqrt{2 \omega_p}}}_{\text{common to all fields}} {u^s_\alpha(\vect{p})}, \qquad \text{initial state polarization (particle)}; \nonumber\\
				\left[(b^-_{\vect{p}})^i_s,\Phi^i_\alpha(x)\right] & = 
				%\underbrace{
								\frac{e^{+i p x}}{(2\pi)^{3/2}\sqrt{2 \omega_p}}
				%} 
								{v^{*s}_\alpha(\vect{p})},\qquad\hspace{0.3cm} \text{final state polarization (antiparticle)}.
\label{eq:Phi_ab_contraction}
\end{align}
and one clearly sees that the factors in the denominators \Eq\eqref{eq:Phi_ab_contraction} are avoided  when the rescaled $\tilde a^\pm$ (or $\tilde b^\pm$) operators \eqref{eq:in_out_states} are used. 

All this machinery can be implemented in a set of \emph{Feynman rules}, which 
are used to draw (and evaluate) \emph{Feynman diagrams}. Every Feynman diagram involves interaction \emph{vertices}, \emph{external} and \emph{internal} lines. Internal lines connect two vertices and correspond to propagators. The expression for propagators can be derived from $\Lag_0$, \eg
												\begin{align}
																\left.
																\begin{matrix}
																				\vphantom{ {\includegraphics{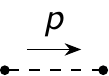}} }
																				\bra{0} T(\phi(x) \phi^\dagger(y) \ket{0}  
																				\\
																				\vphantom{ {\includegraphics{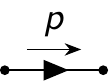}} }
																				\bra{0} T(\psi(x) \bar \psi(y) \ket{0}  
																			  \\
																				\vphantom{ {\includegraphics{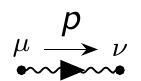}} }
																				\bra{0} T(W_\mu(x) W^\dagger_\nu(y) \ket{0}  
																\end{matrix}
												\right\}
																& = 
																\int \frac{d^4 p}{(2\pi)^4} 
																{\frac{i e^{-i p (x-y)}}{p^2 - m^2 + i \epsilon}}
																\left\{
																\begin{matrix}
																				1  & 
																				\includegraphics{pics/scalar_propagator.pdf}
																			& \phi;
																							\\ 
																				\hat p + m
																				& 	
																				\includegraphics{pics/fermion_propagator.pdf}
																			& \psi;
																				\\ 
																				- g_{\mu\nu} + p_\mu p_\nu / m^2 
																				& 	
																				\includegraphics{pics/vector_propagator.pdf}
																				& W_\mu.
																\end{matrix}
																\right.
																\label{eq:prop_feynman_rules}
												\end{align}
												One can notice that all the dependence on $x_i$ of the integrand in \Eq\eqref{eq:PT_term} comes from 
												either \Eq\eqref{eq:Phi_ab_contraction} or \Eq\eqref{eq:prop_feynman_rules}. As a consequence, it is possible to carry out the integration for \emph{every} $x_i$
\begin{equation}
				\int d^4 x_i e^{-i x_i (p_1 + ... + p_n)} = (2\pi)^4 \delta^4(p_1 + ... + p_n)
\end{equation}
											and obtain a $\delta$-function reflecting energy-momentum conservation at the corresponding vertex.
		
\begin{table}	
												\begin{center}
												\begin{tabular}{cccccc}
															incoming scalar & 1 
																			& \includegraphics{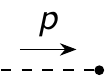} 
																			& \emph{incoming} fermion 
																			& $u_s(\vect{p})$
																			& \includegraphics{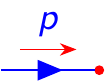} 
																			\\
															outgoing scalar & 1 
																			& \includegraphics{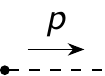} 
																			& \emph{outgoing} fermion 
																			& $\bar u_s(\vect{p})$
																			& \includegraphics{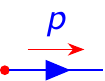} 
																			\\
											incoming vector & $\epsilon^\lambda_{\mu}(\vect{p})$ 
																			& \includegraphics{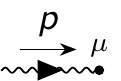} 
																			& \emph{incoming} antifermion 
																			& $\bar v_s(\vect{p})$
																			& \includegraphics{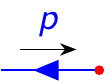} 
																			\\
											outgoing vector & $\epsilon^{*\lambda}_{\mu}(\vect{p})$ 
																			& \includegraphics{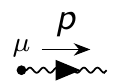} 
																			& \emph{outgoing} antifermion 
																			& $v_s(\vect{p})$
																			& \includegraphics{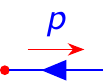} 
												\end{tabular}
\end{center}
\caption{Feynman rules for external states.}
\label{tab:ext_feynman_rules}
\end{table}
												
											Depending on the direction of momenta, the external lines represent incoming or outgoing particles (see Table~\ref{tab:ext_feynman_rules}). Again, the corresponding factors (=polarization vectors) are derived from $\Lag_0$. 
	Notice that we explicitly write the Lorentz indices for vector particles and suppress the Dirac indices for fermions. 
	To keep track of the index contractions in the latter case, one uses \emph{arrows} on the fermion lines.\footnote{There are subtleties when interactions involve Majorana fermions.} 

	Let us turn to interaction vertices. The corresponding Feynman rules can be derived from $\Act_I = \int d^4 \Lag_I$. 
	It is convenient to do this by carrying out a Fourier transform to ``convert'' coordinate derivatives to momenta and considering variations of the action. In the case of   $\Lag_I = - \lambda \phi^4/4!$ we have (all momenta are assumed to be incoming)
\begin{equation}
															\left.	i\frac{\delta^4 \Act_I[\phi]}
																{
																				\delta\phi(p_1) 
																				\delta\phi(p_2)
																				\delta\phi(p_3)
																				\delta\phi(p_4)
																}\right|_{\phi=0} \Rightarrow \underbrace{(2\pi)^4 \delta^4(p_1 + p_2 + p_3 + p_4)}_{\text{conservation of energy-momentum}} \times 
																{\left[-i \lambda \right]}.
												\end{equation}
												In a typical diagram all $(2\pi)^4 \delta(...)$ factors (but \emph{one}\footnote{we factor it out in the definition of $M_{\alpha\beta}$, see \Eq\eqref{eq:s_matrix_element}.}) reflecting the energy-momentum conservation at each vertex, are removed by the momentum integration originating from propagators, \Eq\eqref{eq:prop_feynman_rules}. Due to this, we also omit these factors (see, Table~\ref{tab:vert_feyman_rules} for examples). 
												
												Given Feynman rules, we can draw all possible diagrams that contribute to a process and evaluate the amplitude. We do not provide the precise prescription here (see textbooks \cite{Peskin_Schroeder, Bogoliubov_Shirkov, Ryder:1985wq, Weinberg_12, Zee:2003mt}  for details)  but just mention the fact that one should keep in mind various \emph{symmetry} factors and relative  \emph{signs} that can appear in real calculations.  
											
In order to get probabilities, we have to \emph{square} matrix elements, \eg 
\begin{equation}
\left|M\right|^2 = M M^\dagger \Rightarrow 
\raisebox{-0.45\height}{\includegraphics[width=0.3\linewidth]{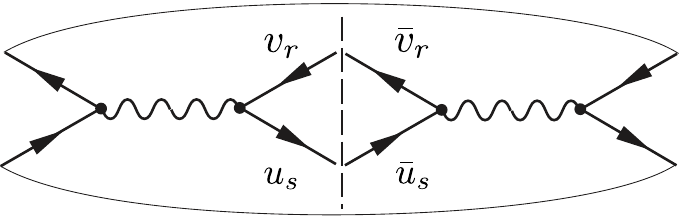}}
\end{equation}
Sometimes we do not care about polarization states of initial or final particles, so we have to \emph{sum} over \emph{final} polarization and  \emph{average} over \emph{initial} ones. 
That is where \emph{spin-sum} formulas,\eg
\begin{equation}
				\sum_s {u_s(\vect{p_1}) \bar u_s(\vect{p_1})} = \hat p_1 + m, \qquad \sum_s 
				       {v_s(\vect{p_2}) \bar v_s(\vect{p_2})} = \hat p_2 - m
\end{equation}
		 become handy 
			\begin{equation}
							M M^\dagger \to \sum_{s,r} ({\bar u_s} A {v_r}) ({\bar v_r} A^\dagger {u_s})  =
							\mathrm{Tr}\left[ ({\hat p_1} + m) A ({\hat p_2} - m) A^\dagger \right].
			\end{equation}
		As a consequence, one can utilize the well-known machinery for gamma-matrix traces to evaluate probabilities  in an efficient way.
\begin{table}	
				\begin{center}
				\begin{tabular}{p{0.27\textwidth}p{0.27\textwidth}p{0.35\textwidth}c}

\centering {$\Lag_I = - y \bar \psi \psi \phi$}
& 
\centering{$\Lag_I = \echarge \bar \psi \gamma_\mu \psi A_\mu$}
& 
\centering{$\Lag_I = i e A_\mu \left( \phi^\dagger {\partial_\mu} \phi - \phi{\partial_\mu} \phi^\dagger\right)$ }
&\\
\centering{\includegraphics{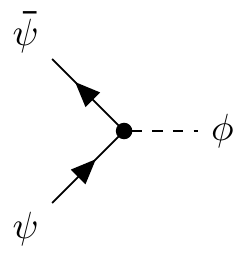}}
& 
\centering{\includegraphics{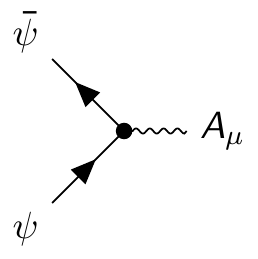}}
& 
\centering{\includegraphics{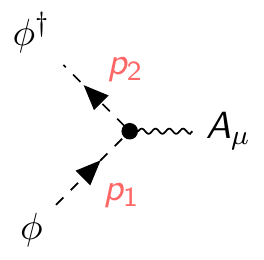}}
&\\
												\centering{$-i y$}
& 
												\centering{$i \echarge \gamma_\mu $}
&
												\centering{$i \echarge ({p_1}+{p_2})_\mu $}

				\end{tabular}
\end{center}
\caption{Vertex Feynman rules. Derivatives in $\Lag_I$ correspond to particle momenta.}
\label{tab:vert_feyman_rules}
\end{table}

			Let us continue by mentioning that only in \emph{tree} graphs, such as
												\begin{center}
												\begin{tabular}{ccc}
																\raisebox{-0.5\height}{ \includegraphics{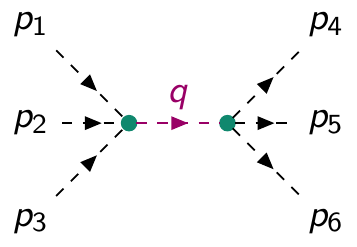} }
& $\Rightarrow$ & 
																				$(2\pi)^4 \delta^4
																				\left(\sum_{i=1}^3 p_i-\sum_{i=4}^6 p_i\right)
																				\left[-i \lambda\right]^2 
																				\frac{  i }{q^2 - m^2 + i\epsilon}$,
\end{tabular}
\end{center}
				all the integrations (due to propagators) are ``killed'' by vertex $\delta$-functions. 
				However, nothing forbids us from forming \emph{loops}. In this case, we have \emph{integrals} over unconstrained momenta, \eg in the $\phi^4$-theory
			\begin{align*}
							\raisebox{-0.5\height}{\includegraphics{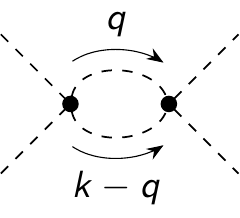}} :\qquad 
							I_2(k) \equiv  \int \frac{d^4 q}{[q^2 + i \epsilon] [(k-q)^2 + i\epsilon]} 
							 \sim \int^\infty \frac{ |q|^3 d|q|}{|q|^4} \sim \ln \infty, 
			\end{align*}
			which can lead to \emph{divergent} (meaningless?) results. This is again a manifestation of \emph{UV} divergences due to \emph{large} momenta (``small distances'').  
			
			A natural question arises: Do we have to abandon QFT? Since we still use it, there are reasons \emph{not} to do this. Indeed, we actually do not know physics up to infinitely small scales and our extrapolation can not be adequate in this case. 
			To make sense of the integrals, we can \emph{regularize} them, \eg 
			introduce a \emph{``cut-off''} $|q|<\Lambda$,
				\begin{equation}
								I_2^\Lambda(k)  = i\pi^2 \left[ 
												\ln \frac{\Lambda^2}{k^2}  + 1
								\right] + \mathcal{O}\left(\frac{k^2}{\Lambda^2}\right)
								= i \pi^2 \left[ 
												{\ln \frac{\Lambda^2}{\mu^2}}
												- \ln \frac{k^2}{{\mu^2}}   + 1
								\right] + \mathcal{O}\left(\frac{k^2}{\Lambda^2}\right)
								\label{eq:int_cutoff_reg}
				\end{equation}
				or use another convenient possibility --- \emph{dimensional} regularization, when  $d=4$ space-time is formally continued to $d=4-2\varepsilon$ dimensions:
				\begin{equation}
								I_2^{4-2\varepsilon}(k)  = \mu^{2\varepsilon} 
								\int \frac{ d^{4-2\varepsilon} q}{q^2 (k-q)^2}	
								= i \pi^2 \left( {\frac{1}{\varepsilon}} - \ln \frac{k^2}{{\mu^2}}   + 2 \right)
				+ \mathcal{O}(\varepsilon).
				\label{eq:int_dreg}
				\end{equation}
			 
				Both the regularized integrals are now convergent\footnote{We do not discuss the issue of possible IR divergences here.} and share the same logarithmic dependence on external momentum $k$. 
				One can also notice a (one-to-one) correspondence between a \emph{logarithmically} divergent contribution $\log\Lambda^2/\mu^2$ in \Eq\eqref{eq:int_cutoff_reg} and the pole term $1/\varepsilon$ in \Eq\eqref{eq:int_dreg}. However, the constant terms are \emph{different}. How do we make sense of this ambiguity?

				The crucial observation here is that the divergent pieces, which blow up when we try to remove the regulators ($\Lambda\to \infty$ or $\varepsilon\to 0$), are \emph{local}, \ie depend polynomially on external kinematical parameters. 
				This fact allows us to \emph{cancel} them by the so-called counterterm (CT) vertices. The latter can be interpreted as new terms in $\Lag_I$. Moreover, in a \emph{renormalizable} QFT model additional (divergent) contributions have the same form as the initial Lagrangian  and thus  can be ``absorbed'	into redefinition of fields and parameters. 

				One can revert the reasoning and assume that the initial Lagrangian is written in terms of the so-called \emph{bare} (unobservable) quantities. 				The predictions of the model are finite since the explicit dependence of Feynman integrals on the cut-off $\Lambda$ (or $\varepsilon$) is actually compensated by the implicit dependence of bare fields and parameters. 
				In some sense these quantities represent our ignorance of dynamics at tiny scales. Physical fields and parameters are always ``dressed'' by clouds of virtual particles.  

				It is obvious that working with \emph{bare} quantities is not very convenient. One usually makes the dependence on
				$\Lambda$ (or $\varepsilon$) explicit by introduction of divergent $Z$-factors
				for \emph{bare} fields ($\phi_B$),  masses  ($m_B^2$), and couplings ($\lambda_B$), \eg
\begin{align}
				\Lag_{full}   & = \frac{1}{2} (\partial \phi_B)^2 - \frac{m_B^2}{2} \phi_B^2  + \frac{\lambda_B \phi_B^4}{4!} = \frac{Z_2}{2} (\partial\phi)^2 - \frac{Z_m m^2}{2} Z_2 \phi^2 + \frac{Z_\lambda \lambda}{4!} (Z_2 \phi^2)^2\\
				& = \frac{(\partial \phi)^2}{2}  - \frac{m^2 \phi^2}{2} + \frac{\lambda \phi^4}{4!} 	
				+ 
				\underbrace{\frac{(Z_2-1)}{2} (\partial\phi)^2 - \frac{(Z_m Z_2 - 1) m^2}{2} \phi^2 + (Z_4 Z_2^2 - 1) \frac{\lambda \phi^4}{4!}}_{\text{counterterms}}.
				\label{eq:bare_lag_phi4}
\end{align}
Here $\phi$, $m$ and $\lambda$ denote \emph{renormalized} (finite) quantities.
				Since we can always subtract something finite from infinity, there is a certain freedom\footnote{Different constant terms in \Eq\eqref{eq:int_cutoff_reg} and \Eq\eqref{eq:int_dreg} are one manifestation of this fact.} in this procedure.
				So we have to impose additional \emph{conditions} on $Z$s, \ie define a \emph{renormalization} scheme.
				For example, in the minimal (MS) schemes  
				we subtract only the divergent terms, \eg only poles in $\varepsilon$, 
				%One the contrary, 
				while in the so-called momentum-subtraction (MOM) schemes we require amplitudes (more generally \emph{Green functions}) 
				to have a certain value at some fixed kinematics.

\begin{figure}
\centering\includegraphics[width=.6\linewidth]{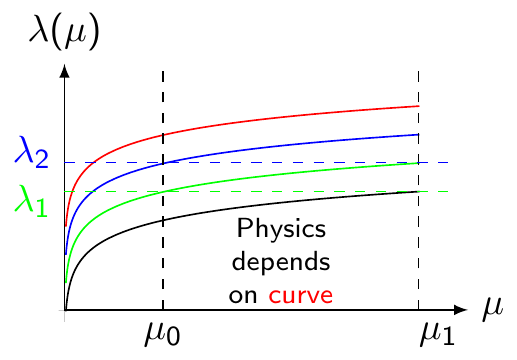}
\caption{Solutions of RGE for different boundary conditions. }
\label{fig:lambda_scale_dep}
\end{figure}

				As an illustration, let us consider a scattering amplitude $2\to2$ in the $\phi^4$ model calculated in perturbation theory: 
\begin{align}
				\raisebox{-0.46\height}{\includegraphics{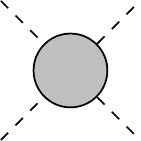}}
				& = 
				\raisebox{-0.46\height}{\includegraphics{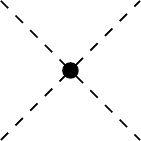}}
				+ 
				\raisebox{-0.46\height}{\includegraphics{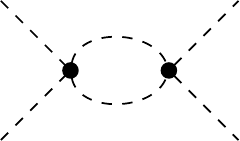}}
				+ \text{permutations}
				+ \text{more loops} 
				\label{eq:phi4_pics}\\
				& = \lambda_B(\Lambda)  
				- \frac{\lambda_B(\Lambda)^2}{2 (16 \pi^2)} \left( \ln \frac{\Lambda^2}{\mu^2}  - \ln \frac{k^2}{\mu^2} + \ldots \right) 
				+ \ldots
				\label{eq:phi4_bare}
				\\
				& =
				\left[ \lambda(\mu) + \frac{3}{2} \frac{\lambda^2(\mu)}{16 \pi^2} \ln\frac{\Lambda^2}{\mu^2} \right]
				- \frac{\lambda(\mu)^2}{2 (16 \pi^2)} \left( \ln \frac{\Lambda^2}{\mu^2}  - \ln \frac{k^2}{\mu^2} + \ldots \right) 
				+ \ldots
				\label{eq:phi4_bare_to_renorm}
				\\
				& = 
				\lambda(\mu)   
				+ \frac{\lambda(\mu)^2}{2 (16 \pi^2)} \left( \ln \frac{k^2}{\mu^2} + \ldots \right) 
				+ \ldots.
				\label{eq:phi4_renorm}
\end{align}
In \Eq\eqref{eq:phi4_pics} the tree-level and one-loop diagrams contributing to the matrix element are shown. The corresponding expression in terms of the bare coupling $\lambda_B(\Lambda)$ that implicitly depends on the regularization parameter $\Lambda$ is given in \Eq\eqref{eq:phi4_bare}. We introduce a renormalized\footnote{We use minimal subtractions here and the factor of three comes from the fact that all three one-loop graphs ($s$, $t$ and $u$) give rise to the same \emph{divergent} term.} coupling $\lambda(\mu)$ in \Eq\eqref{eq:phi4_bare_to_renorm} to make the dependence explicit: %   by introduction of a : 
\begin{equation}
				\lambda_B({\Lambda}) = \lambda(\mu) Z_\lambda = \lambda({\mu}) \left(1 + {\frac{3}{2}} \frac{\lambda({\mu})}{16\pi^2}  \ln \frac{{\Lambda^2}}{{\mu^2}} + ...\right). 
\end{equation}
The final result \eqref{eq:phi4_renorm} is finite (when $\Lambda\to \infty$) and can be confronted with experiment. It seems to depend on an auxiliary scale $\mu$, which inevitably appears in any renormalization scheme. 
The crucial point here is that \emph{observables} (if all orders of PT are taken into account) actually do \emph{not} depend on $\mu$. 
Changing $\mu$ corresponds to a certain reshuffling of the PT series: some terms from loop corrections are absorbed into the rescaled (\emph{running}) couplings. % that appear at lower orders. 
This allows one to improve  the ``convergence''\footnote{Actually, the PT series are \emph{asymptotic} (divergent) and we speak about the behavior of a limited number of first terms here.} of  the series.

The scale-dependence of the \emph{running} couplings is governed by renormalization-group equations (RGE). In the considered case we have
\begin{equation}
				\lambda(\mu_0) \to \lambda(\mu), \quad \frac{d}{d \ln \mu} \lambda = \beta_\lambda(\lambda), \quad \beta_\lambda = {\frac{3}{2}} \frac{ \lambda^2 }{16\pi^2}+ ...
	\label{eq:lambda_run}
	\end{equation}
	The \emph{beta-function} $\beta_\lambda$ can be calculated order-by-order in PT. However,
	the (initial) value $\lambda(\mu_0)$ needed to solve \Eq\eqref{eq:lambda_run} is \emph{not predicted} and has to be extracted from experiment.

\begin{figure}
\centering\includegraphics[width=.5\linewidth]{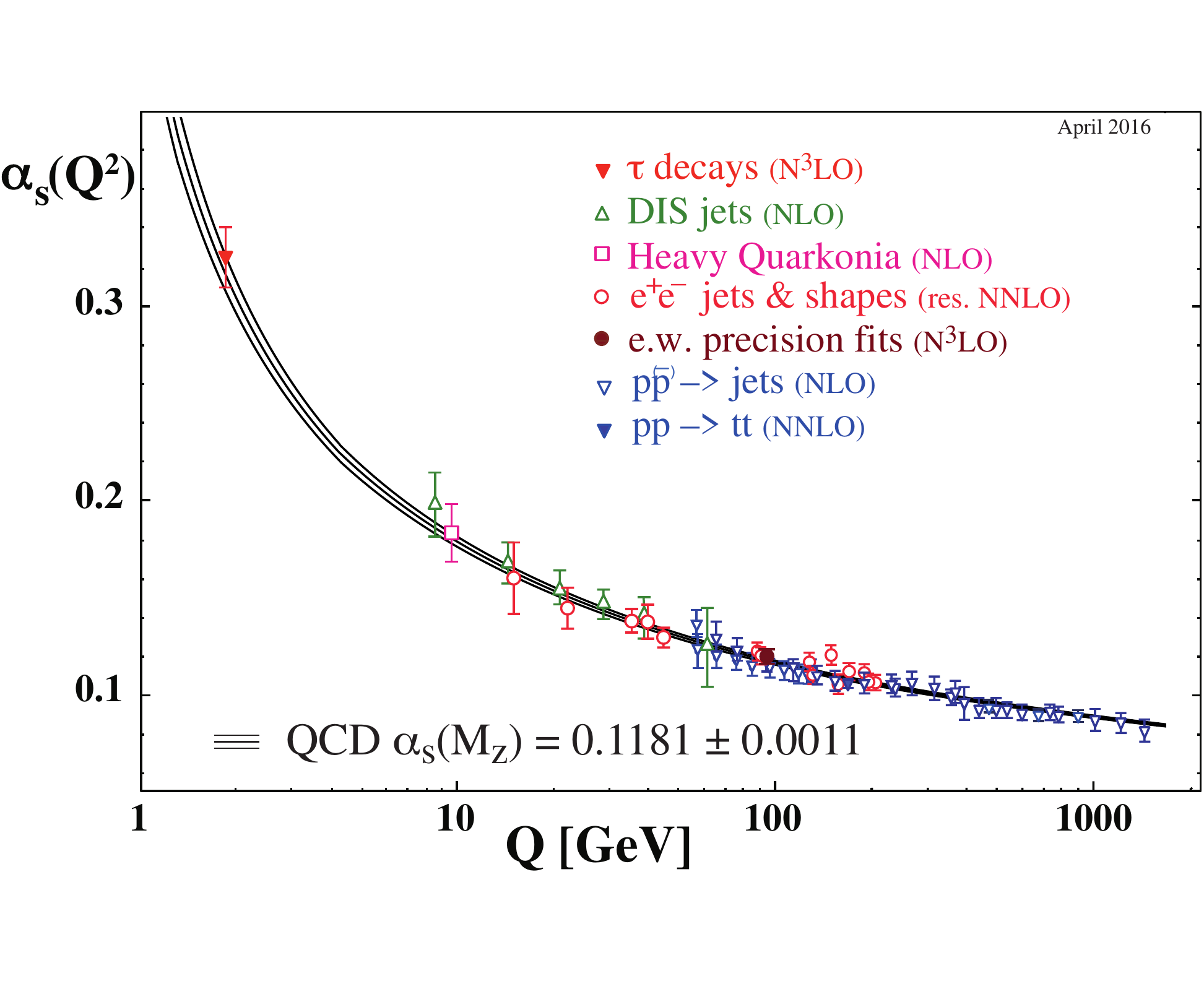}
\caption{The scale dependence of the strong coupling $\alpha_s$.}
\label{fig:as_scale_dep}
\end{figure}

	It is worth pointing out here that two different numerical values of the \emph{renormalized} self-coupling, $\lambda_1$ and $\lambda_2$, 	do not necessarily correspond to different Physics.
	Indeed, if they are fitted from measurements at different scales, \eg $\mu_0$ and $\mu$,
	and are related by means of RGE, they represent the \emph{same} Physics 
	(see Fig.~\ref{fig:lambda_scale_dep}).	
	A prominent example is the running of the strong coupling in Quantum chromodynamics (QCD) described by (see \cite{Tramontano_lect}) 
\begin{equation}
								\beta_{\alpha_s} = {-} \frac{\alpha^2_s}{4\pi} \left(11 - \frac{2}{3} n_f\right) +
								... + \mathcal{O}(\alpha^7_s), \quad n_f - \text{number of flavours}.
\end{equation}
In Fig.~\ref{fig:as_scale_dep} one can see a remarkable consistency between different measurements of $\alpha_s(\mu)$ and 
the scale dependence predicted by perturbative QCD.

\subsection{Renormalizable or non-renormalizable?}
\label{sec:renorm_vs_non_renorm}
Let us stress again that the model is called \emph{renormalizable} if \emph{all} the divergences that appear in loop integrals
can be canceled by local counterterms due to renormalization of bare parameters and couplings from $\Lag_{full}$. 
But what happens if there is a divergent amplitude but the structure of the required subtraction does not have  a counter-part in our initial Lagrangian, \ie we do not have a coupling to absorb the infinity? Obviously, we can modify $\Lag_{full}$ and \emph{add} the required term (and the coupling). 

\begin{figure}
\centering\includegraphics[width=.3\linewidth]{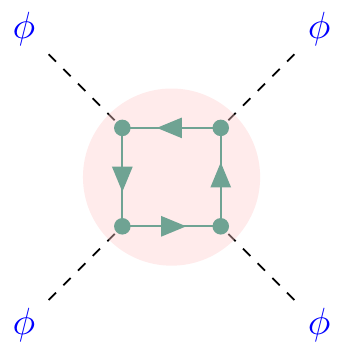}
\caption{One-loop correction to higgs self-interaction.}
\label{fig:phi4_top_contrib}
\end{figure}

An example of such a situation can be found in the model with a scalar $\phi$ (\eg Higgs) coupled to a fermion $\psi$ (\eg top quark) via the Yukawa interaction  characterized by the coupling $y$
\begin{equation}
				\Lag_I \ni \delta \Lag_{Y} = -y \cdot {\bar \psi \psi} \phi.
\end{equation}
Let us assume for the moment that we set the self-coupling to zero $\lambda=0$ and want to calculate the Higgs-scattering amplitude due to top quarks (see, Fig.~\ref{fig:phi4_top_contrib}). We immediately realize that the contribution is divergent and without $ \delta \Lag_{4} = - \lambda {\phi}^4/4!$ we are not able to cancel it. Due to this, we are forced to consider the $\phi^4$ term in a consistent theory. 

Since we modified $\Lag_{full}$, we have to re-calculate all the amplitudes. In principle, new terms in $\Lag_I$ will generate new diagrams, which can require new interactions to be added to $\Lag_I$.  Will this process terminate? 
In the case of \emph{renormalizable} models the answer is positive.  We just need to make sure that $\Lag_I$ include \emph{all} possible terms with \emph{dimensionless} couplings\footnote{Remember the T-shirt Lagrangian?}, or, \emph{equivalently}, local dimension-4 \emph{operators} built from quantum fields and their derivatives.

On the contrary, if one has to add more and more terms to $\Lag_I$, this is a signal of a  \emph{non-renormalizable} model. 
It looks like that we have to abandon such models since we need to measure an infinite number of couplings to predict something in this situation! However, it should be stressed that non-renormalizable models, contrary to renormalizable ones, involve couplings $G_i$ with \emph{negative} mass dimension $[G_i]<0$! Due to this, not all of them are important at \emph{low} energies, such as
\begin{equation}
G_i E^{-[G_i]} \ll 1.	
\end{equation}
This explains the success of the \emph{Fermi model} involving the dimension-6 four-fermion operator  
\begin{equation}
										- \Lag_I =G \bar \Psi_p \gamma_\rho \Psi_n \cdot \bar \Psi_e \gamma_\rho \Psi_\nu + \hc 
\end{equation}
in the description of the $\beta$-decay $n \to p + e^-  + \bar \nu_e$. 
Since the model turns  out to be a harbinger of the modern electroweak theory, let us consider it in more detail and discuss its features, which eventually lead to the construction of the SM.  

In 1957 R. Marshak and G.Sudarshan, R. Feynman and M. Gell-Mann modified the original Fermi theory of beta-decay to incorporate 100 \% violation of Parity discovered by C.S. Wu in 1956:
\begin{equation}
				- \Lag_{\text{Fermi}}  = \frac{G_F}{2 \sqrt 2} (J^+_\mu J^-_\mu + \hc).
				\label{eq:fermi_lag}
\end{equation}
Here the current  
\begin{equation}
				J^-_\rho  = ({V}-{A})_{\rho}^{\text{nucleons}} 
+ \overline\Psi_e \gamma_\rho \left({1}-{\gamma_5}\right)\Psi_{\nu_e}
+ \overline\Psi_\mu \gamma_\rho \left({1}-{\gamma_5}\right) \Psi_{\nu_\mu}
							+ ... 
\end{equation}
is the difference between Vector ($V$) and Axial ($A$) parts.  This kind of \emph{current-current} interactions 
		 can describe not only the proton beta-decay  but also the muon decay $\mu\to e \nu_\mu \bar \nu_e$ or the process of $\nu_e e$ - scattering.
		 Since the \emph{Fermi} constant $G_F\simeq 10^{-5}~\text{GeV}^{-1}$, from simple \emph{dimensional} grounds we have
\begin{equation}
				\sigma(\nu_e e \to \nu_e e) \propto G_F^2 s, \qquad s=(p_e + p_\nu)^2.
\end{equation}
With such a dependence on energy we eventually \emph{violate unitarity}. This is another manifestation of the fact that non-renormalizable interactions are not self-consistent. 

However, a modern view on the Fermi model treats it as an \emph{effective} field theory \cite{Georgi:1994qn} with 
	certain \emph{limits of applicability}. It perfectly describes low-energy experiments and one can fit the value of $G_F$ very precisely (see \cite{Tanabashi:2018oca}). 
\begin{figure}
\centering\includegraphics[width=.3\linewidth]{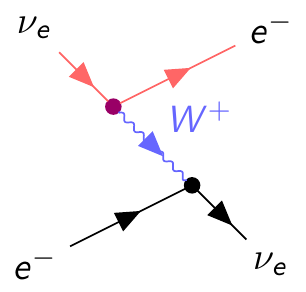}
\caption{A contribution to $\nu e$-scattering due to charged $W$-boson}
\label{fig:nu_e_scattering_W}
\end{figure}
The \emph{magnitude} of $G_F$ tells us something about a \emph{more fundamental} theory (the SM in our case): around $G_F^{-1/2} \sim 10^2-10^3$ GeV there should be some ``New Physics'' (NP) to cure the above-mentioned shortcomings. Indeed, by analogy with (renormalizable) QED we can introduce \emph{mediators} of the weak interactions -- electrically charged \emph{vector} fields $W_\mu^\pm$ (see, \eg Fig.~\ref{fig:nu_e_scattering_W}): 
\begin{equation}
				\Lag_{\text{Fermi}}  = - \frac{G_F}{2 \sqrt 2} (J^+_\mu J^-_\mu + \hc) 
				\to  \Lag_I  = \frac{g}{2 \sqrt 2} ({W^+_\mu} {J^-_\mu} + \hc) 
\end{equation}
with a \emph{dimensionless} coupling $g$. Since we know that weak interactions are \emph{short-range}, the W-bosons should be \emph{massive}. Given $\Lag_I$  we can calculate the tree-level scattering amplitude due to the exchange of $W^\pm$ between two fermionic currents:
\begin{equation}
				T = i(2\pi)^4 \, \frac{g^2}{8} J^+_\alpha \left[\frac{{g_{\alpha\beta}} - {p_\alpha} {p_\beta}/ M_W^2 }{{p^2} - {M_W^2}}\right] J^-_\beta.
				\label{eq:W_exchange_JJ}
\end{equation}
In the limit $|p|\ll M_W$, \Eq\eqref{eq:W_exchange_JJ} reproduces the prediction of the effective theory (Fermi model) if we identify (``match'') 
\begin{equation}
								\text{(effective theory)} \quad \frac{{G_F}}{\sqrt{2}} = \frac{g^2}{8 {M_W}^2}
								\quad \text{(more fundamental theory)}.
								\label{eq:Fermi_matching}
\end{equation}
However, one can see that in the UV region ($|p|\gg M_W$) the amplitude \eqref{eq:W_exchange_JJ} still has bad behavior, leading to all the above-mentioned problems. To deal with the issue, we utilize \emph{gauge} symmetry, which will be discussed in the next section.
	
\section{Gauge symmetries}

\label{sec:gauge_symmetries}
				We are seeking for a model of weak interactions that has good UV-properties.
				Let us revise how the \emph{gauge principle} is implemented in QED. First of all, 
				consider  
\begin{equation}
				\Lag_0 = \bar \psi \left( i \hat \partial - m \right) \psi
				\label{eq:free_dirac}
\end{equation}
			and make the \emph{global} $U(1)$-symmetry of $\Lag_0$
\begin{equation}
				\psi \to \psi' = e^{i e \omega} \psi
\end{equation}
\emph{local}, \ie $\omega \to \omega(x)$. In this case, the Lagrangian ceases to be invariant\footnote{Note that one can use this fact to get an expression for the Noether current $J_\mu$.}:  
\begin{equation}
				\delta \Lag_0 = \partial_\mu \omega \cdot J_\mu, \qquad J_\mu = - e \bar \psi \gamma_\mu \psi, 
\end{equation}
To compensate this term, we add the interaction of the current $J_\mu$ with the photon field $A_\mu$:
\begin{equation}
				\Lag_0 \to \Lag = \Lag_0 + A_\mu J_\mu = \bar \psi \left[i {(\hat \partial + i e \hat A)} - m\right] \psi, \qquad A_\mu \to A'_\mu = A_\mu {-} \partial_\mu \omega.
\end{equation}
The photon $A_\mu$ is an example of \emph{gauge} field. To get the full QED Lagrangian, we should also add a kinetic term for the photon: 
\begin{align}
				\Lag_{QED} & = \bar \psi \left( i \hat D - m \right) \psi - \frac{1}{4} F_{\mu\nu}^2 
			\label{eq:QED_lag}
				\\
								D_\mu & = \partial_\mu + i e A_\mu, \qquad
				F_{\mu\nu}  = \partial_\mu A_\nu - \partial_\nu A_\mu.
\end{align}
Here we introduce a \emph{covariant} derivative $D_\mu$ and a \emph{field-strength} tensor $F_{\mu\nu}$. One can check that \Eq\eqref{eq:QED_lag} is invariant  under
\begin{align*}
				\psi  & \to \psi'  = e^{i e \omega(x)} \psi \\
				A_\mu & \to A_\mu'  = A_\mu - \partial_\mu \omega \\
				D_\mu  \psi & \to D'_\mu \psi'  = e^{i e \omega(x)} D_\mu \psi. 
\end{align*}

The \emph{second} Noether theorem \cite{Noether:1918zz} states that theories possessing \emph{gauge} symmetries are \emph{redundant}, \ie some degrees of freedom are not physical. This makes quantization non-trivial. To deal with this problem in QED, one adds a \emph{gauge-fixing term} to the free vector-field Lagrangian:
\begin{equation}
				\Lag_0(A)  = -\frac{1}{4} F_{\mu\nu}^2 - {\frac{1}{2\xi} \left(\partial_\mu A_\mu\right)^2} \equiv - \frac{1}{2} A_\mu K_{\mu\nu} A_\nu.
\end{equation}
\begin{figure}
\centering\includegraphics[width=.7\linewidth]{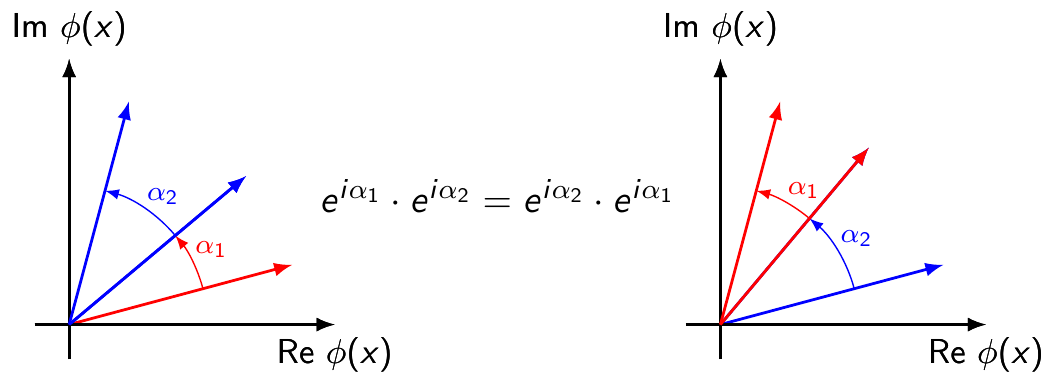}
\caption{$U(1)$ transformations commute with each other.}
\label{fig:u1_transform}
\end{figure}
This term allows one to obtain the photon propagator by inverting\footnote{If we omit the gauge-fixing term, we will not be able to invert the quadratic form.}
				$K_{\mu\nu}$:
\begin{equation}
				\bra{0} T A_\mu (x) A_\nu(y) \ket{0}  = 
				\int \frac{d^4 p}{(2\pi)^4} \frac{-i \left[ g_{\mu\nu} - (1-{\xi}) {p_\mu p_\nu}/p^2\right]}{p^2 + i \epsilon}
				e^{-i p(x-y)}
\end{equation}
The propagator now involves an auxiliary parameter ${\xi}$.  It controls the propagation of \emph{unphysical} longitudinal polarization $\epsilon^L_\mu \propto p_\mu$. The polarization turns out to be  harmless in QED since the corresponding terms \emph{drop out} 
			of physical quantities, \eg due to current conservation
\begin{equation}
				e_\mu^L J_\mu \propto p_\mu J_\mu =  0 \quad [\text{we have no source for unphysical}~\gamma]. 
\end{equation}
			One can see that the propagator has good UV behaviour and falls down as $1/p^2$ for large $p$. 
			The gauge symmetry of QED is $U(1)$. It is \emph{Abelian}  since the order of two transformations is irrelevant (see Fig.~\ref{fig:u1_transform}). 
			However, if we want to apply the \emph{gauge principle} to the case of EW interactions, we have to generalize $U(1)$ to the  \emph{Non-Abelian} case. Let us consider 
the $SU(n)$ group, i.e., unitary $n \times n$ matrices $U_{ij}$ depending on $n^2-1$ parameters $\omega^a$ and having $\det U=1$:
\begin{equation}
																\psi_i \to \psi'_i = U_{ij}(\omega) \psi_j, \quad U(\omega) = e^{i {g} {t^a} \omega^a}.
	\label{eq:non_abelian_transform}
	\end{equation}
	In general, different transformations do not commute in the non-Abelian case. This fact is reflected in commutation relations for the group \emph{generators} $t^a$, which obey the $su(n)$-algebra: 
	\begin{equation}
																[t^a, t^b] = i f^{abc} t^c, \qquad f^{abc} - \text{ structure constants }.
												\end{equation}
												For \emph{constant} $\omega^a$ the transformation \eqref{eq:non_abelian_transform} is a symmetry of  the Lagrangian
\begin{equation}
				\Lag_0 = \bar \psi_i \left( i\hat \partial - m \right) \psi_i, \qquad i=1,...,n
\end{equation}
												describing $n$ free fermions in the \emph{fundamental} representation of $SU(n)$. 

								In order to make the symmetry local, we introduce a 
												(matrix) \emph{covariant derivative} depending on $n^2-1$ gauge fields $W^a_\mu$: 
\begin{equation}
					(D_\mu)_{ij} = \partial_\mu \delta_{ij} - i {g} {t}_{ij}^a W^a_\mu.
					\label{eq:cov_deriv}
\end{equation}
The transformation properties of $W^a_\mu$ should guarantee that for space-time dependent $\omega^a(x)$ the covariant derivative of $\psi$ transforms in the same way as the field itself:
\begin{equation}
			D'_\mu \psi' =   U (\omega) (D_\mu \psi), \quad U(\omega) = e^{i g t^a \omega^a}. 
\end{equation}
		One can find that 
\begin{align}
							W^a_\mu \to W^{'a}_\mu & =  W^a_\mu + {\partial_\mu \omega^a} + {g} f^{abc}  W^b_\mu \omega^c 
						\\
												& = W^a_\mu + (D_\mu)^{ab} \omega_b, \qquad (D_\mu)^{ab} \equiv \partial_\mu \delta^{ab} - i {g} {( - i f^{abc} )}W_\mu^c,
\end{align}
where we introduce the covariant derivative \eqref{eq:cov_deriv} $D_\mu^{ab}$  with generators $(t^c)^{ab} = -i f^{cab}$ in the \emph{adjoint} representation. The field-strength tensor for each component of $W^a_\mu$ is given by the commutator 
\begin{equation}
													\left[ D_\mu, D_\nu \right]  = - i g t^a \mathcal{F}_{\mu\nu}^a, \quad
													\mathcal{F}_{\mu\nu}^a  = \partial_\mu W_\nu^a - \partial_\mu W_\nu^a + g f^{abc} W_\mu^a W_\nu^b. 
\end{equation}
Contrary to the $U(1)$ case, $\mathcal{F}_{\mu\nu}^a$ contains an additional term  quadratic in $W^a_\mu$.  Due to this,  the gauge symmetry predicts not only interactions between fermions $\psi$ (or fields in the fundamental representation of the gauge group) and $W^a_\mu$  but also \emph{self-interactions} of the latter (the gauge fields are \emph{``charged''} under the group).
	
Combining all the ingredients, we can write down the following Lagrangian for an $SU(n)$ gauge (Yang-Mills) theory : % Combining all the in %, which corresponds to \emph{self-interactions} of the gauge fields. 
\begin{align}
				\Lag & = \bar \psi \left( i \hat D - m \right) \psi - \frac{1}{4} \mathcal{F}^a_{\mu\nu} \mathcal{F}^a_{\mu\nu} 
				= \Lag_0 + \Lag_I, \\ 
				\Lag_0 & =  \bar \psi \left( i \hat \partial - m \right) - \frac{1}{4} F^a_{\mu\nu} F^a_{\mu\nu}, \quad F_{\mu\nu}^a = \partial_\mu W^a_\nu - \partial_\nu W^a_\mu, \\
						 \Lag_I & = g \bar \psi_{{\alpha}}^{{i}} 
				\gamma^\mu_{{\alpha\beta}} t_{{ij}}^{{a}} \psi^{{j}}_{{\beta}} W^{{a}}_{{\mu}}
				%\\
				%& 
				- \frac{g}{2} f^{abc} W^b_\mu W^c_\nu F^a_{\mu \nu} 
				 - \frac{g^2}{4} f^{abc} f^{ade} W^a_{\mu} W^b_{\nu} W^d_{\mu} W^e_\nu.
				 \label{eq:lag_YM_int}
\end{align}
For illustration purposes we explicitly specify all the indices in the first term of interaction Lagrangian $\Lag_I$:  the Greek ones correspond to Dirac ($\alpha, \beta$) and Lorentz ($\mu$) indices, while the Latin ones belong to different representations of $SU(n)$: $i,j$ -- fundamental, $a$ -- adjoint.  One can also see that the strength of all interactions in $\Lag_I$ is governed by the single dimensionless coupling $g$.

\begin{figure}
\centering\includegraphics[width=.7\linewidth]{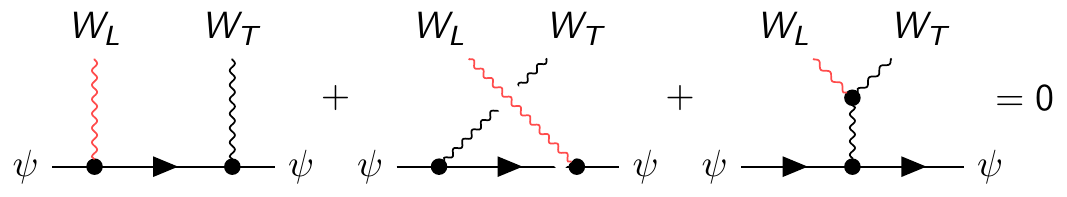}
\caption{Gauge symmetry at work: tree-level  amplitudes with unphysical polarization (L) vanish.}
\label{fig:WL_tree}
\end{figure}

	To quantize a Yang-Mills theory, we generalize the QED gauge-fixing term and write, \eg
\begin{equation}
				\Lag_{gf} = -\frac{1}{2\xi} \left(F^a\right)^2, \qquad F^a = \partial_\mu W^a_\mu 
\end{equation}	
		with $F^a$ being a gauge-fixing function. This again introduces unphysical states in 
		the $W^a_\mu$ propagator. However,  contrary to the case of QED, the \emph{fermionic} current $J_\mu^a = g \bar\psi  t^a \gamma_\mu \psi$ is not conserved and can produce longitudinal $W^a_\mu$. Nevertheless, the \emph{structure} of vector-boson self-interactions guarantees that at \emph{tree} level amplitudes involving unphysical polarizations for external $W^a_\mu$ \emph{vanish} (see, \eg  Fig.~\ref{fig:WL_tree}).

\begin{figure}
\centering\includegraphics[width=.7\linewidth]{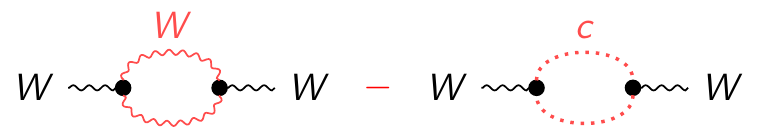}
\caption{Ghosts cancel contributions due to virtual unphysical states.}
\label{fig:WL_loop}
\end{figure}

Unfortunately, this is not sufficient to get rid of unphysical states in loops. To deal with the problem in a \emph{covariant} way, one introduces the so-called \emph{Fadeev-Popov ghosts} $\bar c_a$ and $c_a$. They are \emph{anticommuting} ``scalars'' and precisely cancel the annoying contribution.  The Lagrangian for the fictitious particles is related to the gauge-fixing function $F_a(W_\mu) = \partial_\mu W_\mu^a$ via
\begin{align}
			\Lag_{ghosts}  
			& = -\bar c^a \frac{\partial {F_a}(W^\omega)}{\partial \omega_b} c^b
			= 	
			-\bar c^a \partial_\mu D^{ab}_\mu c^b  \nonumber\\
			& = 
			-\bar c^a \partial^2 c^a - g f^{abc} (\partial_\mu \bar c^a) c^b A_\mu^b. 
\end{align}
The ghosts are charged under $SU(n)$  and interact with gauge fields  in the same way as the unphysical modes. However, there is an additional minus sign for the loops involving anticommuting ghosts (see, \eg Fig.~\ref{fig:WL_loop}) that leads to the above-mentioned cancellations.  
 
\section{Gauge theory of electroweak interactions}
\label{sec:ew_sm}

\subsection{Fermion couplings to gauge bosons}

In the SM we use the gauge principle to introduce EW interactions. Indeed, we utilize  
\begin{equation}
								SU(2)_{L} \otimes U(1)_{Y}
								\label{eq:ew_group}
\end{equation}
gauge group that has four generators or, equivalently, four gauge bosons. Three of them, $W_\mu$,  belong to \emph{weak-isospin} $SU(2)_L$, while the photon-like $B_\mu$ mediates \emph{weak-hypercharge} $U(1)_Y$ interactions. The SM fermions are charged under the group \eqref{eq:ew_group}. To account for the $(V-A)$ pattern only \emph{left} fermions interact with $W_\mu$ and form $SU(2)_L$ doublets: 
\begin{equation}
								L = \begin{pmatrix} \nu_l \\ l^- \end{pmatrix}_L, 
								Q = \begin{pmatrix} q_u \\ q_d \end{pmatrix}_L, \qquad q_u=u,c,t;~q_d=d,s,b;~l=e,\mu,\tau . 
								\label{eq:SU2L_fermions}
\end{equation}

Since the generators of $SU(2)$ are just the Pauli matrices, we immediately write the following expression for the corresponding covariant derivative
\begin{equation}
								D^L_\mu %& = \partial_\mu - i g \frac{\sigma^a}{2} W_\mu^a - i g' \frac{Y_f}{2} B_\mu \\   
								   = 
								\begin{pmatrix}
												\partial_\mu - \frac{i}{2} \left(g {W^3_\mu} + g' Y^f_L {B_\mu}\right) &  -i \frac{g}{\sqrt 2} {W_\mu^+} %\frac{ W_1  - i W_2}{\sqrt 2}  
												\\
												-i \frac{g}{\sqrt 2} {W_\mu^-} %\frac{W_1  + i W_2}{\sqrt 2} 
												& \partial_\mu + \frac{i}{2} \left(g {W^3_\mu} - g' Y^f_L {B_\mu}\right)
								\end{pmatrix}.
								\label{eq:cov_deriv_SU2}
				\end{equation}
				The \emph{right} fermions\footnote{In what follows we do not consider right-handed neutrino and refer again to \Ref\cite{Pascoli_lect}.} are $SU(2)_L$ singlets and do not couple to $W_\mu$: 
				\begin{equation}
								D^R_\mu  = \partial_\mu - i g' \frac{Y^f_R}{2} {B_\mu}.   
				\end{equation}
				The covariant derivatives involve two gauge couplings $g$, $g'$ corresponding to $SU(2)_L$ and $U(1)_Y$, respectively. Different $Y^f_{L/R}$ denote weak hypercharges of the fermions and up to now the values are not fixed. Let us put some constraints on $Y^f_{L/R}$. The first restriction comes from the $SU(2)_L$ symmetry, \ie 
				$Y^u_{L} = Y^d_{L} \equiv Y^Q_{L}$, and $Y^\nu_L = Y^e_L \equiv Y^L_l$.

				One can see that the EW interaction Lagrangian
				\begin{equation}
								\Lag_W = \Lag_{NC} + \Lag_{CC}, 
								\label{eq:WEAK_lag}
				\end{equation}
				in addition to the \emph{charged-current} interactions of the form  
\begin{equation}
				\Lag^l_{CC}  = \frac{g}{\sqrt2} \bar \nu^{e}_{L} \gamma_\mu W^+_\mu e_L + \hc = \frac{g}{2 \sqrt 2}\bar \nu_e 
				\gamma_\mu W_\mu^+ \left(1 - \gamma_5\right) e + \hc
\end{equation}
also involves %there are  
\emph{neutral-current} interactions %in the SM
\begin{align}
				\Lag^l_{NC} & = 
				\bar \nu_L^e \gamma_\mu \left( 
				{ \frac{1}{2}} g W^3_\mu + {\frac{Y^{l}_L}{2}} g' B_\mu \right) \nu^e_L %\\
				%& 
				+
				\bar e_L \gamma_\mu \left( 
				{-\frac{1}{2}} g W^3_\mu + {\frac{Y^{l}_L}{2}} g' B_\mu \right) e_L %\\
				%& 
				+ g' \bar e_R \gamma_\mu {\frac{Y^e_R}{2} } B_\mu e_R.
\end{align}
It is obvious that we have to account for QED in the SM and should predict a photon field that couples to fermions with the correct values of the electric charges. Since both $W^3_\mu$ and $B_\mu$ are \emph{electrically neutral}, they can mix
\begin{align}
				W_\mu^3 & = \phantom{-}Z_\mu \cos \theta_W  + A_\mu \sin \theta_W\nonumber\\
				B_\mu & = -Z_\mu \sin \theta_W  + A_\mu {\cos \theta_W}. 
				\label{eq:W3_B_mixing}
\end{align}
Here we introduce the \emph{Weinberg} angle $\theta_W$. One can try to fix $\sin\theta_W$ and 
various $Y^f_{L/R}$ from the requirement that, \eg  $A_\mu$ has the same interactions as the photon in QED. Indeed, given fermion \emph{electric} charges $Q_f$ (see Fig.~\ref{fig:SM_field_content}) in the units of the elementary charge $\echarge$, one can derive the following relations: 
\begin{align}
				g {\sin \theta_W}  & = \echarge (Q_\nu - Q_e) = \echarge (Q_u - Q_d),  \nonumber\\
				g' Y^l_L {\cos \theta_W} & = \echarge (Q_\nu + Q_e)  = - \echarge, \nonumber\\
				g' Y^Q_L {\cos \theta_W} & = \echarge(Q_u + Q_d)  = \frac{1}{3} \echarge, \nonumber\\ 
				g' Y^f_R {\cos \theta_W} & = 2 \echarge Q_f, \qquad f=e,\,u,\,d. 
				\label{eq:Y_relations}
\end{align}
		As a consequence, $\echarge = g \sin\theta_W$ and, \eg $e = 3 g' Y_L^Q \cos \theta_W$, so that 	
\begin{equation}
				Y_L^l  = - 3 {Y_L^Q},  \quad
				Y_R^e  = - 6 {Y_L^Q}, \quad
				Y_R^u  =   4 {Y_L^Q}, \quad
				Y_R^d  = - 2  Y_L^Q
				\label{eq:Yf_YQ}
\end{equation}
are fixed in terms of one (arbitrary chosen) $Y_L^Q$. It is convenient to normalize the $U(1)_Y$ coupling $g'$ so that $e=g' \cos \theta_W$, so $Y_L^Q = 1/3$. 
As a consequence, the photon field couples to the electric charge $Q_f$ of a fermion $f$. The latter is related to the weak hypercharge %$Y_f$ 
and the third component of weak isospin $T^f_3$ via the Gell-Mann--Nishijima formula:
\begin{align}
				\Lag_{NC} & \ni 
				\bar f 
				\left[ \left( g T^f_3 \sin \theta_W + g' \frac{Y^L_f}{2} \cos \theta_W \right)  P_L +
							 \left( g' \frac{Y^R_f}{2} \cos \theta_W \right)  P_R
			 \right]
								\gamma_\mu f
				A_\mu \\
				& = e \bar f \left( T_3 + \frac{Y}{2} \right) \gamma_\mu f A_\mu  = e Q_f \bar f \gamma_\mu f A_\mu,
				\label{eq:GN_relation}
\end{align}
where in \Eq\eqref{eq:GN_relation} we assume that $T_3$ and $Y$ are operators, which give $T_3^f$ and $Y_L^f$, when 
acting on left components, and $T_3 = 0$ and  $Y_R^f=2 Q_f$ for right fermions.

\begin{figure}
				\begin{center}
								\includegraphics{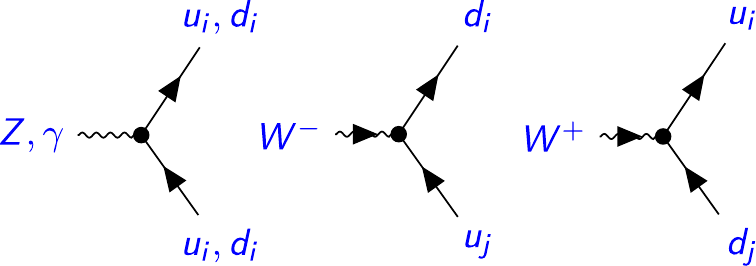}
				\end{center}
				\caption{Gauge-boson--quark vertices. Leptons interact with the EW bosons in the same way.}
				\label{fig:int_ffV}
\end{figure}

The relations \eqref{eq:Yf_YQ} allow one to rewrite the neutral-current Lagrangian as 
\begin{equation}
				\Lag_{NC}  = \echarge J^A_\mu A^\mu   + \frac{g}{\cos \theta_W} J_\mu^Z Z_\mu,
				\label{eq:nc_lag}
\end{equation}
where the photon $A_\mu$ and a new $Z$-boson couple to the currents of the form
\begin{align}
								J^A_\mu & =  \sum_f Q_f \bar f \gamma_\mu f, \qquad 
								J^Z_\mu  = \frac{1}{4} \sum_f \bar f \gamma_\mu \left( {v_f} - {a_f} \gamma_5\right) f, \\
								{v_f} &  = 2 T_3^f - 4 Q_f \sin^2\theta_W, \qquad {a_f} = 2 T_3^f,
								\label{eq:ffZ_couplings}
\end{align}
where $T_3^f = \pm \frac{1}{2}$ for left up-type/down-type fermions. For example, in the case 
of $u$-quarks, $Q_u = 2/3$, $T_3^u = 1/2$, so
\begin{equation}
							v_u = 1 - \frac{8}{3} \sin^2 \theta_W, \qquad a_u = 1.
\end{equation}
For completeness, let us give the expression for  the charged-current interactions in the EW model
\begin{equation}
				\Lag_{CC}  =  \frac{g}{\sqrt 2} \left(J_\mu^+ W^{+\mu} + J^-_\mu W^{-\mu}\right), \quad
				J^+_\mu  = \frac{1}{2} \sum_f  \bar f_u \gamma_\mu \left(1 - \gamma_5\right) f_d,
\end{equation}
where $f_u(f_d)$ is the up-type (down-type) component of an $SU(2)_L$ doublet $f$. 
The corresponding interaction vertices are given in Fig.~\ref{fig:int_ffV}. It is worth emphasizing that in the SM the couplings between fermions and gauge bosons exhibit \emph{Universality}.

It turns out that it was \emph{a prediction} of the electroweak SM that there should be an additional neutral gauge boson $Z_\mu$.  Contrary to the photon, the $Z$-boson also interacts with neutrinos.  This crucial property was used in the experiment called \emph{Gargamelle} at CERN, where in 1973 the discovery was presented (Fig.~\ref{fig:nu_e_scattering_W}). 
About ten years later both $W$ and $Z$ were directly produced at Super Proton Synchrotron (SPS) at CERN. Finally, in the early 90s a comprehensive analysis 
of the $e^+ e^- \to f \bar f$ process, which  was  carried out at the Large Electron Proton (LEP) Collier (CERN) and at the Standford Linear Collider (SLAC) confirmed the SM predictions for the $Z$ couplings to fermions \eqref{eq:ffZ_couplings}.

\begin{figure}[t]
				\begin{center}
								\begin{tabular}{cc}
								\includegraphics[width=.35\linewidth]{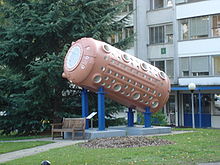}
								& 
								\includegraphics[width=.35\linewidth]{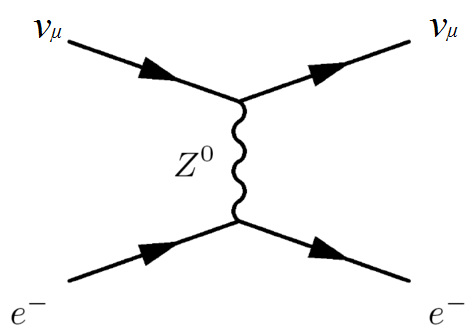}
								\end{tabular}
				\end{center}
				\caption{The chamber of Gargamelle at CERN (left), $\nu_\mu$ scattering 
				due to $Z$-boson (right). From Wikipedia.}
				\label{fig:Z_discovery}
\end{figure}

\begin{figure}[ht]
				\begin{center}
								\includegraphics{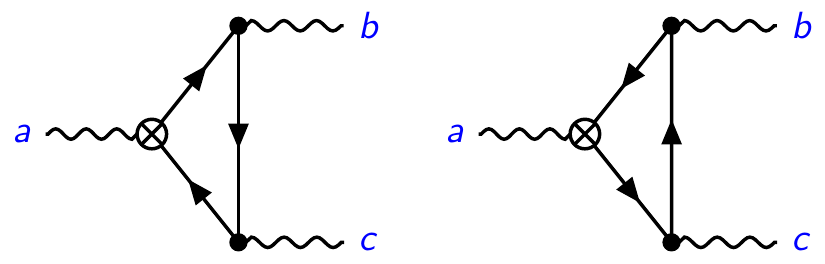}
				\end{center}
				\caption{Diagrams contributing to the anomaly of an axial current (crossed vertex). }
				\label{fig:gauge_anomaly}
\end{figure}

It is also worth mentioning the fact that the (hyper)-charge assignment \eqref{eq:Yf_YQ}
satisfies very non-trivial constraints related to cancellation of \emph{gauge anomalies}.  Anomalies correspond to situations when a symmetry of the classical Lagrangian is violated at the quantum level. 
A well-known example is \emph{Axial or Chiral or Adler--Bell--Jackiw(ABJ)} anomaly when the classical conservation %\footnote{Strictly speaking, even at tree level chiral transformations $\Psi \to e^{i \alpha \gamma_5} \Psi$ become symmetry only in the limit of vanishing masses $m=0$.} 
law for the axial current $J_\mu^A$ is modified due to quantum effects:
\begin{align}
				J_\mu^A & = \bar \Psi \gamma_\mu \gamma_5 \Psi,
				\qquad
				\partial_\mu J^A_\mu  = 2 i m \Psi \gamma_5 \Psi  + \underbrace{\frac{\alpha}{2\pi} F_{\mu\nu} \tilde F_{\mu\nu}}_{\text{anomaly}} , \qquad \tilde F_{\mu\nu} = 1/2 \epsilon_{\mu\nu\rho\sigma} F_{\rho\sigma}.
\label{eq:axial_anomaly}
\end{align}
			The $F\tilde F$-term appears due to loop diagrams
			presented in Fig.~\ref{fig:gauge_anomaly}. 
			%In the limit $m\to 0$ the corresponding amplitude \emph{does not} vanish, when 
			%multiplied by momentum entering the vertex with axial current $J_\mu^A$.

There is nothing wrong when the anomalous current $J_\mu^A$ corresponds 
to a global symmetry and does not enter into $\Lag$.  
It just implies that %classical selection rules are not obeyed in the quantum theory and 
a classically forbidden processes may actually occur in the quantum theory.  For example, it is the anomaly in the \emph{global} axial \emph{flavour} symmetry that is responsible for the decay $\pi\to \gamma\gamma$. 
On the contrary, if an axial current couples to a gauge field, anomalies break gauge invariance, %renormalizability and unitarity, 
thus  rendering the corresponding QFT inconsistent. 
In the SM left and right fermions (eigenvectors of $\gamma_5$) have different $SU(2)_L \times U(1)_Y$ quantum numbers, leaving space for potential anomalies.  
However, since we have to take into account all fermions which  couple to a gauge field, there is a possibility that contributions from different species cancel each other due to a special assignment of fermion charges.  Indeed, in the case of chiral\footnote{that distinguish left and right fermions} theories,  anomalies are proportional to ($\gamma_5 = P_R - P_L$) 
	\begin{align}
	\text{Anom} \propto \text{Tr} [t^a,\{t^b,t^c\}]_L - \text{Tr}[t^a,\{t^b,t^c\}]_R,
	\end{align}
	where $t^a$ are generators of the considered symmetries and the traces are over left ($L$) or right ($R$) fields. In the SM the requirement that all anomalies should be zero imposes the following
	conditions on fermion hypercharges:
	%., $t^a$ collectively denote (8 + 3 + 1) generators of  $SU(3)_c \times SU(2)_L\times U(1)_Y$. 
	\begin{subequations}
	\begin{align}
									0  & =   \,2 Y^Q_L  - Y_R^u - Y_R^d , 
										 & U(1)_Y - SU(3)_c - SU(3)_c,  \label{eq:anom_u1_su3}\\
									0 & =  N_c Y^Q_L + Y^l_L,  
						 		    & U(1)_Y - SU(2)_L - SU(2)_L,  \label{eq:anom_u1_su2}\\
									0 & =  N_c \left[2 (Y^Q_L)^3   - (Y_R^u)^3 - (Y_R^d)^3 \right]
					+  \left[ 2 (Y^l_L)^3 - (Y_R^e)^3\right], 
										& U(1)_Y - U(1)_Y - U(1)_Y, \label{eq:anom_u1}\\
									0 & =  N_c \left[2 Y^Q_L   - Y_R^u - Y_R^d \right]
					      +  \left[ 2 Y^l_L - Y_R^e\right], 
										& U(1)_Y - grav. - grav. \label{eq:anom_u1_grav}, 
	\end{align}
	\label{eq:anomaly_cancellation}
	\end{subequations}
	where, in addition to the EW gauge group, we also consider strong interactions of quarks 
	that have $N_c=3$ colours\footnote{In the SM coloured quarks belong to the fundamental representation of the corresponding gauge group $SU(3)_c$.}. 
	While the first three conditions come
	from the SM interactions, the last one \eqref{eq:anom_u1_grav} is due to the coupling to gravity. Other anomalies are trivially zero. One can see that the hypercharges introduced in \Eq\eqref{eq:Yf_YQ} do satisfy the equations. It is interesting to note that 
	contributions due to colour quarks miraculously cancel those of leptons and the cancellation works within a single generation. This put a rather strong restriction on possible new fermions that can couple to the SM gauge bosons: new particles should appear in a complete generation (quarks + leptons) in order not to spoil anomaly cancellation within the SM. Moreover, the anomaly cancellation condition can select viable models that go beyond the SM (BSM). 

\subsection{Properties of the EW gauge bosons}

\begin{figure}
				\begin{center}
								\includegraphics{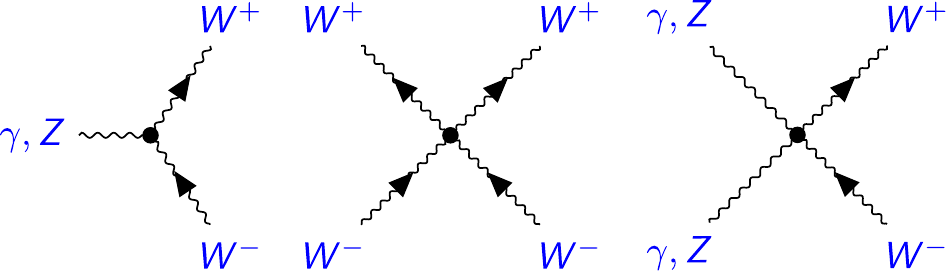}
				\end{center}
				\caption{Gauge-boson self-interaction vertices.}
\label{fig:int_VVV_VVVV}
\end{figure}

Due to the non-Abelian nature of the $SU(2)_L$ group, the gauge fields $W_i$
have  triple and quartic self-interactions (see \Eq\eqref{eq:lag_YM_int}). Since $W_3$ 
is a linear combination of the $Z$-boson and photon, the same is true for $Z$ and $\gamma$.
In Fig.~\ref{fig:int_VVV_VVVV}, self-interaction vertices for the EW gauge bosons are depicted.

\begin{figure}[h]
				\begin{center}
								\begin{tabular}{cc}
	\raisebox{0.33\height}
				{
								\includegraphics[width=.24\linewidth]{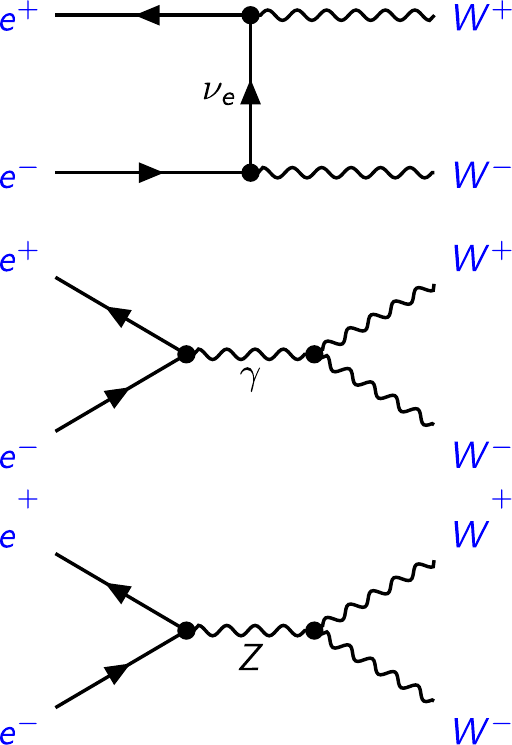}
				}
								& 
								\includegraphics[width=.48\linewidth]{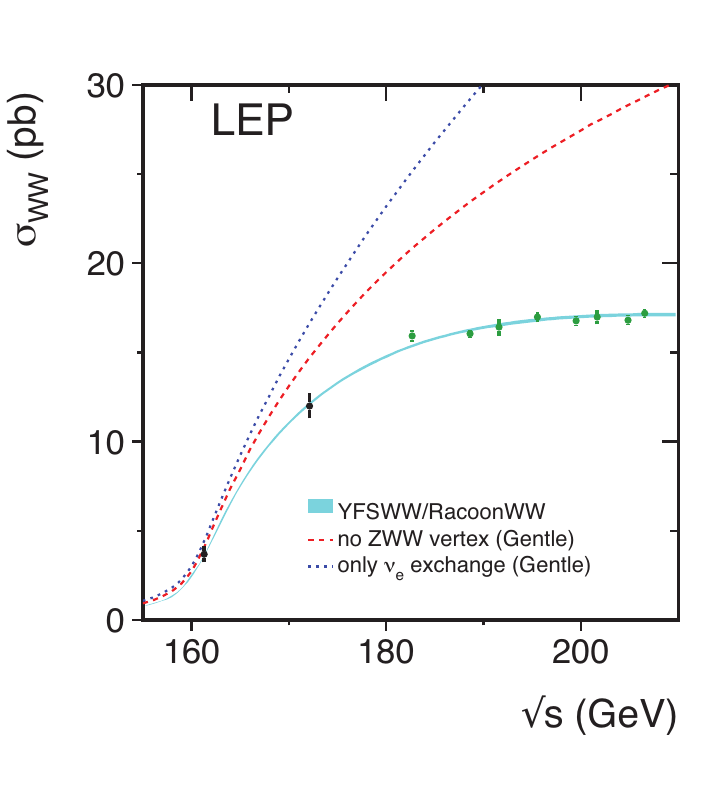}
								\end{tabular}
								\vspace*{-1.0cm}
				\end{center}
				\caption{$e^+e^- \to W^+ W^-$.}
				\label{fig:eeWW_lep}
\end{figure}

The triple vertices $WW\gamma$ and $WWZ$ predicted by the SM were tested at LEP2 in the $e^+ e^- \to W^+ W^-$ process (Fig.~\ref{fig:eeWW_lep}) and agreement with the SM predictions was found. Subsequent studies at hadron colliders  (Tevatron and LHC) aimed at both quartic and triple gauge couplings (QGC and TGC, respectively) also show consistency with the SM and put limits on possible deviations (so-called anomalous TGC and QGC). 

Since we do not observe $Z$-bosons flying around like photons, $Z_\mu$ should have a non-zero mass $M_Z$ and similar to $W^\pm$ give rise to Fermi-like interactions between \emph{neutral} currents $J^\mu_Z$ at low energies. 
The relative strength of the \emph{charged} and \emph{neutral} current-current interactions 
($J^Z_\mu J^\mu_Z)/(J^{+\mu} J^+_{\mu})$ can be measured by the parameter $\rho$:
\begin{equation}
				\rho \equiv \frac{M_W^2}{M_Z^2 \cos^2\theta_W}.
				\label{eq:rho_par}
\end{equation}
Up to now, we do not specify any relations between $M_Z$ and $M_W$. Due to this, the value of $\rho$ can, in principle, be arbitrary. 
However, it is a prediction of the full SM that $\rho \simeq 1$ (see below).

The fact that both $W$ and $Z$ should be massive poses 
a serious problem for theoretical description of the EW interactions. 
First of all, the naive introduction of the corresponding mass terms 
breaks the \emph{gauge} symmetry \eqref{eq:ew_group}. 
For example, $m_W^2 W^+_\mu W^-_\mu$ is forbidden due to $W_\mu \to W_\mu 
+ \partial_\mu \omega + ...$. 
One can also mention an issue with unitarity, which arises in the scattering of longitudinal EW bosons due to gauge self-interactions in Fig.~\ref{fig:int_VVV_VVVV}. 

In addition, the symmetry also forbids \emph{explicit} mass terms for fermions, since 
\eg $m_\mu (\bar \mu_L \mu_R + \hc)$, which accounts for muon mass, 
mixes left and right fields that transform differently under the electroweak group \eqref{eq:ew_group}. 
In the next section, we discuss how these problems can be solved 
by coupling the SM fermions and gauge bosons to the scalar (Higgs) sector (see also \cite{Maltoni_lect}). 

\subsection{Spontaneous symmetry breaking and gauge-boson masses}
We need to  \emph{generate} masses for $W_\mu^\pm$ and $Z_\mu$ (but not for $A_\mu$) without \emph{explicit} breaking of the gauge symmetry. Let us consider for simplicity \emph{scalar} 
electrodynamics:
\begin{equation}
				\Lag = \partial_\mu \phi^\dagger \partial_\mu \phi 
				- V(\phi^\dagger \phi) 
				- \frac{1}{4} F_{\mu\nu}^2 
				+ i \echarge \left(\phi^\dagger \partial_\mu \phi - \phi \partial_\mu \phi^\dagger\right) A_\mu + \echarge^2 A_\mu A_\mu \phi^\dagger \phi \equiv \Lag_1,
				\label{eq:SED_lag}
\end{equation}
which is invariant under $U(1)$
\begin{equation}
					\phi \to e^{i \echarge \omega(x)} \phi, \quad A_\mu \to A_\mu + \partial_\mu \omega.
					\label{eq:U1_transform_sed}
\end{equation}
In \Eq\eqref{eq:SED_lag} a \emph{complex} scalar $\phi$ interacts with the photon $A_\mu$. 
We can use \emph{polar} coordinates to rewrite the Lagrangian in terms of new variables
\begin{align}
				\Lag & = \frac{1}{2} (\partial_\mu \rho)^2 + \frac{\echarge^2 \rho^2}{2} \left( A_\mu - \frac{1}{\echarge} \partial_\mu \theta\right)
				\left( A_\mu - \frac{1}{\echarge} \partial_\mu \theta\right) - V(\rho^2/2) - \frac{1}{4} F_{\mu\nu}^2, \\
						 & = \frac{1}{2} (\partial_\mu \rho)^2 + \frac{e^2 {\rho^2}}{2} B_\mu B_\mu 				- V(\rho^2/2) - \frac{1}{4} F_{\mu\nu}^2(B),
						 \label{eq:SED_lag_B}
\end{align}
where $\rho$ is gauge invariant, while the $U(1)$ transformation \eqref{eq:U1_transform_sed} gives rise to  a \emph{shift} in $\theta$:
		\begin{equation}
						\phi = \frac{1}{\sqrt 2} \rho(x) e^{i \theta(x)},
						\quad \rho \to \rho , \quad \theta \to \theta + e \omega. 
		\end{equation}
One can also notice that $B_\mu \equiv A_\mu - \frac{1}{\echarge} \partial_\mu\theta$ is 
also invariant! Moreover, since $F_{\mu\nu}(A) = F_{\mu\nu}(B)$, we can completely get rid of $\theta$. As a consequence, the gauge symmetry becomes ``hidden'' when the system is described by the variables $B_\mu(x)$ and $\rho(x)$.  

\begin{figure}
				\begin{center}
								\begin{tabular}{cc}
								\includegraphics[width=.3\linewidth]{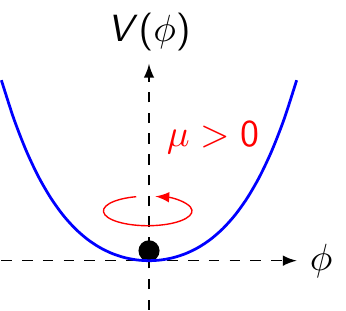}
								& 
								\includegraphics[width=.3\linewidth]{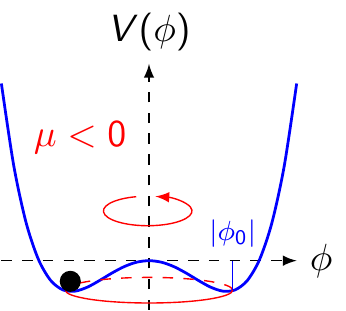}
								\end{tabular}
				\end{center}
				\caption{A symmetric vacuum (left) and degenerate vacua (right).}
				\label{fig:SSB}
\end{figure}

If in \Eq\eqref{eq:SED_lag} we replace our \emph{dynamical} field $\rho(x)$ by a constant $\rho\to v=\text{const}$, we get the mass term for $B_\mu$. This can be achieved by considering %``mexican-hat'' 
the potential $V(\phi)$ of the form (written in terms of initial variables)
\begin{equation}
								V  = {\mu^2} \phi^\dagger \phi + \lambda (\phi^\dagger \phi)^2.
								\label{eq:mexican_hat}
\end{equation}
	One can distinguish two different situations (see Fig.~\ref{fig:SSB}):
	\begin{itemize}
					\item $\mu^2>0$ --- a \emph{single} minimum with $\phi=0$; 
					\item $\mu^2<0$ --- a valley of \emph{degenerate} minima with $\phi\neq 0$.
	\end{itemize}
In both cases we solve EOM for the homogeneous (in space and time) field. When $\mu^2>0$ the 
\emph{solution} is unique and symmetric, \ie it does not transform under $U(1)$. In the second case, in which we are interested here, the potential has non-trivial minima 
\begin{equation}
				\left.\frac{\partial V}{\partial \phi^\dagger}\right|_{\phi=\phi_0} = 0 
								\Rightarrow
				{\phi_0^\dagger \phi_0} = -\frac{\mu^2}{2 \lambda}  = \frac{v^2}{2} > 0
				\Rightarrow {\phi_0} = \frac{v}{\sqrt 2} e^{i {\beta}},
\end{equation}
which are \emph{related} by \emph{global} $U(1)$ transformations \eqref{eq:U1_transform_sed} that change $\beta \to \beta + e \omega$. 
So, in spite of the fact that we do not break the symmetry \emph{explicitly},
it is \emph{spontaneously broken} (SSB) due to a particular choice of our solution ($\beta$). 

In QFT we interpret $\phi_0$ as a characteristic of our \emph{vacuum} state, \ie as a \emph{vacuum expectation value} (vev) or \emph{condensate} of the quantum field%\footnote{In what follows we set $\beta=0$. Why?}
:
\begin{equation}
				\phi_0 = \bra{0} \phi(x) \ket{0}\stackrel{\beta=0}{=} \frac{v}{\sqrt 2}.
\end{equation}
				Since we want to introduce particles as \emph{excitations} above the vacuum, we have to shift the field:
\begin{equation} 
								\phi(x) = \frac{v + h(x)}{\sqrt2} e^{i \zeta(x)/v}, \qquad \bra{0} h(x) \ket{0} = 0, \quad \bra{0} \zeta(x) \ket{0} = 0.
\end{equation}
As a consequence, \Eq\eqref{eq:SED_lag_B} can be rewritten as
\begin{align}
				\Lag & = \frac{1}{2} (\partial_\mu h)^2 + \frac{\echarge^2 v^2}{2} \left( 1 + \frac{h}{v}\right)^2 B_\mu B_\mu 
								%+ \vphantom{\frac{\echarge^2 v^2}{2}}\echarge v h B_\mu B_\mu + \frac{\echarge^2}{2} B_\mu B_\mu h^2   %\left( A_\mu - \frac{1}{e} \partial_\mu \theta\right) \left( A_\mu - \frac{1}{e} \partial_\mu \theta\right) 
								- V(h) - \frac{1}{4} F_{\mu\nu}^2(B) \equiv \Lag_2, \\
								V(h) & = -\frac{|\mu|^2}{2} \left( v + h \right)^2  + \frac{\lambda}{4} (v + h)^4 = 
				\frac{2 \lambda v^2}{2} h^2 + \vphantom{\frac{1}{4}}\lambda v h^3 + \frac{\lambda}{4} h^4 - \frac{\lambda }{4} v^4.
				\label{eq:SED_lag_B_ssb}
				\end{align}
				One can see that the Lagrangian \eqref{eq:SED_lag_B_ssb} describes a massive vector field $B_\mu$ with $m_B^2 = e^2 v^2$ and a massive scalar $h$ with $m_h^2 = 2 \lambda v^2$. We do not break the symmetry explicitly. It is again \emph{hidden} in the relations between couplings and  masses. This is the essence of the \emph{Brout-Englert-Higgs-Hagen-Guralnik-Kibble} mechanism \cite{Englert:1964et, Higgs:1964pj,BEH}.
				
				The Lagrangians $\Lag_1$ \eqref{eq:SED_lag} and $\Lag_2$ \eqref{eq:SED_lag_B_ssb} describe the same Physics  but written in terms of different quantities (variables). Expression \eqref{eq:SED_lag} involves 
				a \emph{complex} scalar $\phi$ with 2 (real) degrees of freedom (DOFs)
				and a \emph{massless} gauge field ($A_\mu$) also having 2 DOFs.  It is manifestly gauge invariant  but not suitable 
				for perturbative expansion ($\phi$ has imaginary mass).	
				
				On the contrary, in $\Lag_2$ the gauge symmetry is hidden\footnote{One can also say that $\Lag_2$ corresponds to the \emph{unitary} gauge, \ie no unphysical ``states'' in the particle spectrum.} and it is written in terms of \emph{physical} DOFs, \ie a \emph{real} scalar $h$ (1 DOF) and a \emph{massive} vector $B_\mu$ (3 DOFs).  In a sense, one \emph{scalar} DOF ($\zeta$) is ``eaten'' by the  gauge field to become massive. It is important to note that the postulated \emph{gauge} symmetry allows us to avoid the consequences of the \emph{Goldstone} theorem, which states that if the vacuum breaks a \emph{global} continuous symmetry there is a \emph{massless} boson (Nambu-Goldstone) in the spectrum\footnote{any non-derivative interactions violate the shift symmetry $\zeta \to \zeta + e v \omega$ for $\omega=\mathrm{const}$}. This boson is associated with `oscillations'' along the valley, \ie in the \emph{broken} direction (see Fig.~\ref{fig:SSB}).  However, due to the local character of symmetry, $\chi$ is not physical anymore, its disappearance (or appearance, see below) reflects the \emph{redundancy}, which was mentioned above. 
						
				In  Sec.~\ref{sec:renorm_vs_non_renorm}, we demonstrated that the massive-vector propagator has rather bad UV behavior and is not very convenient for doing calculations in PT.  It looks like we gain nothing  from the gauge principle. But it is not true. %By utilizing the gauge symmetry one can write down the third form of the model Lagrangian, \ie
				We can write the model Lagrangian in the \emph{Cartesian} coordinates
		$\phi = \frac{1}{\sqrt 2} \left( v + \eta + i {\chi} \right)$:
			\begin{align}
							\Lag_3
  &=
        -\frac{1}{4}F_{\mu \nu}F_{\mu \nu} 
				+\frac{e^2v^2}{2} A_\mu A_\mu 
				+\frac{1}{2}\partial_\mu\chi \,\partial _\mu\chi 
				- \underline{{ev} A_\mu \partial_\mu\chi} 
            +\frac{1}{2}\partial_\mu\eta\,\partial_\mu\eta
            -\frac{2v^2\lambda}{2}\eta^2 
      +   \frac{v^4\lambda}{4} 
			\label{eq:SED_lag_A_ssb_free}
\\& 
       +  e A_\mu \chi\partial_\mu\eta 
       - e A_\mu \eta\partial_\mu\chi
       - {v\lambda}\eta (\eta^2 +\chi^2)
       - \frac{\lambda}{4}(\eta^2 + \chi^2)^2
      + \frac{e^2}{2} A_\mu A_\mu (2 v \eta +\eta^2 + \chi^2).
			\label{eq:SED_lag_A_ssb_int}
			 \end{align}
		The ``free'' part \eqref{eq:SED_lag_A_ssb_free} of $\Lag_3$  seems to describe
		5 real DOFs: a massive scalar $\eta$, a \emph{massless} (would-be \emph{Nambu-Goldstone}) boson $\chi$ and a massive $A_\mu$. However, 
		there is a mixing between the \emph{longitudinal} component of $A_\mu$ and $\chi$ that spoils this naive counting (unphysical $\chi$ is ``partially eaten'' by $A_\mu$) .

		In spite of this subtlety, $\Lag_3$ is more convenient for calculations in PT. To quantize the model, one can utilize the gauge-fixing freedom and add the following expression to $\Lag_3$ 
		\begin{equation}
\delta\Lag_{g.f.} = - \frac{1}{2 \xi} \left( \partial_\mu A_\mu + e v \xi \chi \right)^2 = - \frac{1}{2 \xi} \left( \partial_\mu A_\mu\right)^2  
- \underline{e v \chi \partial_\mu A_\mu} - \frac{e^2 v^2 \xi }{2} \chi^2.
\label{eq:SED_gf}
\end{equation}
It removes %\footnote{after integration by parts.} 
the mixing from \Eq\eqref{eq:SED_lag_A_ssb_free} and introduces a mass for $\chi$, $m_\chi^2 = (e^2 v^2) \xi$. In addition, the vector-boson propagator in this case looks like
\begin{equation}
				\bra{0} T A_\mu (x) A_\nu(y) \ket{0}  = 
				\int \frac{d^4 p}{(2\pi)^4} \frac{-i \left[ g_{\mu\nu} - (1-{\xi}) 
				\frac{p_\mu p_\nu}{p^2- \xi m_A^2} \right]}{p^2 - m_A^2 + i \epsilon}
				e^{-i p(x-y)}, \quad m_A = e v.
\end{equation}
One can see that for $\xi\to \infty$ we reproduce \Eq\eqref{eq:vector_prop_massive}, while for finite $\xi$ the propagator behaves like $1/p^2$ as  $p\to\infty$, thus  making it convenient for PT calculations. 

It should be mentioned that contrary to $\Lag_2$  the full Lagrangian corresponding to $\Lag_3$ involves also unphysical \emph{ghosts}, which do not decouple in the considered case. Nevertheless, it is a relatively small price to pay for the ability to perform high-order calculations required to obtain high-precision predictions.

\begin{figure}
				\begin{center}
								\includegraphics{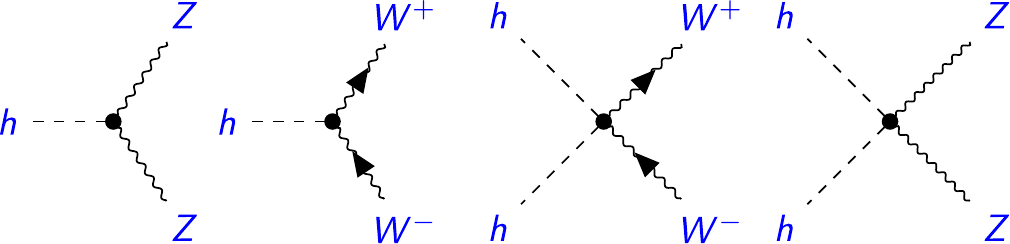}
				\end{center}
				\caption{Gauge-boson--Higgs interactions.}
\label{fig:int_HVV_HHVV}
\end{figure}

Let us switch back to the SM. We have three gauge bosons that should become massive. As a consequence, three symmetries should be broken by the SM vacuum %state 
to feed hungry $W^\pm_\mu$ and $Z_\mu$ with (would-be) Goldstone bosons 
\begin{equation}
				SU(2)_L \times U(1)_Y \to U(1)_{em}. 
\end{equation}
		The photon should remain massless and correspond to the unbroken electromagnetic $U(1)_{em}$.
 This can be achieved by considering an $SU(2)_L$ doublet of scalar fields:
				\begin{equation}
																%\phi = \left(\frac{v + h}{\sqrt2}\right) e^{i \frac{\alert{\zeta}}{v}} & \to 
								\Phi =\frac{1}{\sqrt 2} \exp \left(i \frac{{\zeta_j(x)} \sigma^j}{2v}\right)
																				\begin{pmatrix}
																								0 \\
																								v + h(x)
																				\end{pmatrix}
																,  \qquad \Phi_0 \equiv \bra{0} \Phi \ket{0} = \frac{1}{\sqrt 2} 
																\begin{pmatrix} 0 \\ v \end{pmatrix},
																\label{eq:Higgs_doublet}
												\end{equation}
												where we decompose $\Phi(x)$ in terms of three (would-be) Goldstone bosons $\zeta_j$  and a Higgs $h$. 												 The Pauli matrices $\sigma_j$ represent broken generators of $SU(2)_L$.
												Let $\Phi$ also be charged under $U(1)_Y$:
												\begin{equation}
											\Phi \to \exp \left(i g \frac{\sigma^i}{2} \omega_a + i g' \frac{Y_H}{2} \omega' \right) \Phi.
								\end{equation}
								We do not want to break $U(1)_{em}$ spontaneously so the vacuum characterized by the vev $\Phi_0$ should be invariant under $U(1)_{em}$, \ie has no electric charge $Q$
								\begin{equation}
												e^{i e Q \theta} \Phi_0  = \Phi_0 \to  Q \Phi_0 = 0.
								\end{equation}
								The operator $Q$ is a linear combination of diagonal generators of $SU(2)_L \times U(1)_Y$, $T_3= \sigma_3/2$ and $Y/2$:
\begin{equation}
				Q \Phi_0 = \left(T_3 + \frac{Y}{2}\right) \Phi_0 =  \frac{1}{2} \begin{pmatrix} 1 + Y_H & 0 \\ 0 & -1 + Y_H  \end{pmatrix}
				\begin{pmatrix} 0 \\ \frac{v}{\sqrt{2}} \end{pmatrix} \stackrel{?}{=} 0.
				\label{eq:Q_vacuum}
\end{equation}
As a consequence, to keep $U(1)_{em}$ unbroken, we should set $Y_H = 1$.	Since $\Phi$ transforms under the EW group, we have to
				introduce gauge interactions for the Higgs doublet to make sure that the scalar sector respects the corresponding local symmetry: 
\begin{equation}
\Lag_\Phi = (D_\mu \Phi)^\dagger (D_\mu \Phi)  - V(\Phi), 
\quad \text{with} \quad 
V(\Phi) = m_\Phi^2 %\mu^2 
				\Phi^\dagger\Phi + \lambda(\Phi^\dagger\Phi)^2.
\label{eq:lag_higgs}
\end{equation}
For $m_\Phi^2<0$ the symmetry is spontaneously broken. In the \emph{unitary} gauge (Goldstone bosons are gauged away: in \Eq\eqref{eq:Higgs_doublet} we put $\zeta_j=0$)  the first term in \Eq\eqref{eq:lag_higgs} can be cast into 
\begin{align}
														|D_\mu\Phi|^2 
&=
  \frac{1}{2}(\partial_\mu h)^2 
	+ \frac{g^2}{8}(v+h)^2 |W_\mu^1+iW_\mu^2|^2	+ \frac{1}{8}(v+h)^2 (g W_\mu^3 - g' Y_H B_\mu)^2\\
&=
  \frac{1}{2}(\partial_\mu h)^2 
	+ \frac{g^2}{4}(v+h)^2 W^+ W^-	\quad \left[ \sqrt 2 W^\pm = W^1_\mu \mp i W^2_\mu \right]\nonumber \\
& + \frac{1}{8}(v+h)^2 \left[Z_\mu (g \cos \theta_W + g' \sin \theta_W) 
+ A_\mu ( g \sin \theta_W  - g' \cos \theta_W)\right]^2\\
& = \frac{1}{2}(\partial_\mu h)^2 + M_W^2 \left(1 + \frac{h}{v} \right)^2 W^+ W^- 
	+ \frac{M_Z^2}{2}\left(1 + \frac{h}{v} \right)^2 Z_\mu Z_\mu, 
	\label{eq:Higgs_kinetic_term}
\end{align}
where we \emph{require} the photon to be massless after SSB, \ie 
\begin{equation}
				g {\sin \theta_W}  - g' {\cos \theta_W}  = 0 
				\quad 
				\Rightarrow 
				\quad
				\sin \theta_W = \frac{g'}{\sqrt{g^2 + g'^2}},
				\cos \theta_W = \frac{g}{\sqrt{g^2 + g'^2}}
				\label{eq:sinW_g_gp}
\end{equation}
			and, consequently,
			\begin{equation}	
				g \cos \theta_W  + g' \sin \theta_W  = \sqrt{g^2 + g'^2}, \qquad
				\echarge = g \sin \theta_W = g' \cos \theta_W =  \frac{g g'}{\sqrt{g^2 + g'^2}}.
\end{equation}
The masses of the $Z$ and $W$-bosons are proportional to the EW gauge couplings 
\begin{equation}
				M^2_W = \frac{g^2 v^2}{4},\quad M_Z^2 = \frac{(g^2 + g'^2) v^2}{4}.
				\label{eq:WZ_masses}
\end{equation}
One can see that the Higgs-gauge boson vertices (Fig.~\ref{fig:int_HVV_HHVV}) are related to the masses $M_W$ and $M_Z$.

An important consequence of the SM gauge symmetry and the existence of the Higgs boson is the
 \emph{unitarization} of massive vector-boson scattering. 
By means of simple power counting, one can easily convince oneself that 
the amplitude for (longitudinal) $W$-boson scattering originating from the quartic vertex in Fig.~\ref{fig:int_VVV_VVVV} scales with energy as $E^4/M_W^4$. This kind of dependence will eventually violate unitarity for $E\gg M_W$. 
However, in the SM, thanks to gauge symmetry, QGC and TGC are related. This results in  
$E^2/M_W^2$ behavior when  $Z/\gamma$ exchange is taken into account.  
Moreover, since the gauge bosons couple also to Higgs, we need to include the corresponding contribution to the total amplitude. It turns out that it is this contribution that cancels the $E^2$ terms and 
saves unitarity in the $WW$-scattering. %amplitude from growing with energy. 
Obviously, this pattern is a consequence of the EW symmetry breaking in the SM and can be modified by  the presence of New Physics. 
Due to this, experimental studies of vector boson scattering (VBS) play a role in proving overall consistency of the SM. %First LHC results on VBS appeared only recently and show compatibility with theory predictions.
	\begin{figure}[th]
				\begin{center}
								\begin{tabular}{cc}
	\raisebox{-0.5\height}
				{
								\includegraphics[scale=0.8]{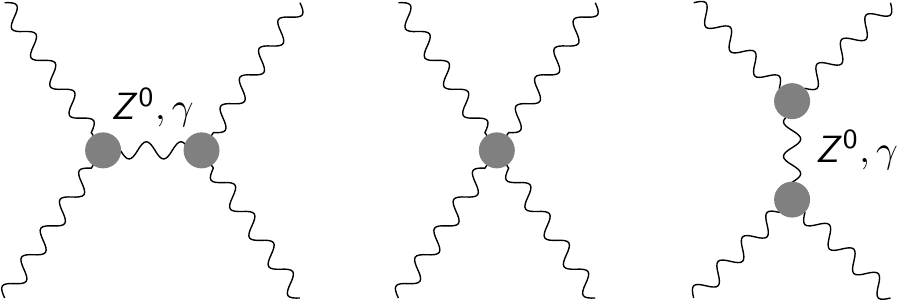}
				}
								& $\mathcal{M} \propto g^2 \frac{E^2}{M_W^2},$
								
								\\[1.5cm]
	\raisebox{-0.5\height}
	{
								\includegraphics[scale=0.8]{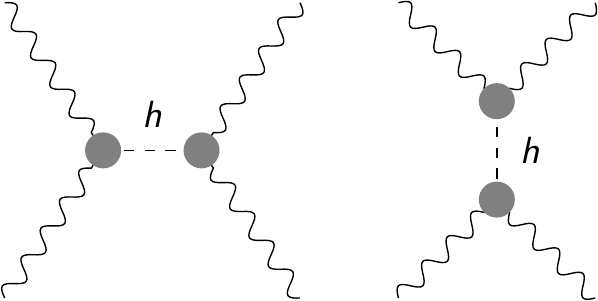}
				}

								& $\mathcal{M} \propto - g^2 \frac{E^2}{M_W^2}$
								\end{tabular}
				\end{center}
				\caption{WW-scattering and Unitarity.}
				\label{fig:WW_unitarity}
\end{figure}

Having in mind \Eq\eqref{eq:Fermi_matching}, one can derive the relation
\begin{align}
				G_F & = \frac{1}{\sqrt 2 v^2}\Rightarrow v \simeq  246~\text{GeV}, 
					\label{eq:GF_vev_relation}
\end{align}
		which gives a numerical estimate of $v$.  
One can also see that due to \eqref{eq:WZ_masses} we have (at the tree level)
\begin{equation}
				\rho =\frac{M_W^2}{M_Z^2 \cos^2\theta_W} = 1.
				\label{eq:rho_par_SM}
\end{equation}
Let us emphasize that it is a consequence of the fact that the SM Higgs is a weak \emph{doublet} with \emph{unit} hypercharge. 
Due to this, $\rho \simeq 1$ imposes important constraints on possible extensions of the SM Higgs sector. 
For example, we can generalize expression \eqref{eq:rho_par_SM} to account for $n$ scalar $(2 I_i + 1)$-plets $(i=1,...,n)$ that transform under $SU(2)_L$ and have hypercharges $Y_i$. 
In case they acquire vevs $v_i$, which break the EW group, we have 
\begin{align}
				\rho = \frac{ \sum_i (I_i (I_i +1) - Y_i^2) v_i^2}{\sum_i 2 Y_i^2 v_i^2}.
				\label{eq:rho_general}
\end{align}
Consequently, any non-doublet (with total weak isospin $I_i\neq 1/2$) vev leads to a deviation from $\rho=1$.

\subsection{Fermion-higgs interactions and masses of quarks and leptons} 
Since we fixed all the gauge quantum numbers of the SM fields, %Given the above-mentioned quantum numbers, 
it is possible to construct the following \emph{gauge-invariant} Lagrangian:   
\begin{equation}
				\Lag_Y = - {y_e}  (\underset{\mathstrut \vphantom{\frac12} +1~}{~\bar L\vphantom{e_R}~} 
												   \underset{\mathstrut \vphantom{\frac12} +1~}{~\Phi\vphantom{e_R}}~) 
													 \underset{\mathstrut \vphantom{\frac12} -2~}{~e_R~} 
									- {y_d} (\underset{\mathstrut -\frac{1}{3}~}{~\bar Q\vphantom{e_R}~}
												   \underset{\mathstrut \vphantom{\frac12}+1~}{~\Phi\vphantom{e_R}~}) 
													 \underset{\mathstrut -\frac{2}{3}~}{~d_R~} 
													 - {y_u} (\underset{\mathstrut -\frac{1}{3}~}{~\bar Q\vphantom{e_R}~} 
												   \underset{\mathstrut \vphantom{\frac12}-1~}{~\Phi^c\vphantom{e_R}~} ) 
													 \underset{\mathstrut \frac{4}{3}~}{~u_R~} 
													 + \hc, 
\label{eq:ffH_diagonal}
\end{equation}
which involves \emph{dimensionless} Yukawa couplings $y_f$. It describes interactions between the Higgs field $\Phi$, left fermion %$SU(2)_L$ 
doublets \eqref{eq:SU2L_fermions} and right %$SU(2)_L$ 
singlets. In \Eq\eqref{eq:ffH_diagonal} we also indicate weak hypercharges of the corresponding fields. One can see that combinations of two doublets,  $(\bar Q \Phi)$ \etc, are invariant under $SU_L(2)$ but have a non-zero charge under $U(1)_Y$. The latter is compensated by hypercharges of right fermions.
In addition, $U(1)_Y$ symmetry forces us to use a charge-conjugated Higgs doublet $\Phi^c = i\sigma_2 \Phi^*$ with $Y=-1$ to 
account for Yukawa interactions involving $u_R$.

\begin{figure}[ht]
				\begin{center}
								\begin{tabular}{cp{1cm}c}
\raisebox{-0.5\height}
				{
								\includegraphics{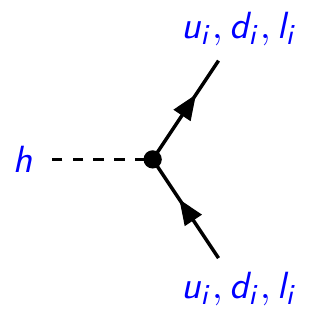}
				}
								&& 
\raisebox{-0.5\height}
				{
								\includegraphics{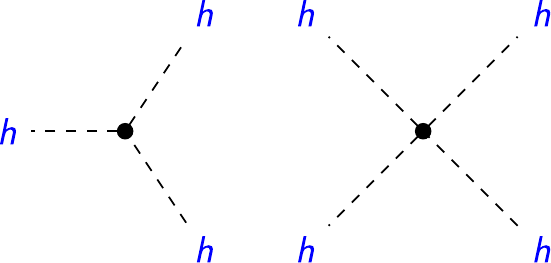}
				} \\
												(a) && (b)
								\end{tabular}
				\end{center}
				\caption{Higgs--fermion couplings (a) and  self-interactions of the Higgs boson (b).}
				\label{fig:int_h3_h4_hff}
\end{figure}

In the spontaneously broken phase with non-zero Higgs vev, the Lagrangian $\Lag_Y$ can be written in the following simple form:
\begin{align}
				- \Lag_Y =  \sum_f \frac{y_f v }{\sqrt 2} \left(1 + \frac{h}{v} \right) \bar f f = 
				\sum_f m_f \left(1 + \frac{h}{v} \right) \bar f f	
				, \quad f = u,~d,~e,
				\label{eq:ffh_ssb}
\end{align}
		where unitary gauge is utilized. One can see that SSB generates fermion masses $m_f$ 
		and, similarly to \Eq\eqref{eq:Higgs_kinetic_term}, \emph{relates} them to the corresponding couplings of the Higgs boson $h$ (see Fig.\ref{fig:int_h3_h4_hff}a).
		
		It is worth noting that \Eq\eqref{eq:ffH_diagonal} is not the most general renormalizable Lagrangian involving the SM scalars and fermions. One can introduce \emph{flavour} indices and non-diagonal \emph{complex} Yukawa matrices $y^{ij}_f$ to account for a possible mixing between the SM fermions, \ie
\begin{equation}
				\Lag_{\mathrm{Yukawa}} = - {y^{ij}_l}  (\bar L_i \Phi) {l_{jR}} - {y^{ij}_d} (\bar Q_i \Phi) {d_{jR}} - {y^{ij}_u} (\bar Q_i {\Phi^c}) {u_{jR}} + \hc.
		\label{eq:ffh_non_diagonal}
\end{equation}	
Substituting $\Phi \to \Phi_0$ we derive the expression for fermion mass matrices $m_f^{ij} = y_f^{ij} \frac{v}{\sqrt{2}}$, which can be diagonalized by suitable unitary rotations of left and right fields.  In the SM the Yukawa matrices~\eqref{eq:ffh_non_diagonal} are also diagonalized by the \emph{same} transformations. This leads again (in the unitary gauge) to \Eq\eqref{eq:ffh_ssb}  but with the fields corresponding to the \emph{mass} eigenstates. 
The latter \emph{do not} coincide with \emph{weak} states, which enter into $\Lag_W$ \eqref{eq:WEAK_lag}.
However, one can rewrite $\Lag_W$ in terms of mass eigenstates. Due to large \emph{flavour symmetry} of weak interactions\footnote{In the SM the symmetry is $U(3)^5$ and corresponds 
				to flavour rotations of left doublets, $Q$ and $L$, and right singlets, $u_R$, $d_R$ and $l_R$. Neutrinos are assumed to be massless.} , 
this introduces a single mixing matrix (the Cabibbo--Kobayashi--Maskawa matrix, or CKM), which manifests itself in the charged-current interactions $\Lag_{CC}$. 
A remarkable fact is that three generations are \emph{required} to have $\mathcal{CP}$ violation in the quark sector. Moreover, a single CKM with only four physical parameters 
(angles and one phase) proves to be very successful in accounting for plethora of phenomena involving transitions between different flavours. 
We will not discuss further details but refer to the dedicated lectures on Flavor Physics \cite{Zupan_lect}.

\section{The SM: theory vs. experiment}
\label{sec:exp_test}

Let us summarize and %We now have all the ingredients and are ready to 
write down the full SM Lagrangian as %in the following form
\begin{align}
				\Lag_{\text{SM}} & = 
	\Lag_{\text{Gauge}}({g_s}, {g}, {g'})  
	%\label{eq:sm_lag_gauge}\\
						%& 
	+ \Lag_{\text{Yukawa}}(y_u, y_d, y_l) %\label{eq:sm_lag_yukawa}\\
						%& 
	+ \Lag_{\text{Higgs}}(\lambda, m_\Phi^2) %\quad \left[-V_{\text{Higgs}}(\lambda, \mu^2)\right] \label{eq:sm_lag_higgs}\\
	+ \Lag_{\text{Gauge-fixing}} 
  + \Lag_{\text{Ghosts}}. \label{eq:sm_lag_full}
\end{align}
The Yukawa part $\Lag_{\text{Yukawa}}$ is given in \Eq\eqref{eq:ffh_non_diagonal}, while $\Lag_{\text{Higgs}} = - V(\Phi)$ is the Higgs potential from \Eq\eqref{eq:lag_higgs}.  After SSB the corresponding terms give rise to the Higgs couplings to the SM fermions (Fig.\ref{fig:int_h3_h4_hff}a) and Higgs self-interactions (Fig.\ref{fig:int_h3_h4_hff}b). The former are diagonal in the \emph{mass} basis.  The kinetic term for the Higgs field is included in 
\begin{align}
				\Lag_{\text{Gauge}} & = -\frac{1}{4} \underbrace{G_{\mu \nu}^a G_{\mu \nu}^a}_{{SU(3)_c}}
				-\frac{1}{4} \underbrace{W_{\mu \nu}^i W_{\mu \nu}^i}_{{SU(2)_L}}
				-\frac{1}{4} \underbrace{B_{\mu \nu}B_{\mu \nu}}_{{U(1)_Y}} 
	+ (D_\mu \Phi)^\dagger (D_\mu \Phi) 
	\label{eq:sm_lag_gauge}
	\\
				& + \underbrace{
				\bar{L_i}\, i \hat D \, L_i 
		 +  \bar{Q_i}\, i \hat D \, Q_i}_{SU(2)_L~\text{doublets}} 
		 + \underbrace{
						 \bar{l}_{Ri} \, i \hat D \, l_{R_i} \ 
	+ \bar{u}_{Ri} \, i \hat D u_{R_i} \ 
	+ \bar{d}_{Ri} \, i \hat D  \, d_{R_i}
	}_{SU(2)_L~\text{singlets}} ,
	\label{eq:sm_lag_gauge_fermions}
\end{align}
where for completeness we also add the colour  group $SU(3)_c$ responsible for the strong force.
	The first three terms in \Eq\eqref{eq:sm_lag_gauge} introduce gauge bosons for the SM gauge groups and in the non-Abelian case account for self-interactions of the latter (Fig.~\ref{fig:int_VVV_VVVV}).
	The fourth term in \eqref{eq:sm_lag_gauge} written in the form \eqref{eq:Higgs_kinetic_term}
	accounts for gauge interactions of the Higgs field (Fig.~\ref{fig:int_HVV_HHVV}).
	Finally, \Eq\eqref{eq:sm_lag_gauge_fermions} gives rise to interactions between gauge bosons and the SM fermions (see, \eg Fig.~\ref{fig:int_ffV}). 
	%The NC processes conserve flavour in the SM 
	%(\ie, remains diagonal in flavour indices both in \emph{weak} and \emph{mass} bases), 
	%while CC interactions allow flavour transitions.

The SM Lagrangian \Eq\eqref{eq:sm_lag_full} depends on 18 physical\footnote{We do not count unphysical gauge-fixing parameters entering into $\Lag_{\text{Gauge-fixing}}$ and $\Lag_{\text{Ghosts}}$.} parameters --- 17 dimensionless couplings (gauge, Yukawa, and scalar self-interactions) and only 1 mass parameter $m_\Phi^2$ (see Table.~\ref{tab:sm_pars}). 
It is worth emphasizing here that there is certain freedom in the definition of \emph{input} parameters.
In principle, one can write down the SM predictions for a set of 18 observables (\eg physical particle masses or cross-sections at fixed kinematics) that can be measured in experiments.  With the account of loop corrections the predictions become non-trivial functions of \emph{all} the Lagrangian parameters. By means of PT it is possible to invert these relations  and express these primary parameters in terms of the chosen measured quantities. This allows us to \emph{predict} other \emph{observables in terms of} a finite set of measured \emph{observables}\footnote{One can even avoid the introduction of \emph{renormalizable} parameters and use \emph{bare} quantities at the intermediate step.}.

However, it is not always practical to strictly follow this procedure. For example, 
due to confinement we are not able to directly probe the strong coupling $g_s$ and usually treat it as a scale-dependent parameter $(4\pi) \alpha_s = g_s^2$ defined in the modified minimal-subtraction ($\overline{\text{MS}}$) scheme. 
It is customary to use the value of $\alpha^{(5)}_s(M_Z)=0.1181\pm0.011$ at the $Z$-mass scale as an input for theoretical predictions. 
A convenient choice of other input parameters is presented in Table.\ref{tab:sm_pars}. 
\begin{table}[t]
\caption{Parameters of the SM.}
\label{tab:sm_pars}
\centering\small
\begin{tabular}{rccccccc}
									18=       & 1     & 1   & 1    & 1         & 1     & 9     & 4        %& 1 
									\\
									primary:  & $g_s$ & $g$ & $g'$ & $\lambda$ & $m^2_{\Phi}$ & $y_f$ & $y_{ij}$ %& $\theta$ 
									\\    
									practical: & $\alpha_s$ & $M^2_Z$  & $\alpha$ & $M^2_H$  & $G_F$ & $m_f$ & $V_{CKM}$ %& $\simeq 0$

\end{tabular}
\end{table}
It is mostly dictated by the fact that the parameters from the ``practical'' set are measured with better precision than the others. 

At the \emph{tree} level one can write % them to f the initial Lagrangian parameters as
\begin{align}
								\begin{matrix}
												\alpha_s  = \frac{g_s^2}{4 \pi}, & (4 \pi) \alpha = g^2 g'^2/(g^2 + g'^2), &  M^2_Z = \frac{(g^2 + g'^2) v^2}{4},  \\[0.4cm]
												{G_F  = \frac{1}{\sqrt 2 v^2}}, & M^2_h = 2 \lambda v^2 = 2 |m_\Phi|^2, &  m_f = y_f v/\sqrt 2.
								\end{matrix}	
								\label{eq:sm_pars_relations}
\end{align}
The relations  are modified at higher orders in PT and perturbative corrections turn out to be mandatory if one wants to confront theory predictions \cite{Bardin:1999ak, Arbuzov:2005ma, Montagna:1998kp} with high-precision experiments. 
A simple example to demonstrate this fact comes from the tree-level ``prediction'' for the $W$-mass $M_W$. 
From \Eq\eqref{eq:WZ_masses} and \Eq\eqref{eq:sm_pars_relations} we can derive 
\begin{align}
				\frac{G_F}{\sqrt 2} = \frac{\pi \alpha}{2 M_W^2 (1 - M_W^2/M_Z^2)}.
				\label{eq:MW_prediction_tree}
\end{align}
Plugging recent PDG \cite{Tanabashi:2018oca} values
\begin{equation}
				\alpha^{-1}  = 137.035 999 139(31), ~ M_Z = 91.1876(21)~\text{GeV}, ~ G_F = 1.166 378 7(6) \times 10^{-5}~\text{GeV}^{-2},
				\label{eq:sm_input}
\end{equation}
in \Eq\eqref{eq:MW_prediction_tree}, one can predict 
\begin{align}
				M_W^{tree}  = 80.9387(25)~\text{GeV},
\end{align}
where only uncertainties due to the input parameters \eqref{eq:sm_input} are taken into account. 
\begin{figure}[h]
\centering\includegraphics[scale=0.7]{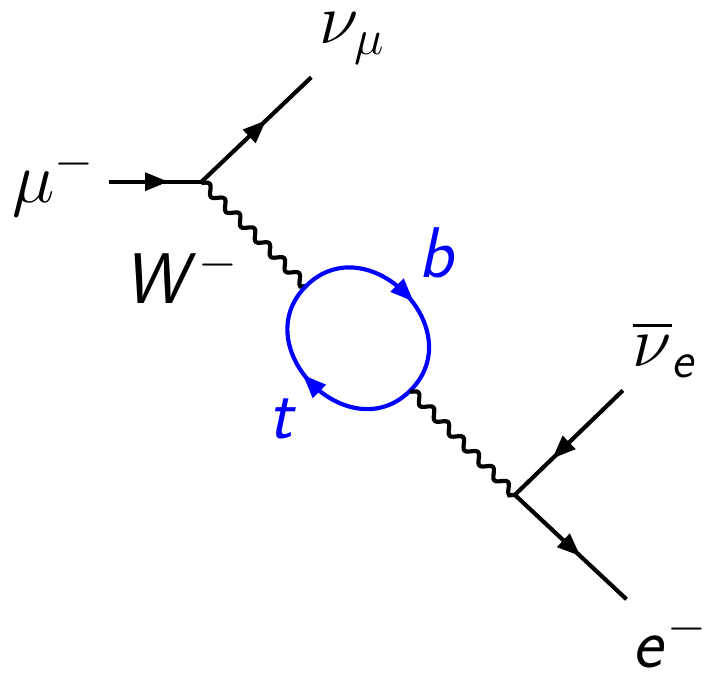}
\caption{An example of loop corrections to the muon decay, which give rise to the modification
of the tree-level relation in \Eq\eqref{eq:MW_prediction_tree}.}
\label{fig:mu_decay_corrections}
\end{figure}
			Comparing $M_W^{tree}$ with the measured value $M_W^{exp} = 80.379(12)$ GeV, one can see that
			our naive prediction is off by about $47\sigma$! Of course, this is not the reason to abandon the SM.
			We just need to take radiative corrections into account (see, \eg Fig.\ref{fig:mu_decay_corrections}). Among other things the latter allows one to \emph{connect} phenomena at 
			different scales in the context of a single model. %and it turns out that the SM indeed provides such a description. 

			A modern way to obtain the values of the SM parameters is to perform a global fit 
			to confront state-of-the-art SM predictions with high-precision experimental data. Due to quantum effects, we can even probe New Physics that can contribute to the SM processes at low energies via virtual states. 
			\begin{figure}[ht]
\centering\includegraphics[scale=0.4]{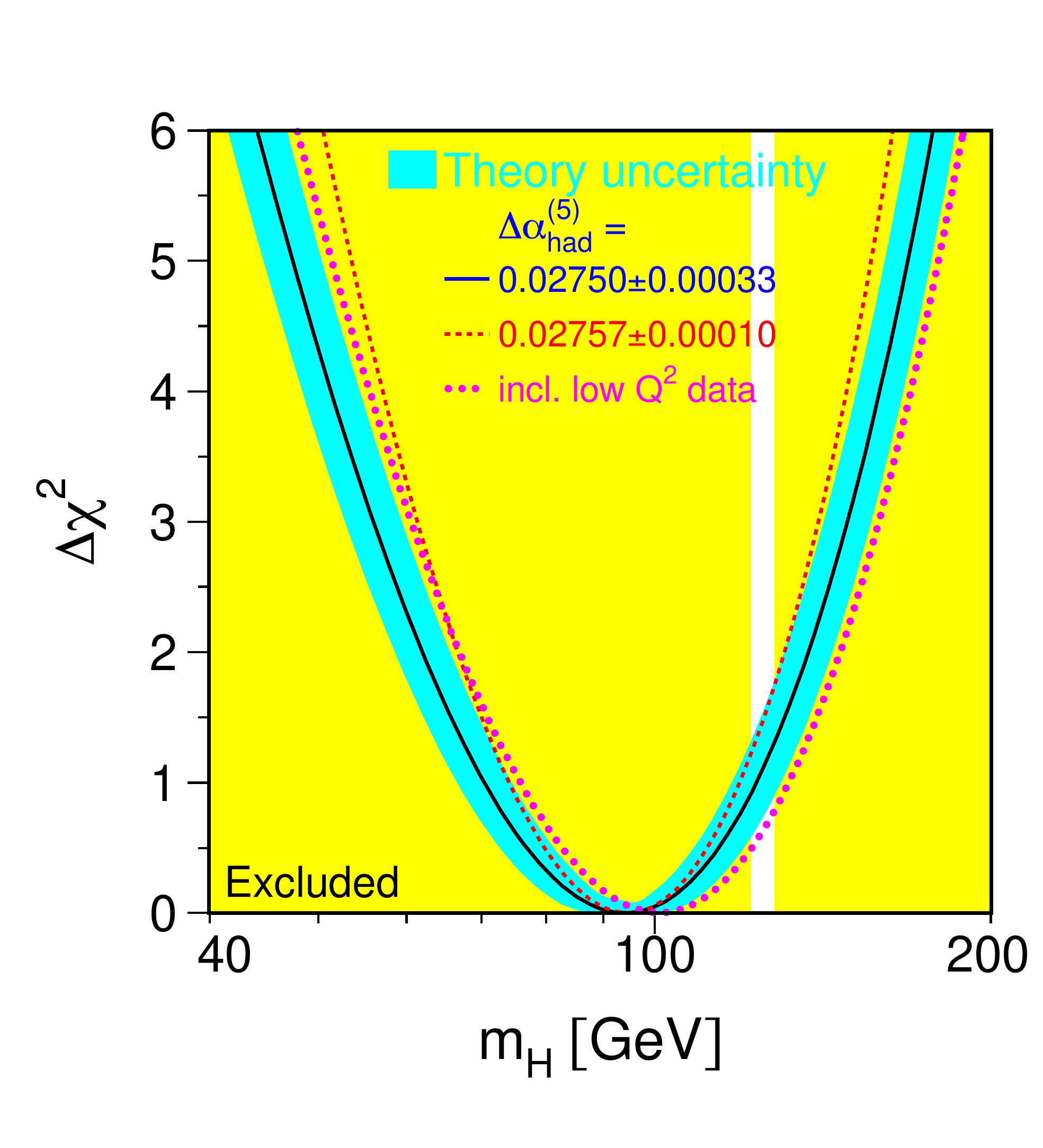}
\caption{The dependence of $\Delta\chi^2_{\rm{min}}(M_H^2)=\chi^2_{\rm{min}}(M_H^2)-\chi^2_{\rm{min}}$
on the value of $M_H$. The width of the shaded band around the curve shows the theoretical uncertainty. 
Exclusion regions due to LEP and LHC are also presented.}
\label{fig:blue_band}
\end{figure}
Indeed, LEP precision measurements interpreted in the context of the SM were used in a multidimensional parameter fits to predict the mass of the top quark $M_t$ (''New Physics''), prior to its discovery at the Tevatron. After $M_t$ was measured it was included in the fit as an additional constraint, and the same approach led to the prediction of a \emph{light} Higgs boson. In Fig.\ref{fig:blue_band}, the famous \emph{blue-band} plot by the LEP Electroweak Working Group (LEPEWWG \cite{LEPEWWG}) is presented. It was prepared a couple of months before the official announcement of the Higgs-boson discovery. One can see that the best-fit value 
			corresponding to $\Delta \chi_{min}^2=0$ lies just about $1\sigma$ below the region \emph{not} excluded by LEP and LHC.
\begin{figure}
				\begin{center}
								\begin{tabular}{cc}
\raisebox{0.07\height}
{
\includegraphics[width=.4\linewidth]{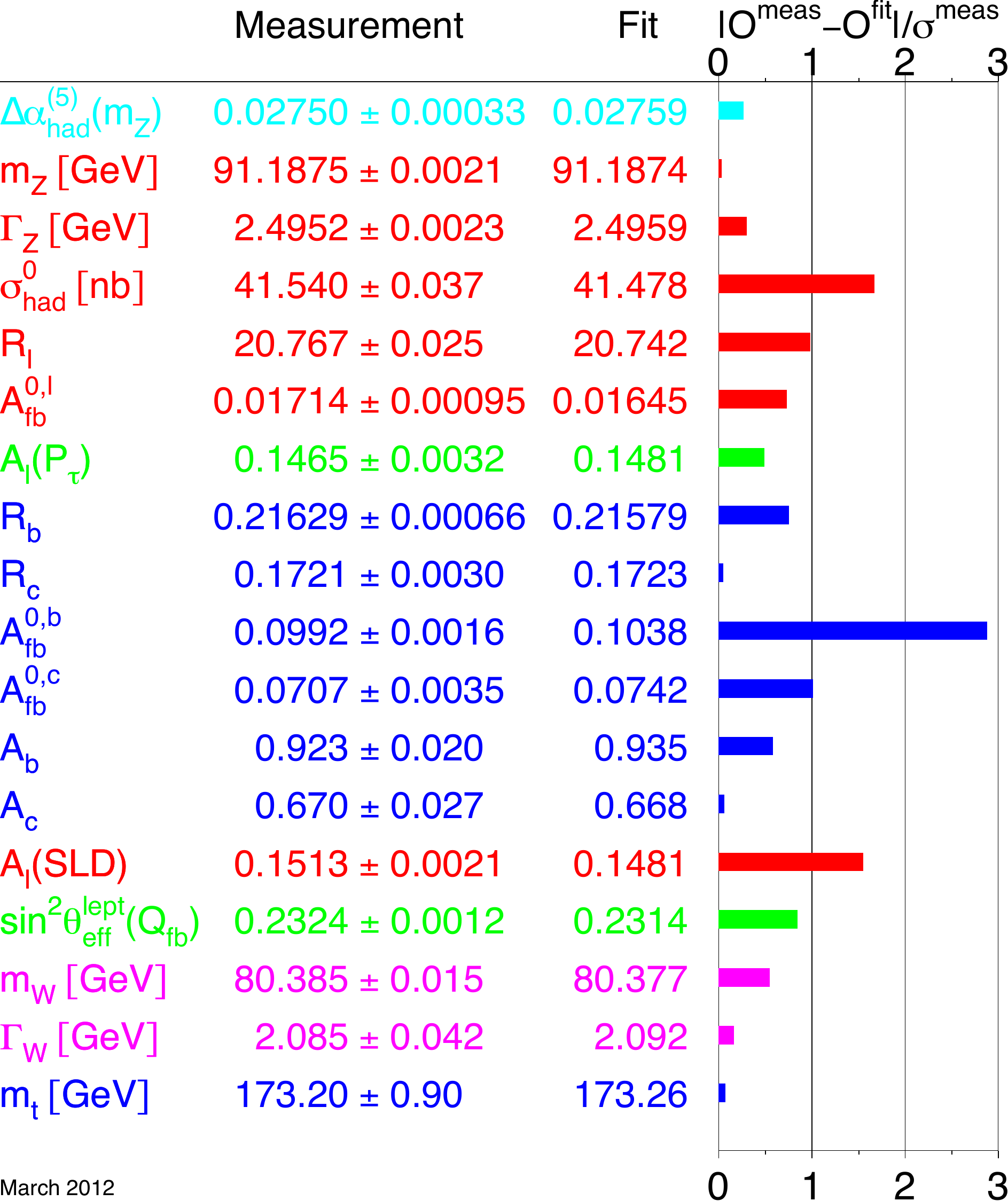}
}
&
\includegraphics[width=.3\linewidth]{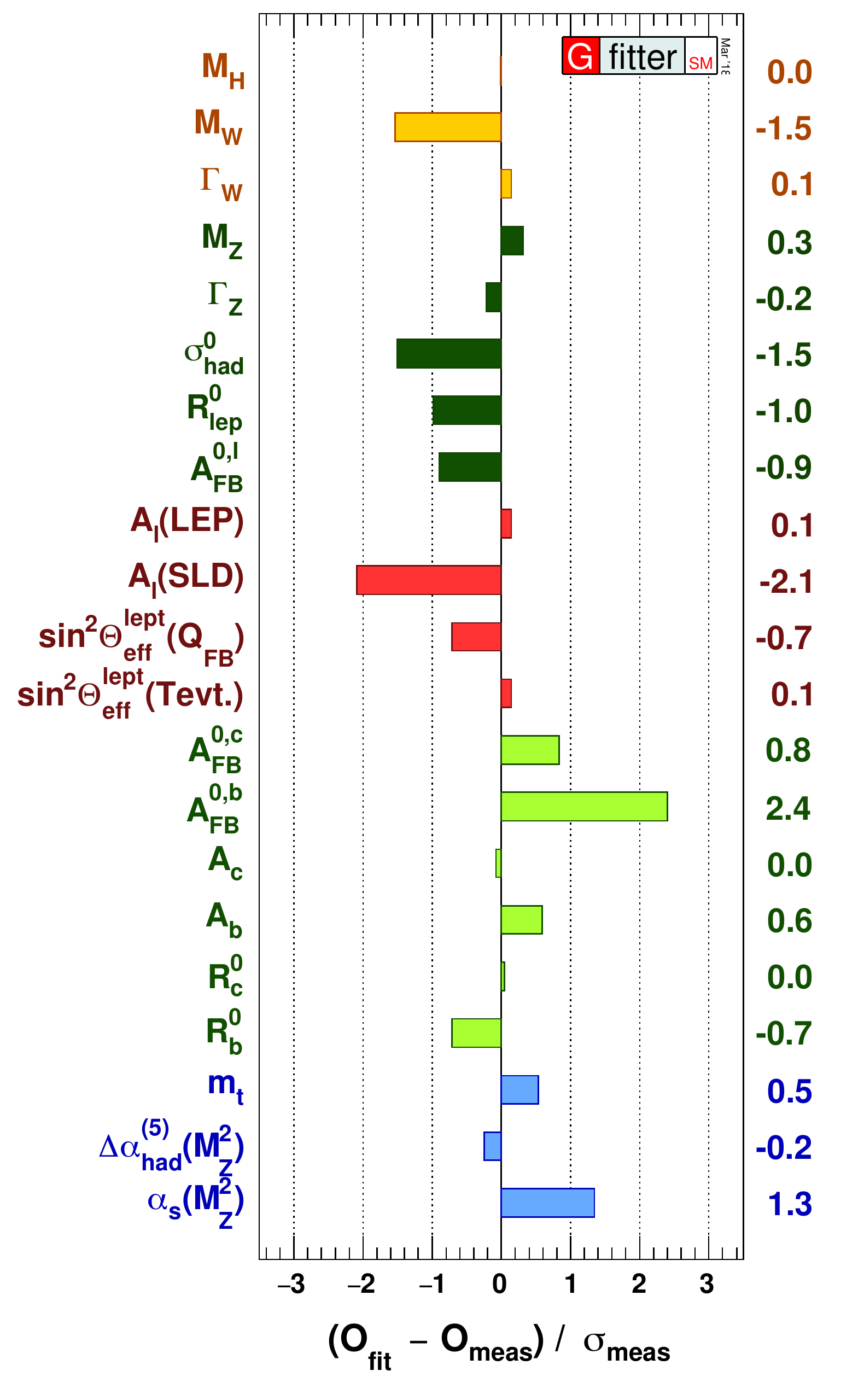}\\
												(a) & (b)
								\end{tabular}
\end{center}
\caption{Pulls of various (pseudo)observables due to (a) LEPEWWG \cite{LEPEWWG} and (b) Gfitter \cite{Baak:2014ora}.}
\label{fig:pulls}
\end{figure}

\begin{figure}[h]
\centering\includegraphics[width=.7\linewidth]{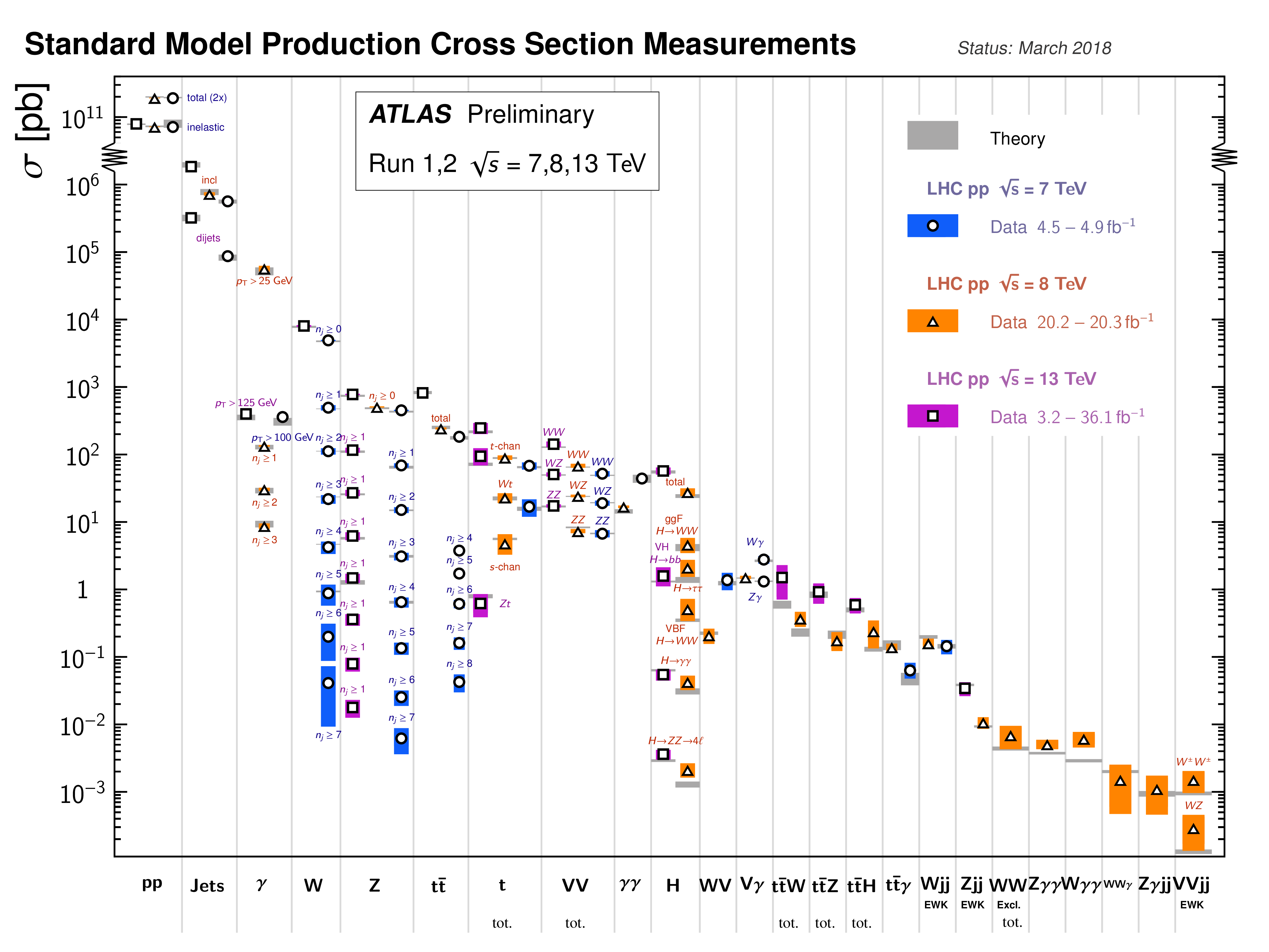}
\caption{ATLAS results of the SM cross-section measurements.}
\label{fig:ATLAS_xsec}
\end{figure}

\begin{figure}[h]
\centering\includegraphics[width=.7\linewidth]{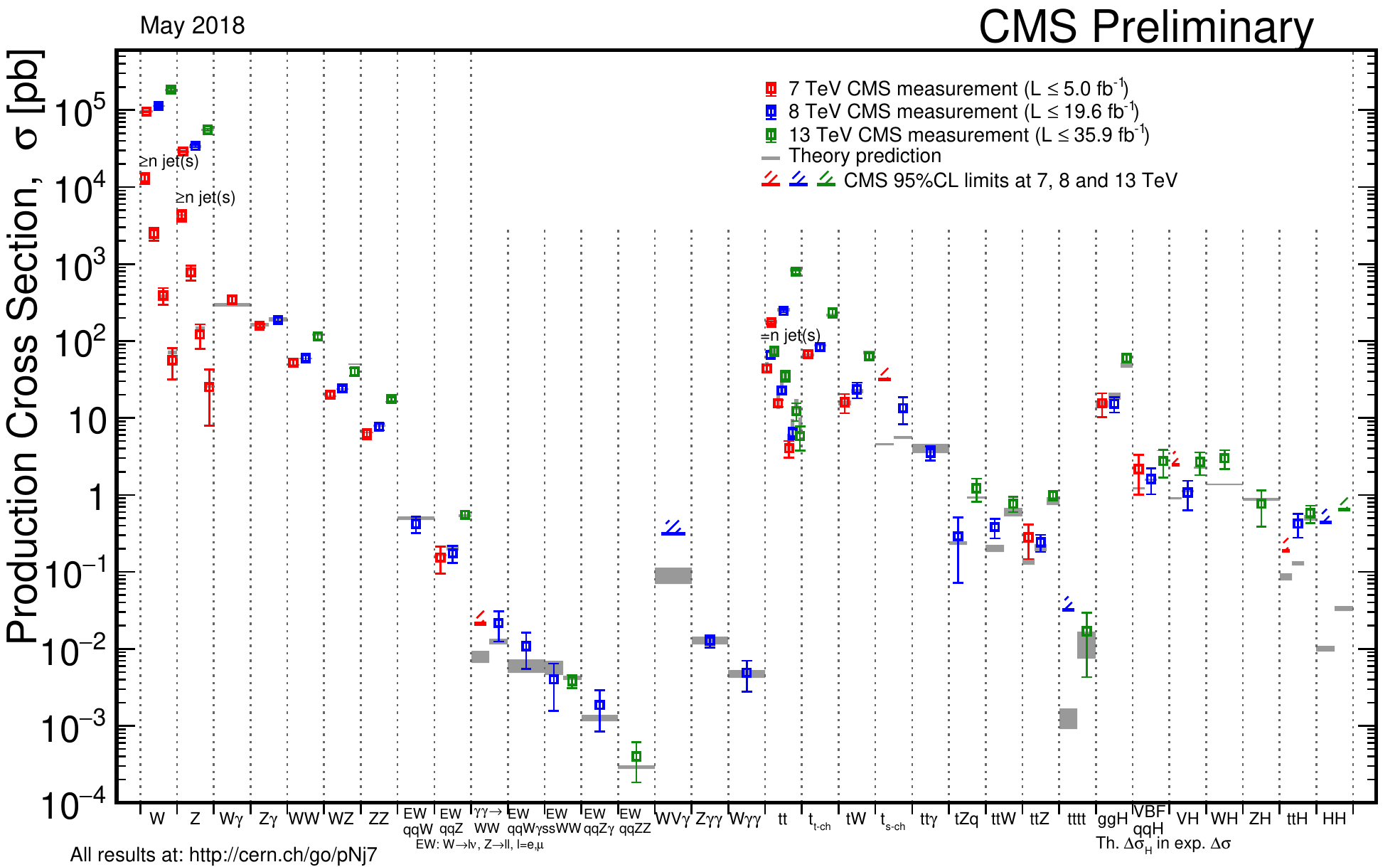}
\caption{SM processes at CMS.}
\label{fig:CMS_xsec}
\end{figure}

\begin{figure}[th]
\centering\includegraphics[width=0.5\linewidth]{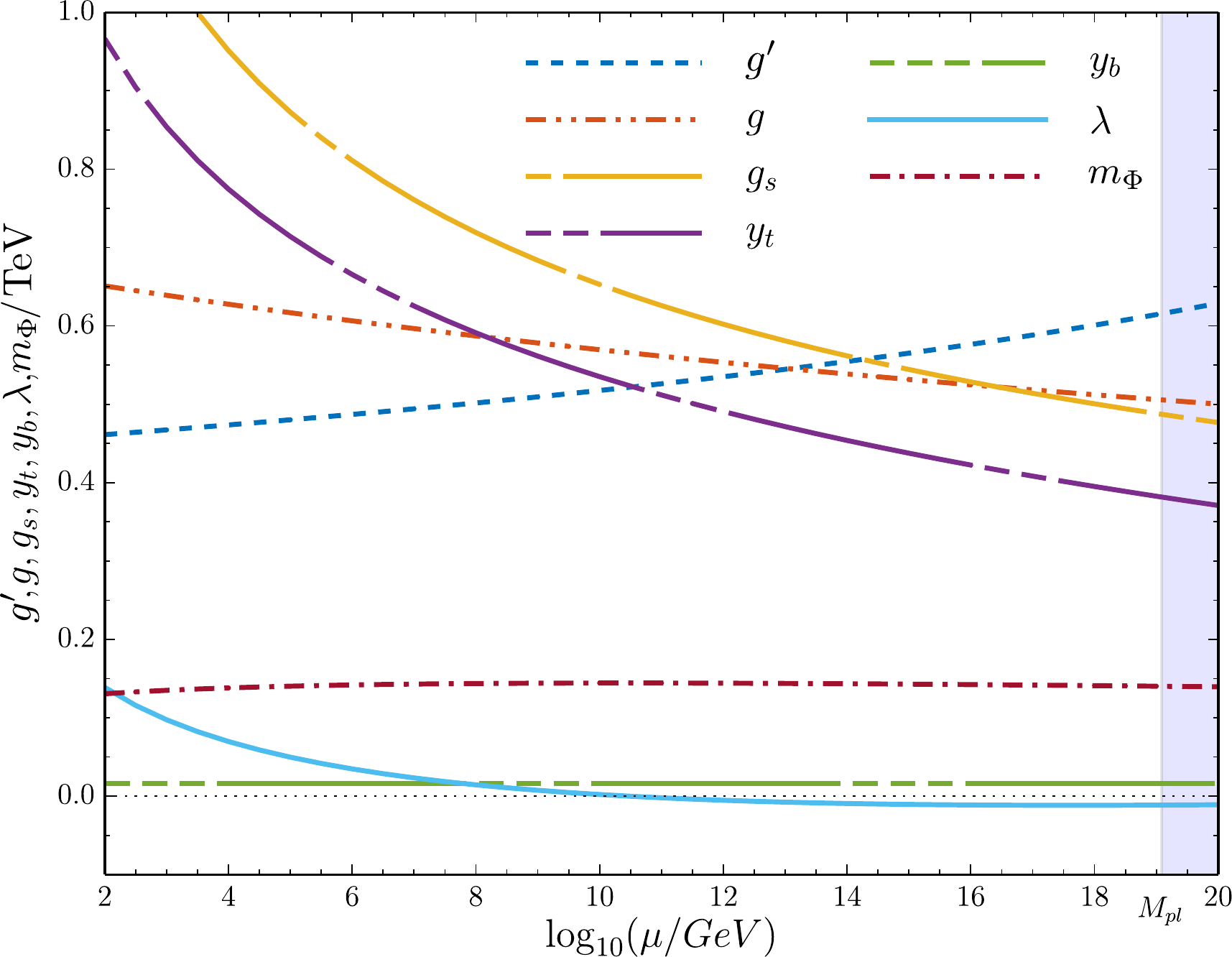}
\caption{Scale dependence of the SM parameters obtained by means of \texttt{mr} package \cite{Kniehl:2016enc}.}
\label{fig:sm_running}
\end{figure}

			Obviously, at the moment the global EW fit is \emph{overconstrained} and can be used to test overall consistency of the SM. In Fig.~\ref{fig:pulls} we present the comparison between measurements of different (pseudo)observables $O^\text{meas}$ and the SM predictions $O^{\text{fit}}$ corresponding to the best-fit values of fitted parameters. 
			Although there are several quantities where \emph{pulls}, \ie deviations between the theory and experiment, reach more than two standard deviations, the average situation should be considered as extremely good. 
			A similar conclusion can be drawn from the recent Figs.~\ref{fig:ATLAS_xsec} and \ref{fig:CMS_xsec}, in which experimental results for various cross-sections measured by ATLAS and CMS are compared with the SM predictions.   
			In case one is interested in the behavior of the SM at ultra-high energies, 
			it is more convenient to get back to the primary parameters and use the renormalization group to estimate how they change with scale. In Fig.~\ref{fig:sm_running}, the scale dependence of the SM parameters  is presented.  
One can see that the gauge couplings tend to converge to a single value at about $10^{13-15}$ GeV, thus  providing a hint for Grand Unification. Another important consequence of this kind of studies is related to the EW vacuum (meta)stability (see, \eg \cite{Bednyakov:2015sca}). In Fig.~\ref{fig:sm_running}, it manifests itself at the scale $\mu\simeq 10^{10}$ GeV, at which the self-coupling $\lambda$ becomes negative, making the tree-level potential unbounded from below.

\section{Conclusions}
\label{sec:conclusion}

Let us summarize and discuss briefly the pros and cons of the SM. 
The Standard Model  has many nice features: 
				\begin{itemize}
								\item it is based on Symmetry principles: Lorentz + ${SU(3)_C}\times {SU(2)_L} \times {U(1)_Y}$ gauge symmetry;
								\item it is renormalizable and unitary;  
								\item the structure of all interactions is fixed (but not all couplings are tested experimentally);
								\item it is an anomaly-free theory;
								\item it can account for rich Flavour Physics (see \cite{Zupan_lect});
								\item three generations allow $\mathcal{CP}$-violation (see \cite{Zupan_lect}); 
								\item it can be extended to incorporate neutrino masses and mixing (see \cite{Pascoli_lect}); 
								\item it allows making systematic predictions for a wide range of phenomena at different scales; 
								\item all predicted particles have been discovered experimentally;
								\item it survives stringent experimental tests.
				\end{itemize}
			Due to this, the SM is enormously successful (\emph{Absolutely Amazing Theory of Almost Everything}).
			Since it works so well, \emph{any} New Physics should reproduce it in the low-energy limit. Unfortunately, contrary to the Fermi-like non-renormalizable theories, 
			the values of the SM parameters do not give us obvious \emph{hints} for a New Physics scale.  But why do we need New Physics if the model is so perfect?  
			It turns out that we do not \emph{understand}, why the SM works so well.  
			For example, one needs to clarify the following:

\begin{itemize}
				\item What explains the pattern behind Flavour Physics (hierarchy in masses and mixing, 3 generations)?
				\item Is there a symmetry behind the SM (electric) charge assignment?  
				\item What is the origin of the Higgs potential?
				\item What is the origin of accidental Baryon and Lepton number symmetries?
				\item Why is there no CP-violation in the strong interactions (strong CP problem)\footnote{The 
								SM Gauge group allows such a term in the SM Lagrangian, 
								$\Lag \ni \theta_{CP} \frac{1}{16\pi^2} F^a_{\mu\nu} \tilde F^a_{\mu\nu}$. But it turns out that $\theta_{CP} = 0$.}?
				\item Why is the Higgs-boson mass so low? (Hierarchy/Naturalness  problem, see \cite{Maltoni_lect})  
				\item Is it possible to unify all the interactions, including gravity?
\end{itemize}
			In addition, there are phenomenological problems that are waiting for solutions and probably require introduction of some New Physics:  
			\begin{itemize}
							\item Origin of neutrino masses (see \cite{Pascoli_lect});
							\item Baryon asymmetry (see \cite{DeSimone_lect});	
							\item Dark matter, Dark energy, Inflation (see \cite{DeSimone_lect});
							\item Tension in $(g-2)_\mu$, $b\to s\mu\mu$, $b\to c l\nu$;
							\item Possible problems with Lepton Universality of EW interactions (see \cite{Zupan_lect,Allanach_lect}).
			\end{itemize}

			In view of the above-mentioned issues we believe that the SM is not an ultimate theory 
			(see \cite{Allanach_lect}) and enormous work is ongoing to prove the existence of some New Physics. 
			In the absence of a direct signal a key role is played by \emph{precision} measurements,
			which can reveal tiny, yet significant,  deviations from the SM predictions.
			The latter should be accurate enough (see, \eg \Ref\cite{Blondel:2018mad}) to compete with modern and future experimental precision \cite{Dam:2016ebi}. 

			To conclude, one of the most important \emph{tasks} in modern high-energy physics is to find the scale  at which the SM breaks down. There is a big chance that some new physical phenomena will eventually manifest themselves in the ongoing or future experiments, thus  allowing us to single out viable model(s) in the enormous pool of existing NP scenarios.  

\section*{Acknowledgements}    
I would like to thank the organizers for the invitation to participate in the School as a lecturer. I am also grateful to the School participants for interesting and illuminating discussions. The support from the Grant of the Russian Federation Government, Agreement No. 14.W03.31.0026 from 15.02.2018 is kindly acknowledged.

\end{document}